\documentclass[11pt,letterpaper]{article}
\pdfoutput=1
\usepackage{braket}
\usepackage{amsmath}
\usepackage{amssymb}
\usepackage{array}
\usepackage{bm}
\usepackage[pdftex,colorlinks,citecolor=blue]{hyperref}
\usepackage{color}
\usepackage[numbers,sort&compress]{natbib}
\usepackage{graphics}
\usepackage{graphicx}
\usepackage{wrapfig}
\usepackage{float}
\usepackage{subcaption}
\usepackage[font=small,labelfont=bf]{caption}
\usepackage{tocloft} 
\usepackage{appendix}
\usepackage{fancyhdr}
\usepackage[utf8]{inputenc}
\usepackage{multirow}
\usepackage{pdfpages}

\def\bea{\begin{eqnarray}}
\def\eea{\end{eqnarray}}

\def\be{\begin{equation}}
\def\ee{\end{equation}}

\def\({\left(}
\def\){\right)}
\def\[{\left[}
\def\]{\right]}

\def\eqref#1{Eq.~(\ref{#1})}

\def\({\left(}
\def\){\right)}
\def\[{\left[}
\def\]{\right]}
\def\<{\left\langle}
\def\>{\right\rangle}

\newcommand{\ba}{\begin{eqnarray}}
\newcommand{\ea}{\end{eqnarray}}

\renewcommand{\v}{{\bf v}}

\newcommand\lo{\mathrel{\raise.3ex\hbox{$<$}\mkern-14mu\lower0.6ex\hbox{$\sim$}}}
\newcommand\go{\mathrel{\raise.3ex\hbox{$>$}\mkern-14mu\lower0.6ex\hbox{$\sim$}}}

\begin{document}

\includepdf[pages=-]{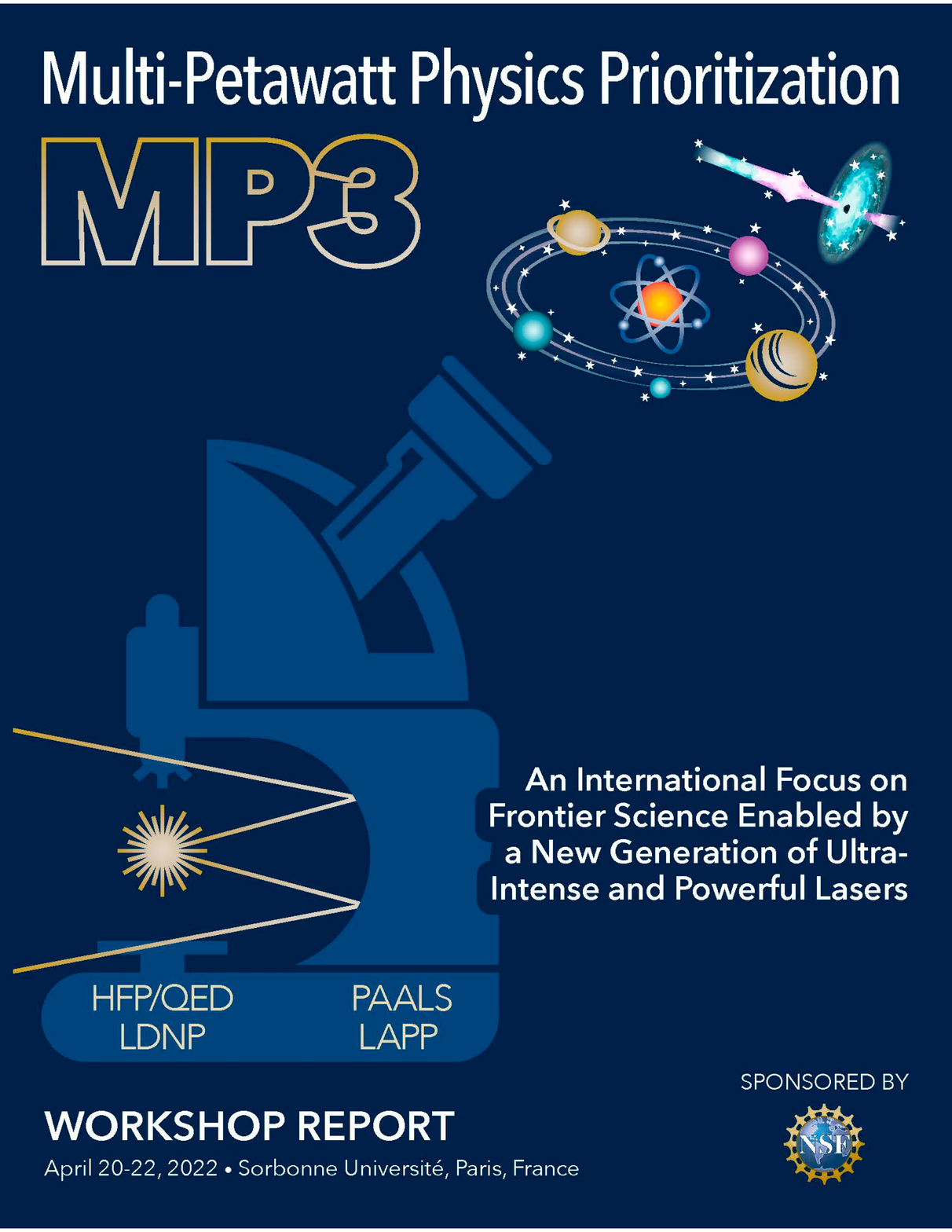}
\includepdf[pages={-}]{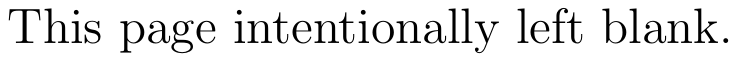}

\thispagestyle{empty}
\centerline{\bf \Large Multi-Petawatt Physics Prioritization}
\vspace{10 pt}
\centerline{\bf \Large Workshop Report}

\vspace{20 pt}
\begin{center}

\textbf{Co-Chairs:}\\
\vspace{11 pt}

Antonino Di Piazza (Max Planck Institute for Nuclear Physics, Germany)\\
Louise Willingale (University of Michigan, USA)\\
Jonathan Zuegel (University of Rochester, USA)\\

\vspace{30 pt}

\textbf{Working Group Leaders:}\\
\vspace{11 pt}
Alexey\ Arefiev (University of California San Diego, USA)\\
Sudeep Banerjee (Arizona State University, USA)\\
Tom Blackburn (University Gothenburg, Sweden)\\
Stepan Bulanov (Lawrence Berkeley National Laboratory, USA)\\
Gianluca Gregori (University of Oxford, UK)\\
Calvin Howell (Duke University, USA)\\
Hye-Sook Park (Lawrence Livermore National Laboratory, USA)\\
Markus Roth (TU Darmstadt, Germany)\\
Gennady Shvets (Cornell University, USA)\\
Klaus Spohr (ELI-NP, Romania)\\
Kazuo Tanaka (ELI-NP, Romania)\\
Dmitri Uzdensky (Colorado University, Boulder, USA)\\
Scott Wilks (Lawrence Livermore National Laboratory, USA)\\
Eva Zurek (University of Buffalo, USA)\\
\end{center}

\pagebreak
\thispagestyle{empty}
\hspace{0pt}
\vfill
\noindent
This study is based on work supported by the National Science Foundation under Award 2112770 awarded 1 February 2021.\\
\\
This report was prepared as an account of work sponsored by agencies of the U.S. Government. Neither the U.S. Government nor any agency thereof, nor any of their employees, makes any warranty, express or implied, or assumes any legal liability of responsibility for the accuracy, completeness, or usefulness of any information, apparatus, product, or process disclosed, or represents that its use would not infringe privately owned rights. Reference herein to any specific commercial product, process, or service by trade name, trademark, manufacturer, or otherwise does not necessarily constitute or imply its endorsement, recommendation, or favoring by the U.S. Government or any agency thereof. The views and opinions of authors expressed herein do not necessarily state or reflect those of the U.S Government or any agency thereof.\\
\\
Any opinions, findings, conclusions, or recommendations expressed in this publication do not necessarily reflect the views of any agency or organization that provided support for the project.
\vfill
\hspace{0pt}
\pagebreak

\newpage
\pagenumbering{roman}
 \setcounter{page}{1}



\setlength{\cftbeforesecskip}{5pt}
{\hypersetup{linkcolor=black}
\tableofcontents
}

\hypersetup{linkcolor=blue}
\hypersetup{urlcolor=blue}



\newpage
\pagenumbering{arabic}
\setcounter{page}{0}
\includepdf[pages={-}]{figures/Blank_Page.pdf}

\part*{Executive Summary}
\label{sec-exec-summary}
\addcontentsline{toc}{part}{Executive Summary and Recommendations}

This Multi-Petawatt Physics Prioritization (MP3) Workshop Report captures the outcomes from a community-initiated workshop held April 20-22, 2022 at Sorbonne Université in Paris, France. The MP3 workshop aimed at developing science questions to guide research and future experiments in four areas identified by corresponding MP3 working groups:
\begin{itemize}
\item high-field physics and quantum electrodynamics (HFP/QED), 
\item laboratory astrophysics and planetary physics (LAPP),
\item laser-driven nuclear physics (LDNP), and
\item particle acceleration and advanced light sources (PAALS).
\end{itemize}
The PAALS working group focused on the unique laser-driven particle and photon sources that will enable the science questions (SQs) identified by other working groups. 

A year-long series of virtual working group and all-hands meetings reviewed concepts, developed science themes, and evolved SQs derived from 97 white papers solicited from 265 registered MP3 community members. The combined in-person and Zoom participation for the MP3 workshop totaled 154 people. Participants enjoyed seeing long-time colleagues and meeting new ones in person after a long hiatus imposed by the COVID-19 pandemic.\\

\begin{figure}
    \centering
    \includegraphics[scale=0.7]{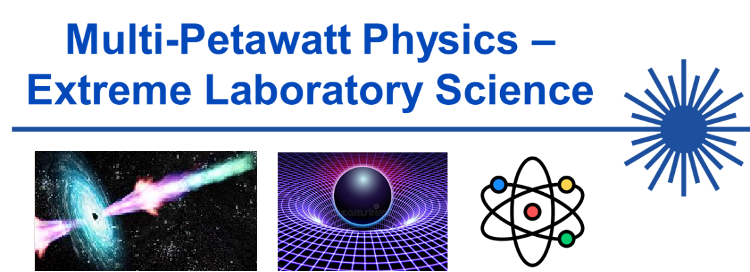}
    \label{fig:MP3_summary_figure}
\end{figure}

\noindent

\noindent
\textbf{Science Theme 1}: Highest-energy phenomena in the universe -- High-power laser facilities now enable unprecedented focused intensities and field strengths in the laboratory. Experiments using multi-petawatt laser pulses promise access to extraordinary field strengths exceeding the threshold where quantum-electrodynamics (QED) effects play an important role in the dynamics of the system.
\begin{itemize}
\item SQ1A – How might multi-petawatt lasers reveal the physical mechanisms that produce the most energetic particles and brightest events in the universe?
\item SQ1B – How does light transform into ``plasma fireball'' composed of matter, antimatter and photons?
\end{itemize}

\noindent
\textbf{Science Theme 2:} The origin and nature of space-time and matter in the universe -- 
Current and future facilities can create matter and radiation at extreme conditions so that on a tiny scale one may replicate a huge variety of environments known to exist in the visible universe.
Some experiments do not produce results predicted from first-principles calculations showing the importance of testing theoretical predictions. Multi-petawatt lasers can generate these conditions of ultra-high temperature, pressure, and acceleration, as well as new diagnostic probes to interrogate them. 
\begin{itemize}
\item SQ2A – How do complex material properties and quantum phenomena emerge at atomic pressures and temperatures relevant to planetary cores?
\item SQ2B – How can multi-petawatt lasers study black hole thermodynamics through the link between gravity and acceleration?
\item SQ2C – How does the electromagnetic interaction behave under extreme conditions?
\end{itemize}

\noindent
\textbf{Science Theme 3:} Nuclear astrophysics and the age/course of the universe -- Nuclear physics using intense lasers can open new frontiers in scientific research. Laser-driven sources of energetic particles, such as protons and neutrons, and photons can induce nuclear reactions, probe nuclear physics, and enable practical applications.
\begin{itemize}
\item SQ3A – What can be learned about heavy-element formation using laser-driven nucleosynthesis in plasma conditions far-from-equilibrium?
\item SQ3B – How can high-flux gamma sources generated from multi-PW lasers be used to explore hadronic physics (low-energy QCD)?
\end{itemize}
High-energy particle and radiation sources based on multi-petawatt lasers with unique capabilities for addressing the science questions posed in this report constitutes a core area of research, as well as enabling new applications based on these sources.

\medskip
Chapters \ref{sec-ST1}, \ref{sec-ST2}, and \ref{sec-ST3} address the MP3 science themes and each of the SQs using a common approach: (1) road mapping science goals and flagship experiments for each SQ using and building on existing capabilities; and (2) identifying any missing critical needs, like diagnostics, theory, computation. Chapter \ref{sec-PAALS} outlines underlying research to produce needed high-energy particle and photon sources, while Chapter \ref{sec-strategies} summarizes collaborative frameworks and joint strategies to realize needs for meeting roadmap goals, and Chapter \ref{sec-future} presents a vision for next-generation facility capabilities.

\section{Introduction to the Multi-Petawatt Physics Prioritization (MP3) Workshop}
\label{sec-intro}

Multi-petawatt laser systems can produce light pressures in the exa-Pascal regime, copious amounts of photons and extremely bright beams of energetic particles up to multi GeV energies including X-rays, electrons, ions, neutrons, or antimatter. These novel capabilities enabled by multi-PW lasers, described in a series of recent reports shown below, open new frontiers in research and development, such as high-field physics and nonlinear quantum electrodynamics (QED), laboratory astrophysics, and laser-driven nuclear physics.


\subsection{MP3 Workshop Charge, Organization and Process}
\noindent

The top-level charge of the Multi-Petawatt Physics Prioritization (MP3) workshop, sponsored by the National Science Foundation, includes:

\begin{itemize}
    \item defining key science goals and flagship experiments;
    \item identifying joint strategies for developing diagnostics; and
    \item discussing a vision for the optimal next-generation facility.
\end{itemize}

The MP3 workshop process employed an atypical approach adapted to the circumstances of the COVID-19 pandemic. It used virtual networking tools to engage a broad range of global experts to prepare for the in-person workshop. The chairs first established the MP3 process and tools, and then recruited and engaged working group (WG) leaders in January to March 2021. The process organized around four topics:\
\begin{itemize}
\item[]\textbf{High-Field Physics and Quantum Electro-Dynamics (HFP/QED)}\\
\textit{Alexey Arefiev (UCSD)}\\
\textit{Tom Blackburn (U. Gothenburg)}\\
\textit{Stepan Bulanov (LBNL)}\\
\textit{Dmitri Uzdensky (CU Boulder)}
\item[]\textbf{Laboratory Astro/Planetary Physics (LAPP)}\\
\textit{Gianluca Gregori (Oxford)}\\
\textit{Hye-Sook Park (LLNL)}\\
\textit{Eva Zurek (U. Buffalo)}
\item[]\textbf{Particle Acceleration and Advanced Light Sources (PAALS)}\\
\textit{Sudeep Banerjee (ASU)}\\
\textit{Gennady Shvets (Cornell Univ.)}\\
\textit{Scott Wilks (LLNL)}
\item[]\textbf{Laser-Driven Nuclear Physics (LDNP)}\\
\textit{Calvin Howell (Duke Univ.)}\\
\textit{Markus Roth (TU Darmstadt)}\\
\textit{Kazuo Tanaka (ELI-NP)}
\end{itemize}

The chairs solicited white papers addressing the WG topics and invited participants to join one or more WGs that aligned with their interests by signing up for the MP3 mailing list. The MP3 mailing list includes 265 individuals self-identifying in the HFP/QED (118), LAPP (91), PAALS (132), and LDNP (71) working groups. The MP3 mailing list includes participants from Europe (135), North America (110), and Asia (22). White paper submissions totaled 97 that aligned with the HFP/QED (51), LAPP (27), PAALS (42), and LDNP (13) working groups.\\

WG leaders led a series of April to June 2021 virtual WG meetings where participants discussed the white papers and identified preliminary science questions that were summarized at a “working group cross-check meeting” in June 2021. Chairs and WG leaders refined these science questions during a summer break in the process and presented distilled questions into three science themes that were presented at an “MP3 all-hands” meeting in early November 2021 and further refined.
\begin{enumerate}
\item \textbf{Highest-energy phenomena in the universe}
\begin{itemize}
    \item How might multi-petawatt lasers reveal the physical mechanisms that produce the most energetic particles and brightest events in the universe?
    \item How does light transform into a ``plasma fireball'' composed of matter, antimatter, and photons?
\end{itemize}
\item \textbf{The origin and nature of space-time and matter in the universe}
\begin{itemize}
    \item How do complex material properties and quantum phenomena emerge at atomic pressures and temperatures relevant to planetary cores?
    \item How can multi-petawatt lasers study black hole thermodynamics through the link between gravity and acceleration? 
    \item How does the electromagnetic interaction behave under extreme conditions?
\end{itemize}
\item \textbf{Nuclear astrophysics to understand the age of the universe}
\begin{itemize}
    \item What can be learned about heavy-element formation using laser-driven nucleosynthesis in plasma conditions far-from-equilibrium?
    \item How can high-flux gamma sources generated from multi-PW lasers be used to explore Hadronic Physics (Low Energy QCD)?
\end{itemize}
\end{enumerate}

COVID-19 pandemic surges led to postponement of the three-day, in-person workshop until April 20-22, 2022. Additional working group meetings prepared material for consideration at the workshop, including summaries of the working group findings and a draft workshop report.

The first day of the MP3 workshop started with a plenary session to kick off the workshop with opening remarks by the chairs. It included plenary sessions to report the WG findings and draft recommendations by WG leaders for all three science themes after an opening keynote talk, as noted in Table \ref{speakers}. 
The second day started with a plenary session summarizing key points from Day 1 and charged breakout groups with the goals for parallel, all-day breakout sessions. Each breakout group reviewed the draft workshop report materials in three areas:
\begin{enumerate}
    \item roadmap each MP3 science question using and building on existing capabilities, and defining any new needs;
    \item identify any missing critical needs (diagnostics, theory, computation, etc.); and
    \item develop collaborative frameworks to realize them for meeting roadmap goals.
\end{enumerate}
The third day started with a plenary session summarizing key points from Day 2 that was followed by an open discussion for any long-term needs beyond what can be realized with existing or upgraded facilities.
\medskip
Workshop chairs and working group leaders met after closing the workshop to define the process for finalizing the workshop report. Working group leaders updated summary material and forwarded it the co-chairs for review. Two co-chairs were assigned primary writing and editing duties for each section. Upon completion, all sections were reviewed by all chairs and distributed to working group leaders for their feedback. Chairs resolved any concerns that were identified during this review. 
\begin{center}
\begin{table}[h]
\begin{tabular}{|m{5cm}|m{10.2cm}|}
\hline
Topic & Speaker\\
\hline \hline
Multi-petawatt science and \newline underlying physics  &  \textbf{Keynote:} \newline Prof.~Stuart Mangles (Imperial College London) \\
\hline
Highest-energy phenomena \newline in the universe  &  \textbf{Summary:} \newline Prof.~Thomas Blackburn,
(Univ.\ Gothenberg, Sweden), \&  \newline Prof.~Dmitri Uzdensky, (Univ.\ Colorado, Boulder, USA) \newline \newline \textbf{Plenary:} \newline Prof.~Mattias Marklund 
(Univ.\ Gothenberg, and Secretary General for Natural and Engineering Sciences, Swedish Research Council, 
Sweden) \\
 \hline
Probing matter and space \newline from the microscopic scale to \newline their emergent (collective) \newline behavior  &  \textbf{Summary:} \newline Prof.~Gianluca Gregori,
(Oxford University, UK)	\newline \newline \textbf{Plenary:} \newline Prof.~Lu\'{i}s Silva
(Inst.\ Superior Técnico, Portugal) \\

\hline
Nuclear astrophysics to \newline understand the age of the \newline universe &  \textbf{Summary:} \newline Dr.~Klaus Spohr
(ELI Nuclear Physics, Romania) \newline \newline \textbf{Plenary:} \newline Prof.~Norbert Pietralla
(TU Darmdstadt and Managing Director of the Inst.\ for Nuclear Physics, Germany)\\

\hline
\end{tabular}
\caption{Speakers who presented in the MP3 workshop plenary sessions}
\label{speakers}
\end{table}
\end{center}

\subsection{Overview of the global MPW status}

\begin{figure}[h]
    \centering
    \includegraphics{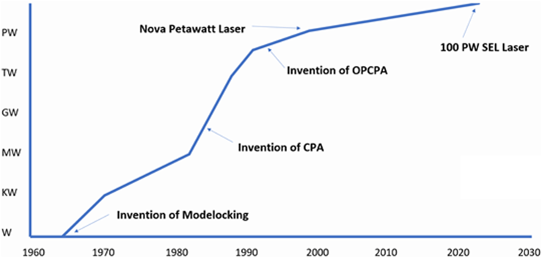}
    \caption{Record ultra-short-pulse laser  peak powers at since the invention of the laser. [From Ref.\ \cite{danson.hplse.2019}]}
    \label{fig:section1_2}
    \end{figure}

Multi-petawatt physics experiments using ultraintense lasers have started emerging now that laser technology has recently crossed the multi-petawatt threshold. Ref. \cite{danson.hplse.2019} provides a recent overview of petawatt- and exawatt-class lasers worldwide and includes Fig. \ref{fig:section1_2} that illustrates the technological progress leading to this new regime. 

\begin{figure}[h]
    \centering
    \includegraphics[width=0.99\linewidth]{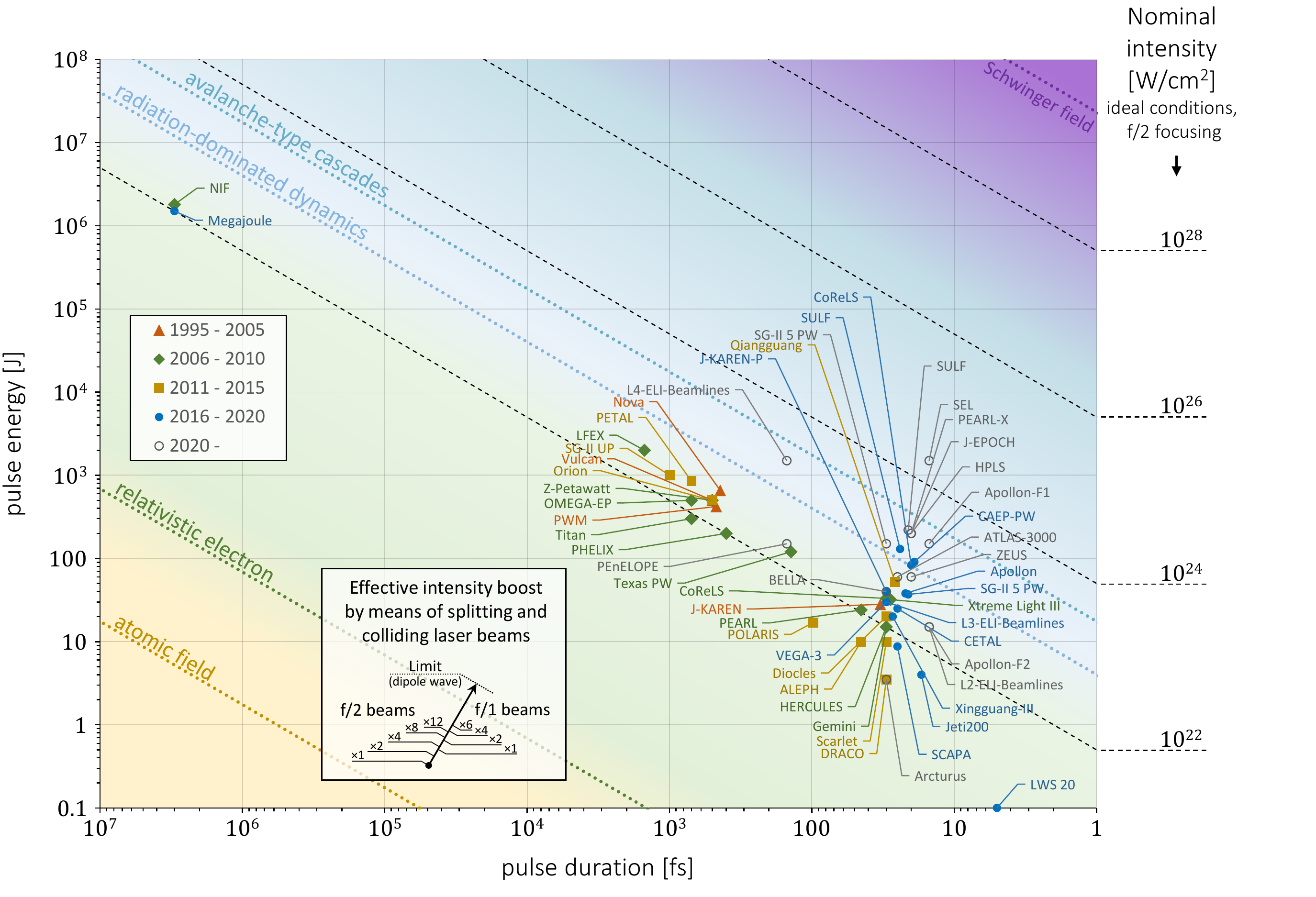}
    \caption{The existing, being built, and planned laser facilities in the (pulse energy, pulse duration) plane, according to Ref.\ \cite{danson.hplse.2019} and the data from individual facilities. In order to illustrate the potential reach of these facilities, the nominal peak intensity levels (marked with dotted lines) are estimated (the laser pulse is assumed to be Gaussian temporally and spatially, with a wavelength of $\lambda = 1 \; \mu$m and focused by an $f/2$ off-axis parabola). The intensification due to the splitting of the laser power among the specified number of beams and colliding them so that the electric field is summed up coherently is shown in the insert. Reproduced from Gonoskov et al., \cite{gonoskov.rmp.2022}.}
    \label{fig:lasers}
    \end{figure}

Numerous petawatt-class lasers have operated around the world (see Fig. \ref{fig:lasers}). 
The development of chirped-pulse amplification (CPA) by Strickland and Mourou \cite{Strickland:1985c} in 1985 enabled rapid progress in the ensuing decades. Implementation of optical parametric chirped-pulse amplification (OPCPA) continued this progress with ever increasing peak powers since it can support broader bandwidth for shorter pulses, as well as higher pulse energies. The pace of advances slowed as technological limits arose, most notably due to the large diffraction gratings required to compress highly energetic, ultrashort laser pulses. Commissioning lasers and experimental systems to full specifications takes careful and deliberate steps, especially as the facility scale increases.

Table \ref{table_MPW_lasers} notes multi-petawatt laser facilities that have come online, started construction, or planned/proposed along with their ultimate MPW peak power and the start or expected year operating. 

Chapters \ref{sec-ST1}-\ref{sec-ST3} of this workshop report present science questions and grand challenges in three research themes enabled by this new generation of ultra-intense and powerful lasers – high-field physics and quantum electrodynamics (HFP/QED), laboratory astrophysics and planetary physics (LAPP), and laser-driven nuclear physics (LDNP). Chapter \ref{sec-PAALS} presents research and development required to produce particle acceleration and advanced light sources (PAALS) using multi-petawatt lasers needed for the grand-challenge experiments. Research will exercise the petawatt and the multi-petawatt laser facilities existing now and those to come.

\begin{center}
\begin{table}[h]
\begin{tabular}{|m{4.7cm}|m{4cm}|m{5cm}|m{1cm}|}
\hline
Facility (country) & MPW Peak Power &	Laser Technology &	Year\\
\hline\hline
\multicolumn{4}{|c|}{Operating (actual start of MPW operations)} \\
\hline
CORELS (South Korea) & 4.2 PW (83 J/19 fs) & Ti:sapphire CPA (0.1 Hz) & 2017\\
SULF (China) & 12.9 PW (551J/23 fs) & Ti:sapphire CPA (shot/min) & 2019\\
ELI-NP HPLS (Romania)	& 10.2 PW (230 J/23 fs) & Ti:sapphire CPA (shot/min) & 2022\\
\hline \hline
\multicolumn{4}{|c|}{Under construction / commissioning (expected full commissioning year)}\\
\hline
Apollon (France) & 10 PW (180 J/18 fs) & Ti:sapphire CPA (shot/mins) & 	2023\\
ELI-ALPS HF (Hungary) & 2 PW (34 J/17 fs) & Ti:sapphire CPA (10 Hz) & 2023\\
NSF ZEUS (USA) & 3 PW (75 J/25 fs) & Ti:sapphire CPA (shot/min) &  2023\\
ELI-Beamlines L4 (Czechia)	& 10 PW (1.5 kJ/150 fs) & OPCPA-glass (shot/mins) & 2023\\
\hline \hline
\multicolumn{4}{|c|}{Planned / Proposed (expected full commissioning year)} \\
\hline
SEL (China) & 100 PW (1.5 kJ/15 fs) & 	All-OPCPA (shot/mins) & $\sim$2027\\
EP-OPAL (USA) & 2 $\times$ 25 PW (500 J/20 fs) & All-OPCPA (shot/5 mins) & $\sim$2029\\
Vulcan 20-20 (UK)	& 20 PW (400 J/20 fs) & All-OPCPA (shot/5 mins) & $\sim$2029\\
\hline
\end{tabular}
\caption{\label{table_MPW_lasers} Multi-petawatt lasers worldwide.}
\end{table}
\end{center}

\section{Science Theme 1: Highest energy phenomena in the universe}
\label{sec-ST1}

\subsection{Science Question 1A: How might multi-petawatt lasers reveal the physical mechanisms that produce the most energetic particles and brightest events in the universe?}
\label{subsec-SQ1A}


\subsubsection{Introduction}
\label{subsubsec-SQ1A-intro}

\smallskip
\noindent {\bf Extremely energetic, ultrarelativistic particles pervade the Universe and produce very high-energy gamma-rays.}
        
Extremely energetic ultrarelativistic particles, often with power-law distributions covering several decades in particle energy, pervade the Universe and emit high-energy gamma-rays.
Astrophysical systems producing these particles usually involve relativistic compact objects -- neutron stars (NSs) and black holes (BHs) --- and their relativistic plasma outflows, such as pulsar magnetospheres and pulsar wind nebulae or spectacular relativistic jets driven by supermassive BHs in active galactic nuclei (AGN), including blazars; they can also involve dramatic cosmic stellar explosions, like supernovae (SN) and Gamma-Ray-Bursts (GRBs).  Perhaps the most notable example of extremely energetic particles is cosmic rays (CRs) --- ultra-relativistic protons with energies up to $10^{20}$~eV. While most CRs with moderate energies $\lesssim 10^{15}$-$10^{16}$~eV are believed to be produced by Galactic sources such as SN shocks, the most energetic particles, including, ultra-high-energy CRs (UHECRs, $E \gtrsim 10^{18}$~eV), require extragalactic origin, with AGN jets and GRBs being the most plausible sources. 

In addition to direct detection of CRs on Earth, extremely relativistic charged particles can also manifest themselves observationally in other ways. For example, accelerated electrons and positrons can produce very high-energy gamma-ray emission, e.g., the TeV emission observed from SN shocks or ultra-rapid TeV flares from blazars, or as MeV-GeV emission, also often rapidly flaring, from many classes of astrophysical NS or BH objects. In addition, multi-PeV CR protons in, e.g., AGN jets are capable of producing, via various hadronic processes, PeV neutrinos that are detected with the NSF's IceCube neutrino observatory. Along with Cosmic Rays (and also gravitational waves), these neutrinos open up the rapidly developing frontier of Multi-Messenger Astronomy.

How these particles are accelerated in Nature to their extremely high energies has been an outstanding scientific question in Plasma Astrophysics over many decades, driving copious observational, theoretical, and computational research. It is generally believed that relativistic particles are accelerated by nonlinear {\it collective plasma processes}; the most important examples are magnetic reconnection, collisionless shocks, and turbulence, taking place in highly dynamic relativistic plasma environments around NSs and~BHs. These plasma processes are complex: they are nonlinear, often involve nontrivial kinetic physics requiring full 6D phase-space treatment, and are typically characterized by huge (many orders of magnitude) scale separations between global macroscales and fundamental kinetic plasma microscales. All these aspects make numerical modeling and developing theoretical physical understanding of these processes and of their associated particle acceleration a profound challenge of modern plasma physics. 

Laboratory studies of these astrophysically relevant collective plasma processes could in principle be of great help, as they could provide invaluable fundamental physics insights into the underlying mechanisms of relativistic particle acceleration and strongly impact astrophysics and fundamental plasma physics research. Dedicated experimental campaigns, including those employing powerful lasers, have started and show great promise, especially in regard to developing the necessary experimental platforms and diagnostics with some  already starting to yield unique physics insights. 

First-principles, kinetic (particle-in-cell, or PIC) numerical simulations and analytical theory have shown that nonthermal particle acceleration to extremely high relativistic energies becomes especially effective (resulting in harder power laws, greater high-energy cutoffs) when the underlying collective plasma processes occur in the so-called {\it relativistic regime}, in which magnetic energy density greatly exceeds the rest-mass energy density of the plasma. Most of today's laser-based experiments studying basic collective plasma processes (such as those using Omega-EP, NIF, Vulcan, etc.) employ relatively long pulses with only modest intensities making it difficult to achieve relativistic plasma conditions, which makes them not very well-suited for studies of relativistic plasma processes.

Experiments using ultra-intense, short-pulse lasers available today already access the relativistic-electron regime and enable pilot laboratory studies of collective plasma processes in semi-relativistic plasmas (relativistic electrons but sub-relativistic ions) under the conditions similar to those in accreting BH coronae, although only at the electron scales.

Direct experimental investigation of truly relativistic collective plasma processes initially requires using a relativistic electron-positron pair plasma as its basic platform, since making ions relativistic in an electron-ion plasma is even more difficult. {\it Creating a relativistic pair plasma in the lab} represents a top priority for High-Energy-Density plasma physics in general and for the next generation of laser facilities in particular. Both collisionless and optically thin relativistically hot pair plasma, and a dense, optically thick, collisional photo-leptonic fireball (see SQ~1B) would be of great interest from the astrophysics point of view.

Just creating a cloud of relativistic electrons and positrons is not enough. In order for such a platform to be useful for studies of collective plasma phenomena, the relativistic pair plasma needs to be ``macroscopic", --- i.e., have a sufficiently large spatial extent and sufficiently long life time, --- so that it could, in fact, be capable of exhibiting collective behavior. 
In particular, the plasma size should exceed the basic plasma kinetic scales such as Debye length, collisionless skin depth, etc., preferably by a factor of 10 to 100. Accessing this demanding regime requires not just a very high ($\gtrsim 10^{23}~{\rm W/cm}^2$) laser intensity, but also a large (kJ) overall laser pulse energy that could enable a sufficiently large transverse spot size (mm) and long duration~(ps). Next-generation multi-PW lasers coming on-line in the next decade can achieve this ambitious goal, as discussed below. 

One particularly interesting key aspect of this quest, especially important for studies of high-energy particle acceleration by astrophysically-relevant plasma processes, is the need to generate very strong magnetic fields that are coherent and long-lasting on macroscopic scales. Indeed, charged particle acceleration is ultimately accomplished by the electric field, whose magnitude, in a macroscopic plasma exhibiting relativistic bulk motions, can approach a sizable fraction (e.g., $\sim$\,0.1 in collisionless relativistic reconnection) of the magnetic field. The maximum attained particle energy, which is limited by the work done by the electric field on the particle, is thus controlled by the product of the electric (and hence magnetic) field strength and its coherence scale. Likewise, the same product controls the maximum energy of particles that can be confined by the magnetic field within the system. For example, accelerating electrons in magnetic reconnection to $\gamma=100$ over a $10\,\mu$m accelerating length by an $E_{\rm rec} \simeq 0.1 B_0$ reconnection electric field would require a reconnecting magnetic field of at least $B_0 \gtrsim 10^6\,{\rm T} = 10^{10}\,{\rm G}$ (ignoring radiation losses).  
Producing coherent fields of such strength is not really feasible now, but should be possible with future next-generation lasers reaching intensities of order~$10^{24}\,{\rm W/cm}^{2}$, requiring laser power [assuming a $(10\,\mu {\rm m})^2$ active spot area] of order~$10^3$\,PW or an exawatt (EW) (see \S~\ref{sec-SQ1B-roadmap-next_gen}).

Strong coherent plasma magnetic fields of 
$10^{5}-10^{7}\,{\rm T} = 10^{9}-10^{11}\,{\rm G}$, which are needed to study relativistic particle acceleration via collective plasma processes and which are expected to be produced in such next-generation multi-PW class laser facilities, are impressive even by astrophysical standards. They exceed the fields expected near accreting stellar-mass BHs in X-ray Binaries such as Cyg X-1 (up to $10^8$\,G) and super-massive BHs in AGN ($10^{4}$\,G). They also exceed the magnetic field at the light cylinder of the Crab pulsar ($10^6$\,G)
or near the surface of weakly-magnetized NSs in X-ray bursters ($10^8-10^9$\,G), although they are still much weaker than the surface magnetic fields of most normal~NSs, such as regular radio- and X-ray pulsars ($10^{12}$\,G), and, by an even larger margin, those near magnetars ($10^{14}-10^{15}$\,G). At the same time, $10^{9}-10^{11}\,{\rm G}$ are expected in the intermediate ($r\sim 10-100 R_{\rm NS}$) zone of magnetar magnetospheres, a hypothesized site of Fast Radio Bursts (FRBs); thus, investigating in the lab relativistic pair plasma behavior in the presence of fields of such strength may be instrumental for understanding this enigmatic phenomenon.

\subsubsection{Roadmap: progress using current facilities}

{\underline{Main idea:} \it Current facilities have the capability, for the first time, to study electron-scale collective plasma processes (like magnetic reconnection) in the semi-relativistic regime, where the ions are subrelativistic but electrons can become ultra-relativistic, with some radiation effects. This regime is of great interest in high-energy astrophysics, specifically, for understanding electron energization processes in accretion flows and coronae around accreting black holes. However, these experiments will not yet be able to achieve QED plasma conditions, necessary for the creation of macroscopic pair plasmas.}

Current state-of-the-art, ultra-intense laser facilities (see Table \ref{table_MPW_lasers}, and a detailed list in Ref.\ \cite{gonoskov.rmp.2022}), employing $\sim 1$~PW-power laser beams tightly focused onto $\sim 1 \; \mu m^2$ (comparable to laser wavelength) spots, achieve intensities of order $10^{23}\,{\rm W cm}^{-2}$. This corresponds to a classical relativistic nonlinearity parameter of $a_0 \sim 200$, corresponding to a laser magnetic field of about~$2\times 10^6$\,T. Assuming for simplicity that the resulting coherent magnetic field generated in the plasma is about 5 times weaker than the laser field, we can reasonably hope to be able to generate plasma fields of $4\times 10^5\,{\rm T} = 4 \times 10^9$\,G.  This may enable one to accelerate electrons in a magnetic reconnection scenario, to moderately relativistic energies with a Lorentz factor of a few; however, the small hot spot size, limited by the available laser power, will not  be large enough to create a macroscopic pair plasma. Nevertheless, these intensities, when used in a laser-solid-target experiments, will allow researchers to produce a small-scale, relativistically hot electron plasma embedded in a sub-relativistic, essentially stationary or slowly moving quasi-uniform ion background. This, in turn, will enable important pioneering studies of collective nonlinear processes like shocks and magnetic reconnection on small (namely, electron kinetic) scales in the semi-relativistic regime [e.g., WP-\ref{MP3_083}], opening new vistas in laboratory plasma astrophysics research and laying down the groundwork for future experiments with relativistic pair-plasma fireballs.

\subsubsection{Roadmap: theoretical understanding}

The growth in theoretical understanding over the next decade will continue benefiting from state-of-the-art 2-D and 3-D simulation studies of collective plasma processes in the relativistic and radiative regimes. Such studies have recently become possible thanks to the advent of novel radiative-PIC and radiative-QED PIC codes, such as EPOCH, OSIRIS, PICLS, Zeltron, Tristan~v2, and Smilei. These codes can simulate complex kinetic plasma processes from first principles, while taking into account all relevant ``exotic" radiation and QED physics. Codes like EPOCH, OSIRIS, Smilei, and Tristan~v2 include the basic QED processes of nonlinear Compton scattering and nonlinear Breit-Wheeler pair production, whereas PICLS and OSIRIS also include effects related to photon-photon scattering and vacuum polarization, respectively. These novel capabilities, combined with ever-growing computing resources, will put understandings of astrophysically-relevant collective plasma processes in relativistic and radiative plasmas on a rigorous, firm theoretical basis. 
The parameter regime of interest, motivated by astrophysical applications, is, however, very broad and multi-dimensional; thus, exploring it comprehensively with rather expensive large-scale 3D kinetic simulations, and then analyzing these simulations, will require very substantial resources, both in terms of computing allocations and human time. This requires the community to develop a systematic, organized approach to such computational studies, which will require coordination between different research groups.

In addition to numerical simulations, further progress will require concerted theoretical (analytical) efforts, to guide both numerical simulation campaigns and experimental explorations. WP-\ref{MP3_009} is an example of new schemes to generate extreme electric and magnetic field conditions. Analytical theory will push the frontiers of understanding plasma-astrophysical particle-acceleration processes into new, so far relatively unexplored, extreme plasma regimes where familiar collective plasma processes interplay in nontrivial ways with radiative and QED physics. Analytical theory will also play a key role in building bridges between laboratory laser-plasma experiments, fundamental plasma physics, and phenomenological models of extreme astrophysical plasma environments around NSs and~BHs.

\subsubsection{Roadmap: flagship experiments requiring new capabilities}
\label{sec-SQ1B-roadmap-new-cap}

{\underline{Main idea:} \it New capabilities at existing facilities will feature more complex experimental configurations involving multiple beams. The resulting increased flexibility will enable studies of diverse physical regimes with a broader coverage of the parameter space. In addition, the new capabilities will open up access to the radiation-dominated regime of collective plasma processes, including some QED effects like radiation recoil.}

In the next few years, upgraded capabilities at existing laser facilities, such as new beam lines, and intensity and/or power upgrades, the ability to shoot fixed plasma targets, will add flexibility to the experimental configurations, such as arrangements involving multiple laser and particle beams. Among other benefits, these enhancements will raise the available laser beam power up to around 10 PW, allowing one to reach intensities of order $10^{23}-10^{24}\,{\rm W\,cm}^{-2}$, depending on the focusing. This, in turn, will open the new radiation-dominated regime of plasma processes to direct experimental exploration [e.g., WP- \ref{MP3_078}]. In this regime, the emitted radiation is not just a passive tracer of the plasma dynamics and energetics (still extremely useful for diagnostics) but also strongly affects them. The interplay between quantum (e.g., quantum recoil on emitting particles) and collective effects will significantly determine the dynamics of the plasma, opening for the first time the possibility to investigate such interactions in the relativistic regime.

In terms of experiments aimed at producing strong coherent plasma magnetic fields, the commissioning of the L4 laser at ELI Beamlines poses an important milestone. The combination of longer pulse duration (150~fs vs. 30~fs) and high intensity ($\sim 10^{23}\,{\rm W\,cm}^{-2}$) will enable generation of volumetric plasma magnetic fields in the MT range, enabling meaningful reconnection experiments in new, strongly radiative regimes relevant to many extreme astrophysical systems described in \S~\ref{subsubsec-SQ1A-intro}. Reconnection experiments require configurations with at least two laser pulses, which introduces additional demands for laser facility capabilities; these demands, however, will be met with the planned upgrades to the above-mentioned facilities in the next few years.

Volumetric laser-plasma interactions leading to generation of strong plasma magnetic fields coherent over relatively large scales are also likely to require specifically designed targets. These might be foam targets with a density that guarantees relativistically induced transparency for the expected laser intensity. Structured targets might be introduced to improve control over the laser-plasma interaction. Currently, such targets are expensive, so new target fabrication capabilities will need to be developed to address the cost. This aspect becomes even more important for high repetition rate laser facilities.

\subsubsection{Roadmap: flagship experiments requiring next-generation facility capabilities}
\label{sec-SQ1B-roadmap-next_gen}

{\underline{Main idea:} \it Flagship experiments on next-generation laser facilities will create a revolutionary experimental platform --- macroscopic (i.e., plasma size exceeding kinetic scales) relativistic pair-plasma fireballs --- used to study collective plasma processes in the truly relativistic regime.}

Next-generation laser facilities (Chapter \ref{sec-future}) --- those reaching laser power of several tens, perhaps up to a hundred, of PW --- will create relativistically hot pair-plasma fireballs via the avalanche-type electromagnetic cascade process using two or more very powerful laser beams [e.g., WP-\ref{MP3_072}, WP-\ref{MP3_078}]. Creating such pair-plasma fireballs of macroscopic size will be a great scientific breakthrough on its own right. How to achieve it using multi-PW-class lasers is described in Science Question~1B (\S~\ref{subsec-SQ1B}), but here we shall briefly discuss how to use these fireballs to build a new experimental platform for laboratory studies of relativistic collective plasma processes [e.g., WP-\ref{MP3_073}], and the associated particle acceleration, and what progress can be expected from such studies.

Most promising schemes will require creating two such fireballs (and hence several, at least 4 beamlines) and make them interact with each other by bringing them into direct contact (e.g., colliding them with each other). This is particularly useful for setting up laboratory studies of magnetic reconnection and shocks --- the two plasma processes that are broadly regarded as the most important astrophysical mechanisms of nonthermal particle acceleration. Lasers have already been used for studies of these processes in traditional, non-relativistic electron-ion plasmas, so some of the key guiding design principles of such experiments are understood. Extending them to studies of similar processes but in the relativistic pair-plasma regime should not cause much difficulty, especially for shock studies, provided that pair plasmas of sufficient size (preferably tens of hundreds of skin depths) can be created. If the two lasers involved in the creation of each fireball are asymmetric, then the resulting fireball can be made to fly with a relativistic speed; colliding two such fireballs can set up a scenario for studying a relativistic shock, mimicking processes that occur at the pulsar-wind termination shock or in the internal shock model of GRB prompt emission.

\subsubsection{Broader Impacts}
The study of relativistic plasma with the parameters relevant to astrophysical phenomena would require significant technological advancement of the high intensity high power laser facilities. Some initial progress can be achieved at current facilities, but in order to move forward, new capabilities at existing facilities need to be introduced and, ultimately, next-generation facilities built. This can not be achieved without concentrated efforts by scientists and engineers, academic institutions, and industry partners. It will require much broader efforts to train a new generation of specialists who will build and then use these facilities to answer the scientific question outlined in this report. New capabilities and new facilities mean not only more powerful lasers, but new and more precise diagnostic tools (Chapter 6) able to cover much wider parameter space, and new advanced targets. Last but not least, the efforts will build a robust collaboration between laser scientists, astrophysicists, and those working in the emerging field of QED plasma.   

\subsubsection{Recommendations}

Since the collective plasma processes in the semirelativistic regime, where ions are subrelativistic but electrons ultra-relativistic with some radiation effects, prove of great interest in high-energy astrophysics for understanding electron energization processes in accretion flows and coronae around accreting black holes, they should be studied at current facilities capable of carrying out such experiments. New capabilities on existing facilities will enable more complex experimental configurations involving multiple beams to provide increased flexibility and enable studies of diverse physical regimes with a broader coverage of parameter space. In addition, the new capabilities will open up access to the radiation-dominated regime of collective plasma processes, including some QED effects like radiation recoil. Flagship experiments on next-generation laser facilities will create relativistic pair-plasma fireballs and use them to study collective plasma processes in the truly relativistic regime. These experimental efforts will require accompanying theoretical and simulation advances.

These conclusions are based on the white papers received: the importance of the study of pair plasmas was emphasized in WPs-[\ref{MP3_026},\ref{MP3_027},\ref{MP3_060},\ref{MP3_065},\ref{MP3_073},\ref{MP3_078}], the study of magnetic reconnection in WPs-[\ref{MP3_066}, \ref{MP3_083}],  the study of instabilities/particle acceleration in WP-\ref{MP3_070}, and shocks in WP-\ref{MP3_075}.

\bigskip
\subsection{Science Question 1B: How does light transform into a ``plasma fireball'' composed of matter, antimatter and photons?}
\label{subsec-SQ1B}

\subsubsection{Introduction}

The creation and acceleration of high-energy particles in astrophysical environments usually require strong electromagnetic fields, highly relativistic plasmas, and interactions governed by quantum electrodynamics (QED). QED is the best verified quantum field theory with astonishing agreement between theoretical predictions and  experiments, and it represents one of the cornerstones of the Standard Model of particle physics; however, one sector of this theory has remained quite undeveloped compared to everything else: the interaction of charged particles and photons with strong electromagnetic fields, which is quite different from the usual single-particle scattering and decay processes \cite{piazza.rmp.2012,zhang.pop.2020,gonoskov.rmp.2022,fedotov.arxiv.2022} and which is usually referred to as strong-field QED (SF-QED). The basic building blocks of the description of such interactions are well known for many years \cite{ritus.jslr.1985}, but the consistent theoretical description of the interaction, which can be verified experimentally, is still under development \cite{zhang.pop.2020}. Moreover, the number of experimental campaigns addressing the challenges of particle interactions with strong electromagnetic fields has been quite limited (see, e.g., \cite{bula.prl.1996,burke.prl.1997,cole.prx.2018,poder.prx.2018}), which hopefully will change soon with the advent of MPW lasers. 

The electromagnetic (EM) field is considered strong in QED when it is on the order of thw critical (Schwinger) field of QED, $E_{cr}$. This field produces work of $m c^2=0.51\,\text{MeV}$ over an electron Compton length $\lambda_c=\hbar/mc=3.9\times 10^{-11}\,\text{cm}$, $E_{cr}=m^2c^3/\hbar e=1.32\times 10^{16}$ V/cm. Such fields can produce an electron-positron pair from vacuum, the so-called Schwinger effect. However, the critical QED field does not only characterize the Schwinger effect, but also provides a scale for the onset of quantum field theory effects in charged particle and photon interactions with electromagnetic fields. For example, the probabilities of the nonlinear Compton scattering ($e\rightarrow e\gamma$) and nonlinear Breit-Wheeler pair production ($\gamma\rightarrow -> e^+e^-$) are characterized by two parameters $\chi_e=\sqrt{|F_{\mu\nu}p^\nu|^2}/mcE_{cr}$ and $\chi_\gamma=\sqrt{|F_{\mu\nu}k^\nu|^2}/mcE_{cr}$, where $F^{\mu\nu}$ is the tensor of the strong EM field and $p^{\nu}$ and $k^{\nu}$ are the four-momenta of the electron and photon, respectively. These probabilities acquire optimal values at $\chi_{e,\gamma}\sim 1$. Both $\chi_e$ and $\chi_\gamma$ are the products of EM field tensor and the particle momentum normalized to the critical field. An electron in a strong field provides a straightforward interpretation of $\chi_e$, which is the EM field strength in electron rest frame normalized to $E_{cr}$: 

\begin{equation}
            \chi_e = \frac{\gamma E}{E_cr} =
                0.3
                \left( \frac{E}{500~\mathrm{MeV}} \right)
                \left( \frac{I}{10^{22}~\mathrm{W}\mathrm{cm}^{-2}} \right)^{1/2},
            \end{equation}
where the laser intensity and the amplitude of the vector-potential of the EM field are connected in the following way:
\begin{equation}
            a_0 = \frac{e E}{m c \omega} = 85\frac{\lambda}{\mu\mathrm{m}}\left( \frac{I}{10^{22}~\mathrm{W}\mathrm{cm}^{-2}} \right)^{1/2}.
            \end{equation}
Thus, these parameters literally compare EM field strength to the critical one, indicating the importance of quantum corrections to the particle dynamics, radiation emissions, and pair production in EM fields.

Strong EM fields can be found in different interaction setups, including close proximity of compact astrophysical objects, such as magnetars and central engines of GRBs, high-Z nuclei, dense particle beams in the interaction points of high energy particle accelerators, aligned crystals, and in the foci of high-power lasers. Whereas some of these environments provide fields of critical value, others provide fields that are below and need to be combined with high-energy particle beams (below critical fields may appear of critical strength in the reference frame of the relativistic particle beam) or fixed plasma targets to probe SF-QED effects. For example, the case for aligned crystals and in the foci of high power lasers. The latter seems to be the most attractive option due to its flexibility from the point of view of providing different interaction configurations, especially, at the multi-PW level.

While many phenomena in SF-QED have attracted a lot of attention over the last several decades, one stands really apart from the others: the so-called electromagnetic cascade, which is a multi-staged process of initial particle and/or laser pulse energy transformation into secondary electrons, positrons, and photons. Here the dynamics of particles is dominated by photon emission and pair production, the laser pulse properties might be severely affected by emerging electron-positron-photon plasma. This is the intersection of SF-QED and relativistic plasma physics. It is exactly here, where SQ1B comes from: ``How does light transform into plasma fireball composed of matter, antimatter and photons?" The most general answer to that would be ``through the electromagnetic cascade." There are two types of cascades, usually associated with either longitudinally or transversely dominated motion of electrons, positrons, and photons with respect to the laser pulse propagation direction. The former is referred to as a shower-type cascade, while the latter as an avalanche-type cascade. 

A typical setup for a shower-type cascade is a head-on collision of a high energy particle beam with a laser pulse with $\gamma\gg a_0$. In this case the laser works as a target for the particle beam. As the particles interact with the laser field they experience Compton and Breit-Wheeler effects leading to beam energy transformation into secondary electrons, positrons, and photons, which are mainly streaming in the same direction as the initial particle beam.

A typical setup for an avalanche-type cascade is the interaction of two or more colliding laser pulses with a fixed target. In this case the lasers not only accelerate charged particles, but also provide conditions for Compton and Breit-Wheeler effects, and then re-accelerate electrons and positrons either after a photon emission or after being produced in a Breit-Wheeler process. Due to the re-acceleration, the avalanche-type cascade is significantly more effective in producing secondary particles, when comparing to the shower-type cascade. This process can produce a plasma fireball consisting of matter, antimatter, and photons. Theoretical estimates show that an avalanche-type cascade starts when the interaction enters the radiation dominated regime, which can be defined as the electron (or positron) being able to emit almost all of its energy via a single Compton process. This occurs at laser intensities approaching $10^{24}$ W/cm$^2$. Multi-PW laser facilities are needed to reach such intensities; however, as was repeatedly emphasized in the literature (see \cite{gonoskov.rmp.2022} for details), the availability of tens of PW of laser power is not enough to observe the cascade. The electromagnetic field should be structured in a way that enhances the emission of photons and production of electron-positron pairs, as well as the re-acceleration. The multiple colliding laser pulses seems to offer the most advantageous and robust scheme to satisfy this requirement.  

\subsubsection{Roadmap: progress using current facilities}

As mentioned above, strong EM fields can be found in different interaction setups. Here we consider those that need to be combined with high-energy particle beams to observe quantum behavior. Experiments using aligned crystals were reported recently \cite{Wistisen_2018,Wistisen_2019}. Experiments using the beams of particle colliders are being proposed \cite{yakimenko.prl.2019} but can not be carried out yet due to the lack of accelerators with necessary parameters. Therefore, experiments involving high-power lasers provide the best immediate path forward for studying SF-QED effects. It was already demonstrated at SLAC (E144) \cite{bula.prl.1996,burke.prl.1997} and at CLF (GEMINI) \cite{cole.prx.2018,poder.prx.2018} that a high-energy electron bunch interacting with a counter-propagating laser will not only loose a lot of energy due to radiation but this radiation needs to be described in the SF-QED framework. The parameter $\chi_e$ was approximately $0.3$ for the E144 and $0.2$ for GEMINI experiments. The main difference between the SLAC and CLF experiments was that the former used the conventional SLAC accelerator to produce high energy electrons, while the latter resorted to the all-optical scheme, where the lasers are used to both accelerate electrons via laser wakefield acceleration (LWFA) and to scatter off the accelerated electrons.

Building on these results, many facilities are planning experiments that aim to increase $\chi_{e}$ or $\chi_{\gamma}$ well above unity to reach into previously inaccessible quantum regime and track the transition of the radiation reaction from classical to quantum description. For example, SLAC is running the E320, and DESY is planning the LUXE experiment \cite{Abramowicz_2019} using conventionally accelerated $10$ or $17.5$ GeV electron beams in collision with tens of terawatt laser pulses. The  University of Michigan ZEUS facility will use two laser pulses (with 2.5~PW and  0.5~PW), one to accelerate electrons (either $\gtrsim 10$ GeV or several GeV) and one to provide the electromagnetic  field (intensity $10^{21}$ W/cm$^{2}$ or $10^{23}$ W/cm$^{2}$). Other laser facilities with active SF-QED study program include J-Karen in Japan, Apollon in France, CORELS in Korea, CALA in Germany, ELI-NP in Romania, and ELI-BL in Czech Republic (for an expanded list see Ref.\ \cite{gonoskov.rmp.2022} and Fig.\ \ref{fig:lasers}).

These experiments at current facilities all share the same features. First, the experimental setup utilizes PW-class lasers colliding with GeV-class electron beams. Second, the transition of the radiation reaction description from classical to quantum will be studied. Third, the proposed experiments are supposed to test the validity and the applicability limits of different theoretical and computational models used now. Other possible applications of the laser pulse collision with a high-energy electron beam include but are not limited to gamma-ray sources, positron sources, and strong magnetic field generation for astrophysical studies (see Fig.\ \ref{fig:RoadmapSFQED}). 

\subsubsection{Roadmap: theoretical understanding}

Theoretical understanding of SF-QED processes is based on a number of approximations, including plane wave, external field, and local constant field ones. Most of them result from an inability to obtain solutions of Dirac's equation in any EM field configuration other than a plane wave, which prevents us from carrying out analytical calculations to quantize the strong field to account for the backreaction of the SF-QED processes on the field itself, and to calculate radiative corrections and multi-staged processes. In addition, one can identify the limits where SF-QED methods should fail, but were not able to formulate a self-consistent approach how to go beyond these limits.

In most cases we rely on the separation of scales, assuming that SF-QED processes happen instantaneously, i.e., their formation lengths are much smaller than the characteristic temporal and spatial scales of strong fields. This allows to consider the motion of charged particles and photons in strong fields according to the classical equations of motion between SF-QED processes (either a photon emission by a charged particle, or an electron-positron pair production by a photon). Such approach turned out to be extremely advantageous for computer modeling of SF-QED processes in different field configurations with the majority of the results obtained with the help of PIC-QED codes. 

That is why any new high power laser facility should be accompanied by a robust theoretical/simulation effort. The immediate theoretical goals are to go beyond the mentioned above approximations and develop a theoretical framework for the analysis of shower- and avalanche-type cascades, which should involve the ability to calculate rates for multi-staged processes and to take into account the backreaction of the SF-QED processes on strong EM fields. 

\subsubsection{Roadmap: flagship experiments requiring new capabilities}

It is well understood that the current facilities are quite limited in their studies of the SF-QED effects. That is why many of these facilities plan to acquire new capabilities, like new beamlines, intensity and/or power upgrades, the ability to shoot fixed plasma targets, etc. From the point of view of EM cascades development studies new capabilities should involve a multi-GeV, laser-driven source of electrons to collide with a second intense laser pulse to study shower-type cascades. The avalanche-type cascades would require approximately 20 PW of laser power in the form of colliding laser pulses to become observable. So this task might be more appropriate for the next generation laser facilities; however, the avalanche ``pre-cursors'' can be observed at lower power levels corresponding to current facilities with new capabilities \cite{magnusson.pra.2019}. These ``pre-cursors'' are characterized by the enhanced energy absorption from the laser pulse (or pulses) by charged particles, which hints towards energy loss due to photon emission and energy gain due to the re-acceleration. 

\begin{figure}[h]
    \centering
    \includegraphics[width=0.9\linewidth]{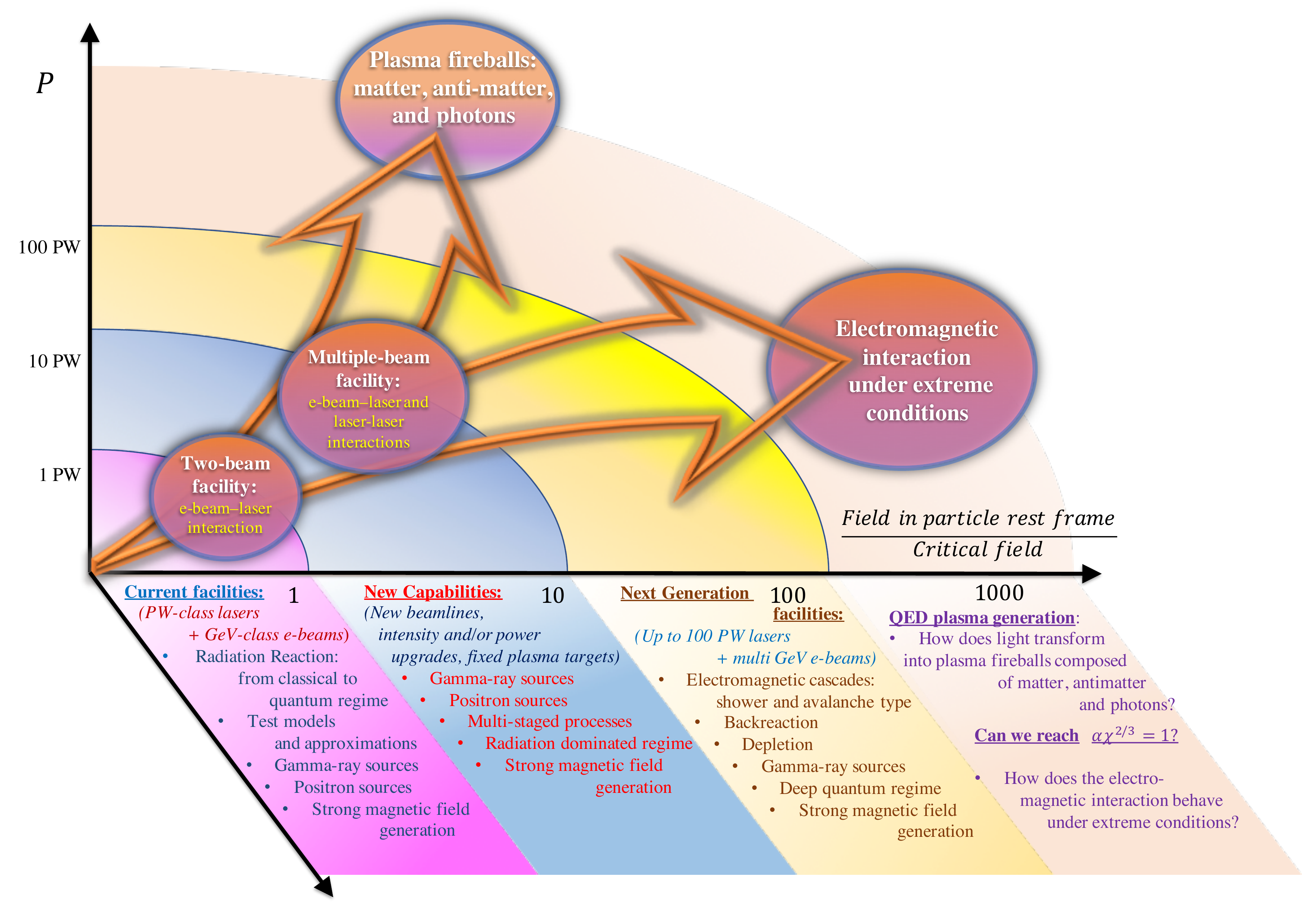}
    \caption{%
             Timeline of the HFP/QED studies leading to answering SQ1B and SQ1C envisioned as a three-stage process starting with ``current facilities" operating at PW level, then using ``new capabilities" to reach new regimes of interaction at up to 10 PW level, and then ``next generation'' facilities to study the most fundamental SF-QED phenomena at laser power level approaching 100 PW.}
    \label{fig:RoadmapSFQED}
    \end{figure}

The interaction setups relevant for the cascade studies can also be used to generate gamma-ray sources, positron sources, and strong magnetic fields (see Fig.\ \ref{fig:RoadmapSFQED}).

\subsubsection{Roadmap: flagship experiments requiring next generation laser facilities.}

\begin{figure}[h]
    \centering
    \includegraphics[width=0.45\linewidth]{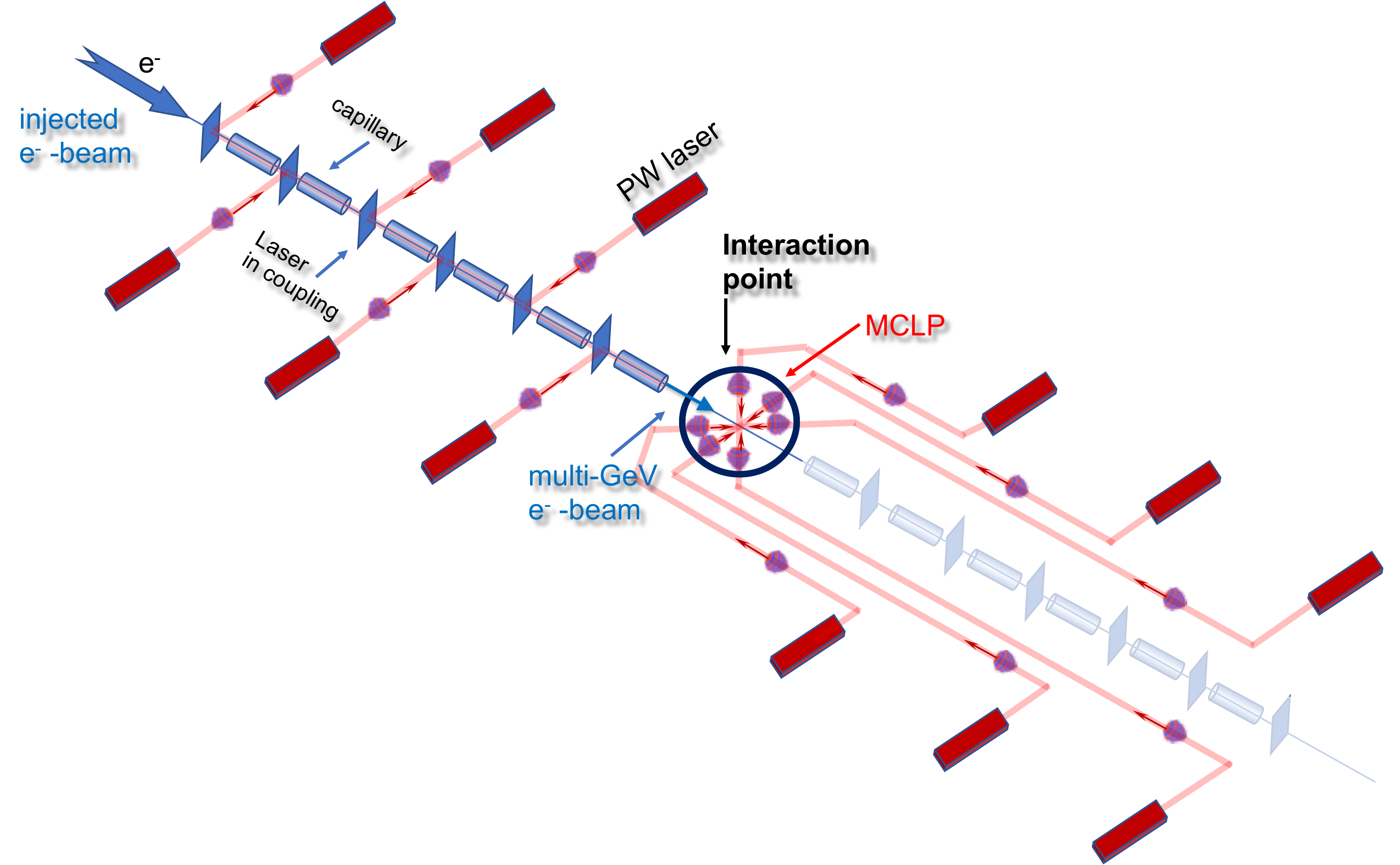}
    \includegraphics[width=0.45\linewidth]{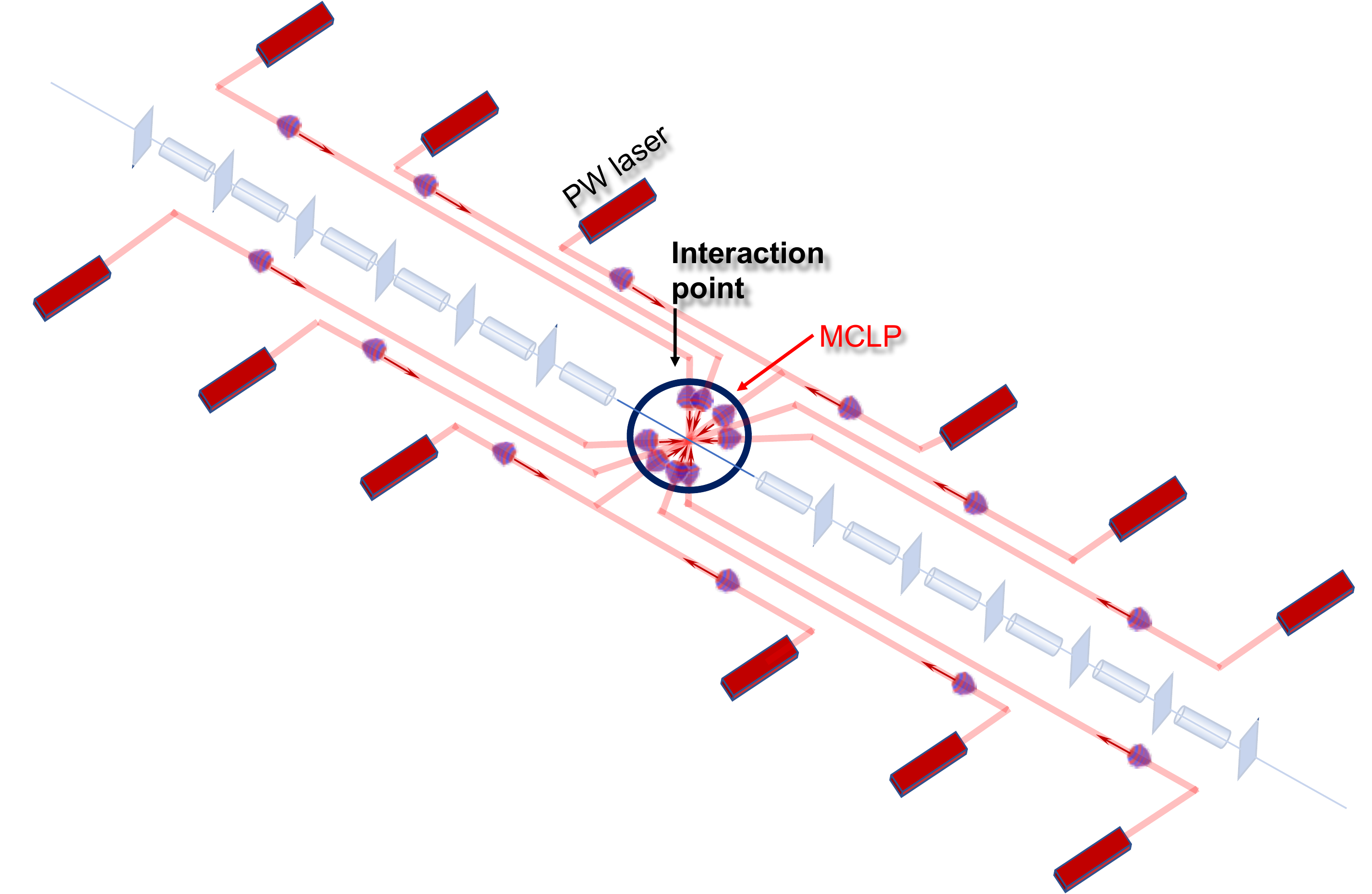}
    \caption{%
    A multi PW laser facility, which can operate in two modes:  electron-beam interaction with high intensity laser pulses for the study of shower-type cascades (left), and ultimately SQ1C and all the laser pulses are brought to the interaction point to generate highest intensity possible for the study of avalanche-type cascades and ultimately SQ1B (right). Reproduced from Zhang \textit{et al}. \cite{zhang.pop.2020}}
    \label{fig:facilitySFQED}
\end{figure}

The main experimental goal of the next-generation laser facilities from the point of view of SQ1B will be the observation and study of the avalanche-type cascade in order to advance our understanding of ``how does light transform into plasma fireball composed of matter, antimatter and photons?'' This would require a multi-beam, multi-PW facility of the type shown in Fig.\ \ref{fig:facilitySFQED}.

Such a facility design arises naturally from laser-driven particle acceleration. It is due to the fact that one of the more efficient ways of accelerating electrons and positrons is laser wakefield acceleration, which uses a staged approach \cite{Esarey_2009}. This means the utilization of many laser pulses, each powering its own acceleration stage. Alternately, these pulses can instead be simultaneously focused to produce a multi-beam configuration yielding the highest intensity. 

Due to the exponential growth of the number of electrons, positrons, and photons, the avalanche-type cascade provides a natural environment for the study of backreaction of produced lepton-photon plasma on the laser electromagnetic field and subsequent laser energy depletion. This is the domain of QED plasma physics, which is highly relevant for the astrophysical studies and almost unexplored by theory, simulations, and experiments. The coupling of high-energy physics processes with collective plasma effects will likely challenge our understanding of this state of matter and lead to significant scientific discoveries.

\subsubsection{Broader Impacts}
Significant scientific and technological efforts must be undertaken to answer the question, ``how does light transform into plasma fireball composed of matter, antimatter, and photons?'' To reach the required laser parameters, next-generation facilities need to be built that are equipped with advanced targetry, laser controls, detectors, and computational capabilities. They will also require the next generation of scientists and engineers trained at existing and upgraded facilities. 

The obvious connection of this scientific program is to the astrophysical studies (WP-\ref{MP3_089}), since the parameters of generated electromagnetic fields, plasma densities, and particle energies are of relevance to that field. Different particle and radiation sources can become available as we explore the parameter space of laser plasma interactions towards high energies and high intensities. 

Last but not least, the possible design of a next generation, multi-PW laser facility (see Fig.\ \ref{fig:facilitySFQED}) can lead to an electron-positron collider for high energy physics studies.

\subsubsection{Recommendations}

The scientific question \emph{``how does light transform into plasma fireball composed of matter, antimatter, and photons?''} lies at the heart of QED plasma physics studies. Though it can be addressed with the help of multi-PW laser facilities able to reach $10^{24-25}$ W/cm$^2$ intensities in a matter of tens of femtoseconds and accelerate charged particles to the energies exceeding 10's and 100's of GeV, this requires much progress. Current facilities need to study individual SF-QED processes to build up understanding of particle behavior in strong fields. These facilities need to be upgraded with new capabilities to make the study of the electromagnetic cascades possible. It will open up possibilities to study the plasma dynamics in a QED/radiation-dominated regime, where collective effects of the produced electrons, positrons, and photons begin to manifest themselves for QED plasma studies at next-generation, multi-PW laser facilities. The experimental effort should be accompanied by theoretical and computational advances, which would require to significantly expand the existing workforce and the resources available to it.

This conclusion is based on multiple MP3 white papers received: WPs-[\ref{MP3_013},\ref{MP3_024},\ref{MP3_025},\ref{MP3_032},\ref{MP3_041},\ref{MP3_055},\ref{MP3_061},
\ref{MP3_068},\ref{MP3_069},\ref{MP3_071},\ref{MP3_079},\ref{MP3_081}] emphasize the importance of the study of particle dynamics in strong fields,  WPs-[\ref{MP3_008},\ref{MP3_038},\ref{MP3_062},\ref{MP3_073},\ref{MP3_076}] consider the study of plasma dynamics in strong fields, and WP-\ref{MP3_029} investigates the theoretical possibility of investigating the ``duration'' of the process of pair production using muti-petawatt lasers.

The production of particles from light is probably the most striking manifestation of the interaction among electromagnetic fields in vacuum, as predicted by QED. The community has also shown interest on another still untested aspect of light-light interaction in vacuum: Intense electromagnetic fields alter the dielectric properties of the vacuum, which, according to the predictions of QED, behaves as a birefringent medium. For instance, the WPs-[\ref{MP3_018}, \ref{MP3_020} and \ref{MP3_063}] propose different setups involving either optical or x-ray photons, as probes of the birefringence properties of the vacuum as induced by a super-intense optical laser beam  (WPs[\ref{MP3_018}, \ref{MP3_020}), and interferometric techniques already tested to study the dielectric properties of materials (WP-\ref{MP3_063}). The elementary process responsible for vacuum birefringence is photon-photon scattering and the WPs-[\ref{MP3_031}, \ref{MP3_087}] aim at measuring the corresponding cross section using multi-petawatt laser beams (WP-\ref{MP3_031}), as well as at improving our theoretical understanding by using novel analytical techniques (WP-\ref{MP3_087}).

The mentioned WPs agree on the fact that, according to existing and soon-available detectors, such processes cannot be observed without multi-petawatt laser beams. Moreover, experiments aiming at measuring such small cross sections, like that of photon-photon scattering and/or related processes require a high-quality vacuum (better than $10^{-7}$ mbar) and high-sensitivity photon detectors. Another key requirement from the detection point of view is the spatial/angular and temporal resolution,to be able to discern the photon-photon scattering from unavoidably high background. WP-\ref{MP3_018} discusses the possibility and the challenging aspects of co-locating a multi-petawatt laser with an x-ray free electron laser, a setup suitable also to observe vacuum birefringence.
\section{Science Theme 2: The origin and nature of space-time and matter in the universe}
\label{sec-ST2}

\subsection{Science Question 2A: How do complex material properties and quantum phenomena emerge at atomic pressures and temperatures relevant to planetary cores?}
\label{subsec-SQ2A}

The interiors of planets (e.g.\ silicates, oxides, iron, ices and hydrogen) are subject to extreme pressures that dictate their crystal structures, and also alter the electronic shell structures of atoms. Such pressures therefore impact phase transitions, the dynamics of planetary interiors, and their evolution. Moreover, these conditions may be harnessed to create materials with new chemistry and unprecedented properties. Multi-petawatt laser facilities allow one to use convergent direct- or indirect-drive techniques to access pressures higher than those currently available using shock and ramp compression, as well as to employ advanced diagnostics including: femtosecond x-ray and electron diffraction, spectroscopy, and broadband reflectivity. New developments at multi-petawatt laser facilities will enable the experimental study of matter under the unchartered extreme pressures at which most of the known mass in the universe resides [WPs-\ref{MP3_090},\ref{MP3_091}].

\subsubsection{Introduction}

The quantum unit of pressure, 29.4 TPa, is the pressure required to disrupt the shell structure of atoms, engage core electrons in bonding, and unlock a new regime where correlations of electrons and ions can grow to the macroscale at high temperatures.  Most of the recently discovered extrasolar planets and stars have internal pressures approaching or exceeding such conditions. First-principles calculations have predicted remarkable behavior of matter at these conditions including transforming simple metals into transparent insulators, superionic phases, hot superconductors, unexplained bonding in dense plasmas, and more. We seek to harness the power of multi-petawatt lasers to create these quantum materials here on Earth, and to probe their structure and electronic structure via advanced diagnostics. The results will make it possible for scientists to model and understand the behavior of Earth and exoplanets. 

One can create ultrahigh pressures using high-power laser facilities, such as the National Ignition Facility (NIF), to conduct material studies under extreme conditions. In the direct-drive configuration, the lasers irradiate the ablator material directly producing an ablation plasma.  An approximate formula to calculate the ablation pressure in direct-drive laser experiments is \cite{Lindl_1995}:

\begin{eqnarray}
&P_{DD}(\text{Mbar})=40\left(I_{15} / \lambda_{\mu m}\right)^\frac{2}{3}
\label{eq:Eq1}
\end{eqnarray}
where $P_{DD}$ is the direct-drive pressure in Mbar, $I_{15}$ is the laser intensity in units of 10$^{15}$~W/cm$^2$, and $\lambda_{\mu m}$  is the laser wavelength in $\mu$m. As an example, if one has a 351~nm ($\lambda=1.053/3 \; \mu$m) wavelength and deposits 10 kJ of laser energy onto a 1 mm diameter spot in 1 ns ($I=2.5 \times$10$^{15}$W/cm$^2$), the ablation pressure is  $\sim$100 Mbar, i.e.\ 10 TPa. 

To reach ultrahigh pressures without melting the sample, the compression must be ramped up along a quasi-isentropic path. The easiest way to compress a sample is to apply a strong shock; however, this method quickly reaches its limit as the sample follows the Hugoniot curve and melts if the shock strength is too high. Melting can be avoided by either staging the shocks or ramping the compression following a quasi-isentropic path. In this way, as the sample is gradually compressed to higher pressures, compression is achieved with minimal entropy generation; thus, the sample stays in a solid-state while reaching high pressure. 

Another interesting line of study is to shock melt the sample first and then apply a ramp compression. There are several ways of creating a ramped compression: In gas gun experiments, one uses a graded-density impactor to shape the impact pressure profile; in laser experiments, one uses the laser pulse shape to control the pressure profile; or one can use a reservoir-gap configuration where a strong shock in a reservoir material releases a plasma across the gap and then stagnates on the sample, creating a ramp compression profile on the sample \cite{park_2021}.

While the drive capability exists on existing facilities, the diagnostic capability is very limited especially probing lattice spacing using diffraction techniques. Laser-generated x-ray sources are commonly used in diffraction experiments at laser facilities. A subset of the lasers is used to heat a backlighter foil to generate helium-like, quasi-monoenergetic x-rays \cite{rygg_2020}. These laser-generated x-ray sources convert $\sim$1\% of the laser energy into x-rays and usually include different emission lines and continuum bremsstrahlung components. The most dominant source is the He$_\alpha$ line emission with the spectral resolution of: \mbox{$\Delta$E$_x$/E$_x$ $<$$\sim$0.6\%}, where E$_x$ is the backlighter x-ray energy.

Current capabilities limited to generating only $\sim$10 keV x-ray sources prevents probing very high pressure planetary core conditions. Diffraction of up to $\sim$MeV electrons or $\sim$20-100 keV x-rays would provide information about the long-range order of phase transitions at TPa pressures.


\subsubsection{Roadmap: progress using current facilities}
Recent exciting results have been obtained at the National Ignition Facility at LLNL and the Omega Laser Facility that are revealing the complexity of matter in extreme environments at which most of the known mass of the universe resides. These experiments show that the results of first-principles calculations do not necessarily reveal the phases that will be adopted at these conditions because they typically do not consider kinetic effects, or finite temperatures. For example, ramp compression of carbon \cite{Lazicki:2021}, the fourth-most abundant element in the universe, probed by nanosecond-duration time-resolved XRD showed that it retains the diamond structure up to 2~TPa, in contrast to theoretical predictions. Other examples include measuring the melting curve of iron at conditions similar to those of super Earth cores \cite{Kraus_2022}, spectroscopic signatures of superionic phase of ice that may be the predominant form of H$_2$O throughout the universe \cite{Millot:2019}, and phase behaviour of H-He mixtures \cite{Brygoo_2021}.

\subsubsection{Roadmap: theoretical understanding}
First-principles calculations have revealed that molecules and their mixtures at extreme pressures exhibit entirely new and unprecedented behavior \cite{Zurek:2019k}. They have shown that the long-held belief that all matter, when sufficiently compressed, will assume a Thomas-Fermi-Dirac state where a sea of electrons surrounds ionic cores, and simple compact structures are assumed, is too simple. Elements that are metallic at 1 atm (such as sodium) become insulating with accumulation of charge in interstitial regions, which can be thought of as ``quasi-atoms'', the energy ordering of atomic orbitals is affected so that normally unoccupied orbitals become valence, and core or semi-core orbitals mix with the valence states. This allows core electrons to participate in bonding, and atoms may assume unprecedented oxidation states. Theoretical studies of hydrogen, a major constituent of planetary interiors, turns out to be extremely challenging because they must consider complex phenomena, such as quantum nuclear and anharmonic effects. Unsolved predictions for hydrogen include a high-temperature superconducting phase, a dense plasma ground state, Wigner crystallization, and a high-temperature plasma state of hydrogen isotopes. Density-functional theory has predicted ``hot superconductivity'' in hydrogen-rich alloys. Many first-principles calculations on complex materials performed at 0~K do not take into account anharmonic or quantum nuclear effects, nor strong correlations of the electrons. As computer power increases, and algorithms implemented to treat these effects efficiently, routine modelling of these effects will reveal new phenomena to be searched for experimentally.

\subsubsection{Roadmap: flagship experiments requiring new capabilities}
Recent advances in generating radiation sources using laser wakefield acceleration (LWFA) have led to the possibility of using them for diffraction \cite{Albert_2021, He_2016}. LWFA-driven electron or x-ray (from betatron or Compton scattering processes \cite{Albert_2021, Albert_2018}) probe beams can provide high flux over short femtosecond time scales. Collocation with terawatt, long-pulse ($>$30 ns) lasers with precise laser pulse shaping can ramp, shock, or shock-ramp compress materials to study planetary materials at the pressures, densities, and temperatures needed to make a meaningful impact on our understanding and modeling of terrestrial and gas giant planets. 

LFWA-based probes produced by ultrashort-pulse lasers prove advantageous when x-ray free electron lasers (XFELs) or synchrotron sources cannot be collocated with the required long-pulse lasers drivers. The probe beams should be quasi-monochromatic ($\delta E/E \approx$ 1) and have a duration of several femtoseconds \cite{Albert_2021}. Approximate requirements for the electron beam are energies of $\sim$100 keV to several
MeV, a divergence of $<0.1^{\circ}$, and flux of $\sim$1 pC/pulse. Significant target design and development efforts will be necessary when using electron probes given their short and may require collimation of the electron beam to obtain the desired energy and spot size \cite{Sakabe_2011}. The x-ray beam should have energy of several 10's to 100 keV to investigate the long-range coordination and $>10^{11}$ photons/pulse to be competitive with XFELs. Extending these tools  would benefit studies of ionization balances, line shapes, and opacities of warm dense matter (WP-\ref{MP3_028}, WP-\ref{MP3_021}) and would allow validation of theory and simulations (WP-\ref{MP3_081}).

\subsubsection{Roadmap: flagship experiments requiring next-generation facility capabilities}

Current dynamic diffraction experiments in the HED realm cannot determine the structure of complex solids (10's to hundreds of atoms per unit cell),  or warm dense matter, which is now thought to have significant atomic or partially bonded coordination. Next-generation facilities, like EP-OPAL (Fig.\ \ref{Fig_7.1a}), when combined with a compression facility, such as the Omega Laser Facility, could enable a high-energy, x-ray source that could reveal the first dynamic, complex solid-and-fluid structure determination for HED matter. Such a short-pulse, high-intensity source will open the door to very large Q scattering, thus revealing this long-range structural complexity. Much like the difference between graphite and diamond, this atomic arrangement will likely play a pivotal role in the behavior of such solids. Also, recent observations of partially bonded complex fluids in CO$_2$, SiO$_2$, MgSiO$_3$, carbon, all suggest these materials are highly viscous, nominally electrically conducting, and structurally complex, in a pressure-temperature regime where scientists imagined the atoms were more or less random and unbonded until recently.

Other flagship experiments that require a significant compression facility coupled to an intense, short-pulse laser capability, like EP-OPAL, include determining the most extreme structures of electrides (compounds composed of positively charged ionic cores, where localized electrons serve as anions), the quantum states of electron pairs (such as those comprising quasiatoms within electrides), the viscocity of warm dense matter, the chemistry of extreme matter (keV chemistry - where core electrons take part in bonding), the multiscale evolution of matter (going from cascade to turbulence), testing for Hawking radiation, and the breakdown of the vacuum continuum.

\subsubsection{Broader Impacts}

The findings will impact the field of chemistry, where it is traditionally assumed that only valence electrons are involved in chemical bonding, resulting in the development of new periodic tables for specific pressure regimes. Within materials science and energy-related research, the techniques will be useful to probe the long-range order and extraordinary properties of high-pressure, high-temperature superconducting and topological materials that contain a substantial fraction of light-element atoms. Higher flux x-ray or more sensitive electron diffraction probes could aid in their structural determination. The experimental observables will be used to benchmark theoretical developments in next-generation density functionals, methods that can be employed to treat anharmonic and quantum nuclear effects, and evolve crystal structure prediction techniques so they may be used at finite temperatures. Moreover, the explored matter states could help determine if white dwarf stars are in a glassy or crystalline state, which would affect their thermal conductivity, luminosity, and determine the age of the halo and disk in galaxies.

\subsubsection{Recommendations}

Upgrade existing high-compression facilities by implementing high-intensity lasers to produce relativistic particle beams and advanced light sources for creating and probing matter at atomic pressures and temperatures. These new capabilities will extend and expand forefront science already performed at major facilities, such as high-energy-density physics and laboratory astrophysics. 
Next-generation multi-petawatt lasers envisioned in Chapter \ref{sec-future} that exercise newly developed methods described in \S~\ref{electrons}, \S~\ref{protons}, and \S~\ref{photons} will enable even deeper experimental studies of extreme material properties and quantum phenomena found across the universe.

\bigskip
\subsection{Science Question 2B: How can multi-petawatt lasers study black hole thermodynamics through the link between gravity and acceleration?}
\label{subsec-SQ2B}

\medskip
In pursuit of a ''theory of everything," gravitationally driven particle productions through quantum processes may hold keys to new breakthroughs. The physical mechanisms involved bring together general relativity, particle physics, and thermodynamics to the limits of our understanding of fundamental physics. Multi-petawatt laser facilities will enable one to access and control the unprecedented large acceleration that may lead to the laboratory observation of particle production in an accelerating frame for the first time. This would include the experimental confirmation of Unruh radiation, considered by many to be equivalent to Hawking radiation as a consequence of black hole thermodynamics and a pointer to quantum gravity.
WP-\ref{MP3_052} introduced ideas on how high-power lasers can access physics beyond the Standard Model.
WP-\ref{MP3_058} and WP-\ref{MP3_030} considered how to detect Unruh radiation in a laser experiment.


\subsubsection{Introduction}


It is well known that particle-production phenomena can occur in a curved or dynamic spacetime \cite{Birrell:1982q}. For example, thermal radiation can arise from particle production near the event horizon of a black hole, an effect commonly known as Hawking radiation \cite{r7,r8}. The expansion of the universe also occurs in curved spacetime described by the Friedmann-Lema\^{i}tre-Robertson-Walker metric. Particle production occurs due to the varying gravitational field, 
\cite{Parker:1968pc, Parker:1969qf,Mensky:1980ki, Parker:2012at}. 
Understanding particle production \cite{Campos:1991ff, Zeldovich1974c, Zeldovich:1971mw} during inflation \cite{linder1990particle, Lyth:1998xn}
will help to address fundamental questions in cosmology and may also be relevant for (non-thermal) production of supermassive dark matter in the early universe.

Particle production is a consequence of the mappings between the Fock (particle number) states associated with different reference frames, called the Bogoliubov transformation, a technique widely used in condensed matter, particle physics, and cosmology. The Bogoliubov transformation between a non-inertial frame to another reference frame results in particle non-conservation, which may arise from a curved spacetime or an accelerating frame in flat spacetime, both of which can be interpreted as gravitational particle productions through the equivalence principle. While these techniques are generally accepted, gravitational particle production has not yet been observed in the laboratory.

\subsubsection{Roadmap: progress using current facilities}
Recent advances in ultra-high intensity lasers \cite{Strickland:1985c} have stirred interest in the possibility of detecting both the Schwinger effect and testing non-perturbative QED effects \cite{Mourou:1998u,Bulanov:2003zz,Ahlers2008,Piazza:2012e} using facilities, such as the European Extreme Light Infrastructure \cite{eli}, which will provide radiation beams of intensities exceeding $10^{23}\,\mathrm{\:W/cm^{2}}$. An electron placed at the focus of such beams would experience an acceleration comparable to what it would feel if placed near the event horizon of a $6 \times 10^{18}$ kg ($=3 \times 10^{-12} \,\,\,  M_{\rm sun}$) black hole. For such low-mass black holes, the surface gravity is strong enough that pairs of entangled photons can be produced from the vacuum, with one of the pair escaping to infinity. The black hole can then radiate, and the spectrum of such radiation is a blackbody at the Hawking temperature.

\subsubsection{Roadmap: theoretical understanding}
Given the insurmountable difficulties in directly observing Hawking radiation, Unruh proposed, by virtue of the
equivalence principle (of gravitational and inertial accelerations), that a similar effect could be measured by an accelerated observer \cite{Unruh:1976db}.

\begin{figure}[h]
    \centering
    \includegraphics[width=0.9\linewidth]{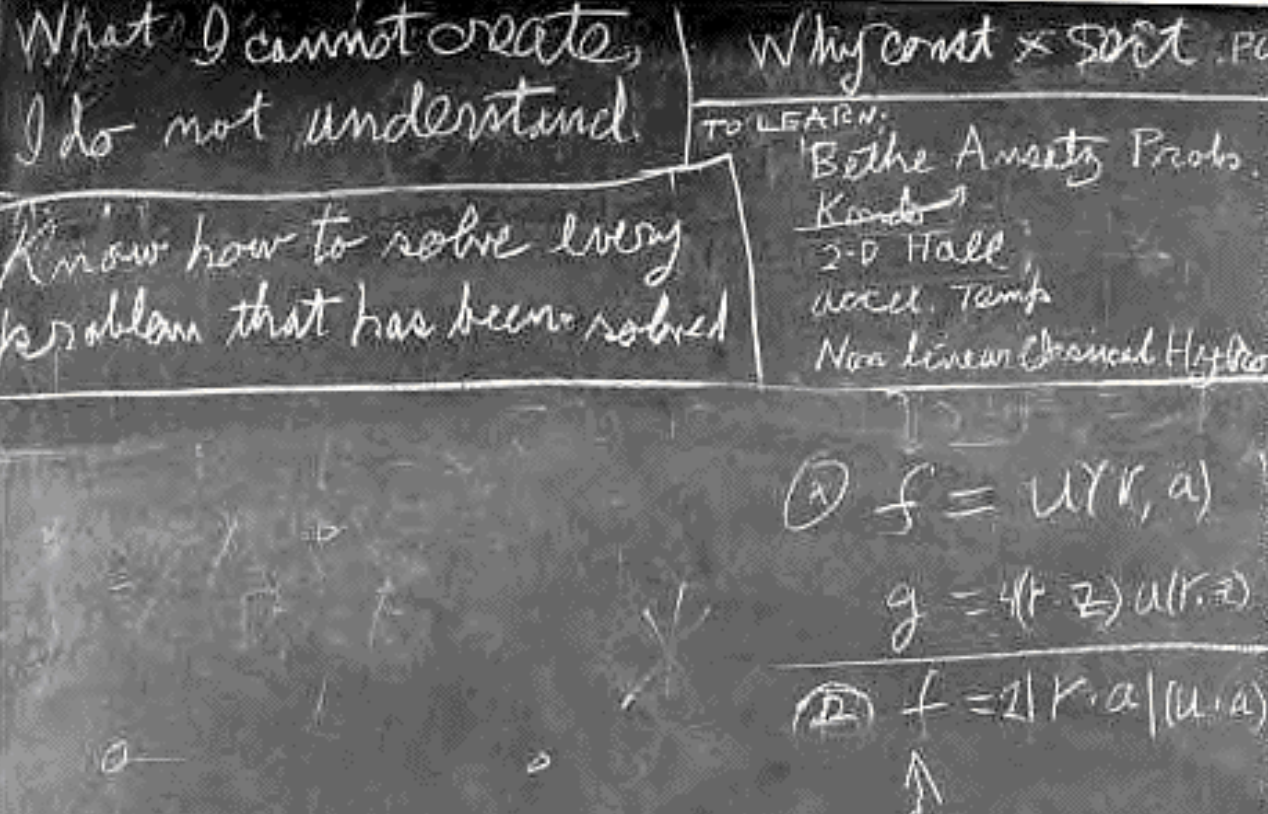}
    \caption{%
        Photo of Feynman's blackboard left at time of his death. Note the section ``To learn: Accel Temp''. Courtesy from the California Institute of Technology.}
    \label{fig:Unruh}
    \end{figure}

While the scientific community generally believes that the derivation of Hawking radiation is sound, this is nevertheless made possible by several approximations that have not been tested. Similarly, while an accelerated observer experiences the equivalent Unruh radiation, it remains unclear what a detector in the laboratory frame will ultimately measure. Views are split, with some researchers believing the answer only comes from ordinary quantum field theory, while others pointing out that additional effects due to the acceleration must be included. A reason for these many opposing views is that an experimental proof of Hawking and Unruh radiation requires a deep knowledge of seemingly disparate fields such as gravity, non-equilibrium quantum field theory in curved space-time, and high-intensity lasers.
Figure \ref{fig:Unruh} is emblematic. Acceleration temperature is still a mystery now as it was then. High-power lasers may finally unlock 
what is needed to learn and address the meaning of acceleration temperature.

The exact derivation of the Unruh effect is far from simple, but some physical intuition can be gained by considering an idealized two-level atom that is subject to a constant acceleration \cite{scully2018,Ben-Benjamin.2019}. In the accelerated state, higher-order processes by which a photon is simultaneously emitted and absorbed can also be considered. Since the acceleration incrementally changes the velocity of the atom, the frequency at which the photon is emitted can be slightly different than the frequency at which it is absorbed. If this change in frequency exceeds the linewidth of the atomic transition, the emitted photon disentangles with the absorbed one,
and we are left with an atom in the excited state and a real photon being radiated away.


If the acceleration continues for a sufficiently long time, the process of 
emission and absorption of photons by the atom reaches a steady state. Equilibration requires that the duration of the acceleration to be at least comparable to the inverse of the Einstein coefficient for spontaneous emission (for the case of natural broadening of the lineshape).
Unruh showed that the flux of emitted photons can be represented in terms of a blackbody function at a temperature
\begin{equation}
T_U = \frac{\hbar \dot{v}}{2 \pi k_B c},
\end{equation}
which is the Unruh temperature, where $\dot{v}$ is the acceleration.

From the above considerations, we see that the key element in the realization of the Unruh effect is the atom having discrete energy levels. This can be understood as a system that can change its state or its internal energy. We call this system a {\it detector}. An elementary particle with internal structure, such as a proton or a neutron, would also be considered as a detector.
The situation is less trivial for a 
particle without any internal structure, such as an electron. In this case, the interaction with the photon bath in the accelerated frame occurs through a continuous change in its momentum via Thomson scattering \cite{Landulfo.2019,hegelich2022}. These energy changes can be infinitely small, meaning
that those small continuous transitions may require infinitely long times to equilibrate and consequently the power radiated by the Unruh effect at those frequencies is very small.
Note, however, that a flip in the electron spin can also be considered as a change of its internal structure. In  fact, the residual depolarization   
of electrons in storage rings has been claimed to be the result of the Unruh effect \cite{Bell.1983o45}.

The power emitted by Unruh radiation is given by\cite{Lynch2021}
\begin{equation}
P_\nu = \frac{2}{3} \left(\frac{\hbar}{c^2}\right) \alpha^2 \dot{v}^2,
\end{equation}
where $\alpha$ is the fine structure constant.  
Assuming that the electron had sufficient time to thermalize, we indeed reproduce the classical Larmor formula for electromagnetic radiation, which can be seen as the limit power achieved by accelerated detectors with no internal structure  \cite{Cozzella.2020}. Thus, the Unruh effect reduces to the classical Larmor radiation for accelerated electrons. This is correct if we consider only the lower order (classical) limit. 
In fact, the above formula was derived assuming that the electron carries no recoil; however, as photons are exchanged with the detector, the latter experience a finite recoil. 
Adding this effect, the power emitted by Unruh radiation now becomes \cite{Lynch2021}
\begin{equation}
P_\nu = \frac{2}{3} \left(\frac{\hbar}{c^2}\right) \alpha^2 \dot{v}^2 \left(1 - \eta \frac{k_B T_U}{2 m c^2} \right),
\end{equation}
and $\eta$ is a coefficient that depends on the details of the detector. For electrons, $\eta=24$ \cite{Lynch2021}. The above formula shows that quantum corrections to the Larmor formula contain terms of order
$\dot{v}^2 k_B T_U/m c^2$ \cite{Lin2005,Lynch2021}.
This is what should be understood as the Unruh effect for an accelerated electron (that is, a detector with no internal structure).

\subsubsection{Roadmap: flagship experiments requiring new capabilities}
Large accelerations can already be realised in the laboratory using existing high-intensity lasers. For example, lasers with intensity $I \sim 10^{19}$ W/cm$^2$ can accelerate electrons to a Unruh temperature of about 1 eV \cite{Chen:1998kp}. While, in principle, this can be measured in the laboratory, it is very challenging to separate the effect of Unruh radiation against other classical and quantum radiation processes involving acceleration of charged particles.
For this reason, Unruh and others have instead adopted an alternative approach to exploit the mathematical analogy between trans-sonic flowing water \cite{Weinfurtner}, Bose-Einstein
condensates \cite{Nova} and other analogue systems \cite{Drori}, and the behaviour of quantum fields in the vicinity of a black hole
horizon. While these experiments have successfully demonstrated the mathematical soundness of Hawking's
solution, they may have fallen short in proving that the radiation is actually emitted by non-inertial bodies and
that the underlying theory is indeed physically correct \cite{Crowther}.

\subsubsection{Roadmap: flagship experiments requiring next-generation facility capabilities}

Unruh radiation has been a controversial subject in the literature \cite{Ford.2006}. For the case of accelerated electrons, distinguishing Unruh radiation from other processes is essential. Since Unruh radiation is thermal, it is very different from non-thermal Larmor radiation. But how to do this in practice is unclear. Likely, higher accelerations are needed. For intensities $I > 10^{23}\,\mathrm{\:W/cm^{2}}$, $T_U > 100$~eV, making it simpler to separate Unruh radiation from the optical background. A more direct observation of the Unruh effect would involve the acceleration of atoms; however, acceleration of atoms and ions is extremely challenging. Using radiation pressure acceleration \cite{Gelfer.2016}, even at intensities $I > 10^{25}\,\mathrm{\:W/cm^{2}}$, we only expect $T_U > 10^{-3}$~eV. This temperature is already high enough to affect the ionization state of the bound levels in the accelerated ion. Measurement of the ionization balance or line emission spectral changes as a function of the acceleration, particularly for rotational/vibrational spectra of molecules, would provide possible avenues for the detection of Unruh radiation.

The detection of thermal radiation from an accelerated oscillator is not, by itself, sufficient to prove the Unruh effect. We will need to develop a unified new theoretical framework, based on which a convincing experimental test can be constructed and a reliable data analysis and interpretation can be carried out.
Such a framework must be able to address the fact, for example, that in the laboratory acceleration is not constant (as assumed in Unruh's derivation) \cite{Unruh:1976db}, and compare Unruh radiation against conventional QED processes \cite{ROSU_1994}. 

\subsubsection{Broader Impacts}
The significance of the proposed research could hardly be overstated. It will lead to the first experimental test of the Hawking-Unruh radiation. The ability of doing so will already be a major scientific milestone marking a new era of experimentally testing quantum gravity that would have required an inconceivable $10^{19}$ GeV high-energy collider based on the standard particle physics approach.

This will shed important light on a wider range of fundamental scientific issues related to the (im)possibility of information loss and the smallest possible scale in nature, with crucial implications on the foundation of physics to justify whether quantum evolution is indeed the correct framework for all laws of physics and what would be the size of the basic building blocks of space and time. The project may further impact on application areas such as quantum computing (as a quantum process under unscreenable gravitational fluctuations) to drive the next generation of artificial intelligence.

The proposed research will yield important implications for classical and quantum theories of gravity by testing the equivalence principle. High-order acceleration with Unruh radiation as a special case are manifestations of gravitational particle productions rooted in the equivalence principle. Therefore, their direct detection are crucial in fundamental physics.

Moreover, the recently discovered double copy correspondence between QED/QCD and classical and quantum gravity \cite{Bern2010,alfonsi2020} can be employed to effectively perform strong-gravity experiments relating to black holes and graviton scatterings using new intense lasers.

\subsubsection{Recommendations}

The extreme conditions required for accelerating atoms or ions to directly observe the Unruh effect and address controversies surrounding Hawking-Unruh radiation depends on two next steps:  (1) developing a new, unified theoretical framework for devising experimental tests and analyzing data; and (2) establishing next-generation multi-petawatt laser facilities that can produce intensities greater than  $I > 10^{23}\,\mathrm{\:W/cm^{2}}$, $T_U > 100$~eV,

\subsection{Science Question 2C: How does the electromagnetic interaction behave under extreme conditions?}
\label{subsec-SQ2C}



\subsubsection{Introduction}

The creation of ultrastrong electromagnetic fields would enable exploration of QED, the most precisely tested component of the Standard Model, in an entirely new region of parameter space. The theory of QED has had great success in predicting interactions at increasing energies, thanks to perturbative approaches that exploit the small size of the fine-structure constant $\alpha$; the same cannot yet be said for increasing \emph{intensities}, where the number of particles participating in a single interaction is large and perturbative approaches fail. It is convenient to imagine the physical electromagnetic field divided into two components: a strong classical background and a quantized radiation field.
In the region $a_0 \gtrsim 1$, it is necessary to take all orders of the interaction with the background field into account, using what we have already indicated as SF-QED, because the probability of a single interaction is increased by the high photon flux to $P \sim \alpha a_0^{2 n}$, which is not necessarily small for larger $n$.
Investigation of this regime is the subject of SQ1B. If the field strength is increased even further, the interaction with the radiation field itself also ceases to be perturbative. Ever higher orders of interaction, featuring virtual electron-positron loops, absorption and re-emission of photons, become as likely as lower orders of interaction.
Perhaps even the identification of individual electrons or photons becomes impossible.
No existing theory describes this regime, although encouraging progress is being made (see \S~\ref{sec:SQ2CTheory}).
Experimental investigations could shed light on the fundamental structure of the electromagnetic interaction and the dynamics of extreme environments in the early Universe.

\begin{figure}[h]
    \centering
    \includegraphics[width=0.9\linewidth]{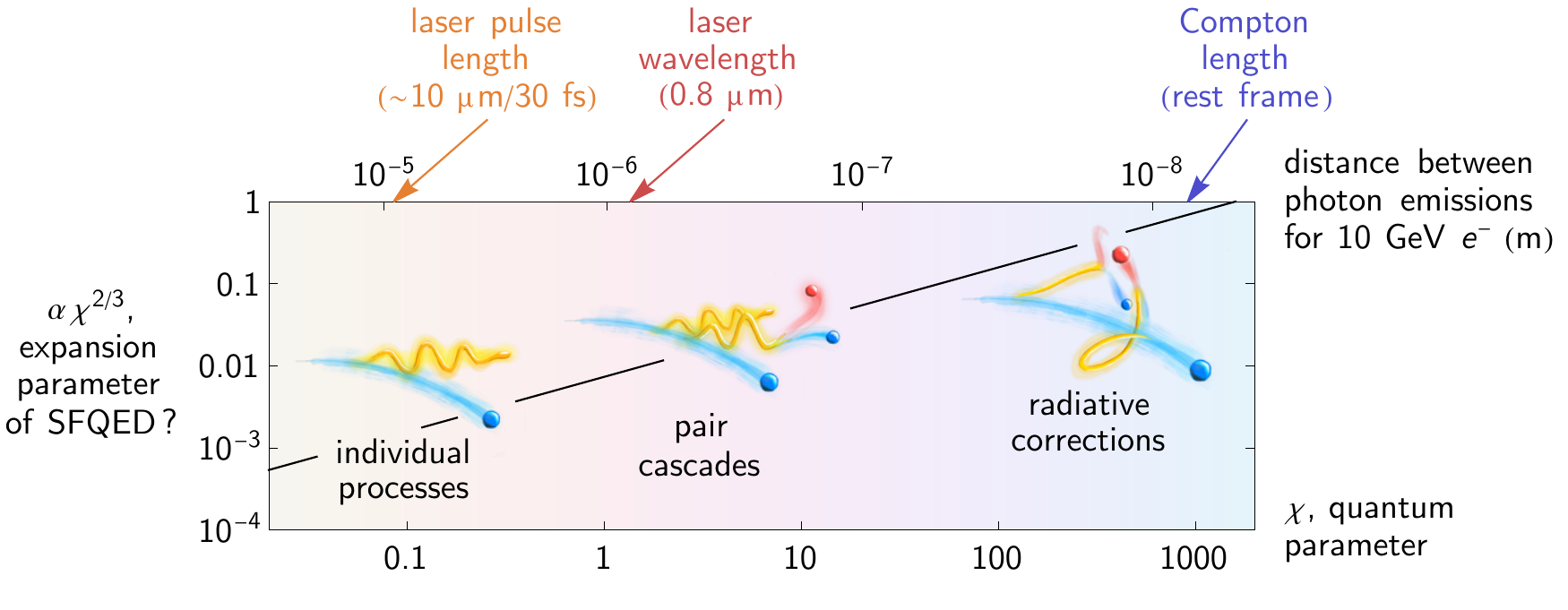}
    \caption{%
        The acceleration induced by an electromagnetic field on an electron is quantified by the quantum parameter $\chi = E_\text{rf} / E_\text{cr}$, where $E_\text{rf}$ is the background electric field in the rest electron frame. As the acceleration increases in strength, the distance travelled by an electron before it emits a high-energy photon decreases, allowing us to explore new kinds of physics with ultrashort lasers.
        At $\chi \simeq 0.1$, where one photon is emitted per laser pulse duration, it is possible to explore single-vertex processes of strong-field QED in detail.
        At $\chi \gtrsim 1$, the electron emits multiple photons per laser wavelength, leading to an electromagnetic cascade (see SQ1B).
        For the most extreme parameters, $\chi \gtrsim 1000$, the distance between QED events collapses to the scale of the Compton length and individual electrons and photons can no longer be identified.}
    \label{fig:IncreasingChi}
    \end{figure}

The theoretical argument underlying these points is the \emph{Ritus-Narozhny conjecture}, which posits that the `true' expansion parameter of strong-field QED is $\alpha \chi^{2/3}$~\cite{Ritus:1970,Narozhnyi:1980}, where $\chi$ is the quantum non-linearity parameter.
This may be understood in the following way (see Fig.~\ref{fig:IncreasingChi}).

\begin{figure}[h]
    \centering
    \includegraphics[width=0.5\linewidth]{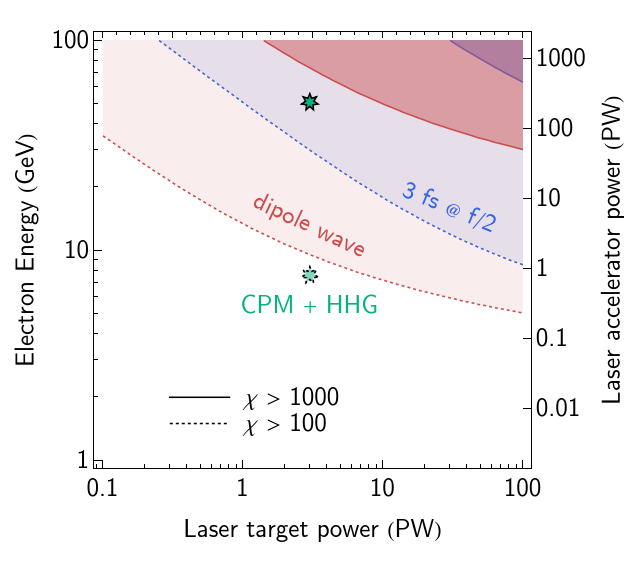}
    \caption{%
        Requirements on the electron-beam energy and the laser power required to reach a quantum parameter of 100 (dashed lines) and 1000 (solid lines).
        The electron beam is considered to collide with: (blue regions) a single-cycle optical pulse ($f/2$ focusing); (red regions) a dipole wave ($4\pi$ focusing, or the coherent combination of multiple laser pulses); or (green stars) the focused attosecond pulse from a curved plasma mirror (CPM+HHG)~\cite{vincenti.prl.2019}.
        The bounds account for radiative energy losses during the collision, as estimated in~\cite{blackburn.njp.2019}.
        The right-hand axis gives the laser power necessary to reach the given electron energy in single-stage LWFA acceleration, estimated using a plasma density of $3 \times 10^{17}$~cm$^{-3}$~\cite{lu.prstab.2007}.
        More sophisticated field geometries and particle-acceleration schemes could lead to lower requirements on the necessary power: see chapter \ref{sec-PAALS}.
        }
    \label{fig:CollisionParameters}
    \end{figure}

At large values of $\chi$, perturbative strong-field QED predicts that photon emission rate is given by $P = \alpha m \chi^{2/3}$.
Thus at $\alpha \chi^{2/3} \simeq 1$, the mean free path between emission events collapses to the scale of the Compton length $\lambda_c = 1/m$ and theory becomes strongly coupled (see \cite{Fedotov:2016afw,fedotov.arxiv.2022}).
Theoretical analysis suggests that the conjecture holds, at least in the strong-field limit $a_0^3 / \chi \gg 1$~\cite{Podszus:2018hnz,Ilderton:2019kqp}, as would be achieved with multipetawatt lasers.
In order to explore this region experimentally, it is necessary to reach $\alpha \chi^{2/3} \gtrsim 1$ or $\chi \gtrsim 1600$.
The quantum parameter achieved in the collision of an electron beam of energy $E$ and a laser with peak intensity $I$ is
    \begin{equation}
    \chi \simeq 1600 \left( \frac{E}{200~\mathrm{GeV}} \right) \sqrt{\frac{I}{10^{24}~\mathrm{W}\mathrm{cm}^{-2}}},
    \end{equation}
and therefore the region $\alpha \chi^{2/3}$ is well beyond the capability of currently existing laser facilities, and indeed that of currently existing conventional accelerators.
Furthermore, the above expression optimistically ignores the effect of radiative energy losses: at such extreme intensities, electrons would radiate away a significant fraction of their energy in a single laser period, reducing their $\chi$ below the target value.
Overcoming this means concentrating the electromagnetic field into as small a spatiotemporal region as possible (in principle, below the mean free path of a single photon emission).
Various methods have been proposed to do so, including the use of beam-beam collisions~\cite{yakimenko.prl.2019}, oblique-incidence electron-laser collisions~\cite{blackburn.njp.2019}, collisions with the attosecond pulses emitted during laser irradiation of a plasma mirror~\cite{baumann.sr.2019}, or aligned crystals \cite{Di_Piazza_2020_c}.
We remark also the highest field strength for fixed input power is achieved using a dipole wave ($4\pi$ focusing via coherent combination of multiple laser pulses)~\cite{bulanov.prl.2010,gonoskov.prl.2013} and that this too could be used as the target of a high-energy electron beam~\cite{magnusson.prl.2019}.
In Fig.~\ref{fig:CollisionParameters} we show the electron-beam energy and laser power necessary to reach $\chi = 100$ and $1000$, assuming a collision with a single-cycle optical pulse, a dipole wave, or an attosecond pulse from a plasma mirror.

The bounds we show account for radiative energy losses, using the scaling given in \cite{blackburn.njp.2019}, and we also estimate what laser power would be necessary to reach a given electron-beam energy in LWFA~\cite{lu.prstab.2007}.
It is clear that probing the extreme-field regime, $\chi > 100$ requires several lasers of multipetawatt power that also have (or can be used to generate secondary radiation with) ultrashort duration or XUV wavelengths.


\subsubsection{Roadmap: progress using current facilities}

It is possible at present to probe the onset of SF-QED effects at $0.1 < \chi < 1$, using `all-optical' collisions between LWFA-electron beams and lasers at multi-laser facilities~\cite{cole.prx.2018,poder.prx.2018}, see review~\cite{gonoskov.arxiv.2021}.
Collocation of an intense laser with a conventional electron accelerator facility~\cite{bula.prl.1996,burke.prl.1997} is also being revisited~\cite{meuren.exhilp.2019,abramowicz.epjst.2021} (see WP-\ref{MP3_076}).
Such experiments will build our understanding of the basic processes of SF-QED, including tree-level processes of nonlinear Compton scattering and nonlinear Breit-Wheeler pair creation (see WP-\ref{MP3_013}), and test the approximations used in simulations, including the locally constant field approximation (LCFA),  an important component of the Ritus-Narzohny conjecture.

\subsubsection{Roadmap: theoretical understanding}
\label{sec:SQ2CTheory}
Our theoretical understanding of electrodynamics in strong fields takes as its starting point the separation of the electromagnetic field into a fixed, classical background and a fluctuating, quantized radiation field~\cite{gonoskov.arxiv.2021,fedotov.arxiv.2022}.
It is possible to find exact, all-order, solutions for the interaction with the background for fields of particular symmetry, including plane electromagnetic waves~\cite{volkov.zp.1935}.
What follows is a perturbative expansion of the interaction with the radiation field, in the so-called Furry picture~\cite{furry.pr.1951}, using the Volkov solutions to construct Feynman rules for the tree-level processes of photon emission $e \to e\gamma$ and electron-positron pair creation $\gamma \to e^- e^+$.

The non-trivial spacetime dependence of the basis states makes it difficult to obtain analytical results for higher order processes, although there has been much progress in the study of trident pair creation ($e \to e e^- e^+$), loop contributions, and chains of bubble diagrams~\cite{Piazza:2012e,gonoskov.arxiv.2021,fedotov.arxiv.2022}.
If the interaction with the radiation field does indeed become nonperturbative for $\chi \gg 1$, as conjectured, then it may be necessary to develop new `all order' methods for this interaction as well.
For example, resummation techniques have been used to study the analogous breakdown of perturbation theory in the classical regime~\cite{Heinzl:2021mji}, to construct high-order predictions of quantum radiation reaction~\cite{torgrimsson.prl.2021}, and to compute the probabilities of nonlinear Compton scattering and nonlinear Breit-Wheeler pair production by taking into account that electrons, positrons and photons in a plane wave are unstable (electrons and positrons under radiation of photons) \cite{Podszus_2021,Podszus_2022}.
How these techniques could be applied to more general field configurations or combined with numerical simulations requires research.

\subsubsection{Roadmap: flagship experiments requiring new capabilities}
\label{SQ2C_new_capabilities}

Probing $\chi > 10$, while mitigating radiative energy losses, requires not only 10-GeV class electron beams but also ultrashort-duration electromagnetic pulses.
One method to achieve very high fields is to use a relativistic plasma mirror as an optical-to-XUV converter (see \S \ref{RHHG}).
Intense, optical laser light incident on such a mirror drives nonlinear, relativistic oscillations of the plasma surface.
These relativistic oscillations in turn temporally compress (over tens of attoseconds) and intensify the reflected light at each laser cycle, producing a broadband spectrum of high harmonics~\cite{landecker.pr.1952,lichters.pop.1996}.
The reduced diffraction limit for higher frequency radiation means that this secondary emission can be focused more tightly~\cite{baeva.pre.2006,bulanov.prl.2003,naumova.prl.2004,gordienko.prl.2005}.
So far, the proof-of-principle experiments that have been carried out have not demonstrated light intensification~\cite{kando.prl.2009,kiefer.ncomm.2013,kim.ncomm.2012}.

Proposed implementations of relativistic plasma mirror focusing suffer from two major impediments.
The first is a lack of robustness to laser-plasma imperfections: in realistic conditions, it was predicted that the maximum intensity achievable would actually be limited to $10^{24}$~W/cm$^2$, at most, with PW-class lasers~\cite{solodov.pop.2006}.
The second comes from the high degree of experimental control required in most implementations, which may mandate sub-micron and/or sub-fs laser stability/alignment. 
Nevertheless, high-resolution numerical solutions now show that the natural curvature of the plasma mirror surface induced by the incident laser radiation pressure can very tightly focus the high harmonics at an unprecedented optical quality~\cite{vincenti.prl.2019}.
For a PW-class laser, intensities close to $10^{25}$~W/cm$^2$ could be reached at the plasma mirror focus.
In combination with a 10-GeV electron beam~\cite{Gonsalves_2019}, it would be possible to reach $10 < \chi < 100$.
Testing SF-QED predictions would, however, require careful characterisation of the field's spatiotemporal structure, such that differences between theoretical predictions and measured spectra could be attributed with confidence to the underlying theory, as opposed to uncertainties in the interaction conditions.

\subsubsection{Roadmap: flagship experiments requiring next-generation facility capabilities}

Reaching the point where $\alpha \chi^{2/3} \simeq 1$ requires 100-GeV class electron beams (as shown in Fig.~\ref{fig:CollisionParameters}) and ultrashort electromagnetic pulses, in order to mitigate the extreme radiative energy losses expected.
Only a next-generation facility would be able to achieve the electron energies required, either by colocation of an ultraintense laser (or laser-driven secondary source) with a new linear collider, or by the construction of a staged laser-driven plasma accelerator facility, as shown in Fig.~\ref{fig:facilitySFQED}.
The electromagnetic field with which such an electron beam would be made to collide would need to be highly confined.
Apart from the laser energy being used to drive a curved plasma mirror of the type describe in the previous section, it could be split and recombined into a dipole-wave structure, which would confine the electromagnetic field to a $\lambda^3$ volume.
The essential idea is as follows~\cite{gonoskov.arxiv.2021}.
If focusing a single laser pulse leads to a peak amplitude of $a_0^{(1)}$, splitting the laser into two beams of equal power and focusing them to the same point (from opposite directions) yields an increased field amplitude of $a_0^{(2)} = \sqrt{2} a_0^{(1)}$.
Extending this concept to $n$ `multiple colliding laser pulses' (MCLP), arranged in counterpropagating pairs in a common plane (Sec. 7.1, Pf. 15), leads to an even higher amplitude of $a_0^{(n)} = \sqrt{n} a_0^{(1)}$~\cite{bulanov.prl.2010}.
The theoretical limit, with $4\pi$ focusing, provides the strongest possible field strength for fixed total power $P$, $a_0 \simeq 780 (P~[\text{PW}])^{1/2}$~\cite{bassett.oa.1986,gonoskov.pra.2012,jeong.oe.2020}.
Realizing a dipole wave by the use of six or twelve beams is discussed in the Appendix of \cite{gonoskov.prl.2014}.
Colliding a 10-GeV electron beam with such a field structure has already been proposed as a means of generating a high-flux of multi-GeV photons~\cite{magnusson.prl.2019}.
A 100-GeV class electron beam would ensure that the $\chi \gtrsim 1000$ region could be reached.

\subsubsection{Broader Impacts}

There may be connections between QED in electromagnetic fields of extreme intensity and other nonperturbative quantum field theories, such as QCD, although the theory of this regime of QED is at a much earlier stage of development. From a theoretical point of view, the long experience acquired in QCD can be exploited to investigate analytically and numerically also the fully non-perturbative regime of QED at $\chi\gtrsim 1000$. The experimental tools required to reach such high values of $\chi$ themselves represent significant technological development, from the creation and control of the ultrastrong electromagnetic fields themselves, to the generation of 10 to 100-GeV electron beams.

\subsubsection{Recommendations}

QED has proven the best-tested and most accurate physical theory so far, but unprecedented regimes will enable deeper investigations. 
Reaching the `fully nonperturbative' regime ($\chi \gtrsim 1000$) demands entirely new methods of generating and focusing light at the highest intensity, as well as accelerating electrons to energies of the order of 10-100 GeV. 
Current and next-generation multi-petawatt laser facilities will play a decisive role to achieve the required conditions. 
For the sake of a clean signal, one needs to produce ultra-short, ideally few-cycle, pulses and it may be necessary to work in the XUV domain because optical lasers of such strength might drive pair cascades (see SQ1B) of such density as to shield the focus from incoming radiation.
Accurate synchronization of laser and electron beams proves crucial so that the electrons cross the laser field at the maximal intensity. 
Experiments will require the detection of photons and/or electrons with energies on the scale of 100 GeV, which will require methods and know-how from the high-energy particle physics community. Corresponding advancements in analytic (resummation) and numerical (strong-field QED in a lattice) techniques are also required in order to possibly compare theory with experiments.

A proposal to reach ultra-high intensity and ultimately the Schwinger limit has been put forward in WP-\ref{MP3_012}, whereas electron energies of the order of 20-30 GeV are envisaged in WP-\ref{MP3_023} by employing multi-petawatt lasers. WP-\ref{MP3_055} proposes an experiment on quantum radiation reaction employing a 60-PW laser beams a 10-GeV electron beam. Although a tight focusing is not required in this proposal, if the 60-PW beam can be focused down to a spot 1-2 wavelengths in size, sufficiently high values of the nonlinear quantum parameters could be reached so as to detect signatures of the fully-nonperturbative regime. Finally, the importance of studying plasma dynamics in the presence of supercritical electromagnetic fields has been pointed out in WP-\ref{MP3_080}.

\includepdf[pages={-}]{figures/Blank_Page.pdf}

\section{Science Theme 3: Nuclear Astrophysics and the age/course of the universe}
\label{sec-ST3}

Nuclear physics using Intense lasers can open new frontiers in nuclear physics research. Laser-driven sources of energetic particles, such as protons and neutrons can induce nuclear reactions, probe nuclear physics, and enable practical applications. 
Exploring nuclear science at high temperature conditions is important to test models of astrophysical reaction rates (WP-\ref{MP3_056}, WP-\ref{MP3_059}).
Scaling laser-driven Compton sources to MeV photon energies with extremely high fluxes and using their unique properties can excite and manipulate atomic nuclei. 
The MP3 workshop identified two frontier science areas enabled by new secondary radiation sources generated by multi-petawatt lasers: studies of nuclear astrophysics using intense neutron sources; and studies of the strong nuclear force in low-energy quantum chromodynamics (QCD) using tunable intense gamma sources.


\subsection{Question 3A: What can be learned about heavy-element formation using laser-driven nucleosynthesis in plasma conditions far-from-equilibrium?}
\label{subsec-SQ3A}

\subsubsection{Introduction}

All elements other than hydrogen result from fusing lighter elements into successively heavier elements, like carbon, oxygen, aluminum, iron, and nickel, which are synthesized in the cores of stars, like the sun. For example, two hydrogen atoms fuse to form helium, two helium atoms fuse to form beryllium, and three helium atoms fuse to form carbon.

Heavier elements, like copper, silver, gold, or uranium, form via nucleosynthesis in the helium- and carbon-burning shells of massive stars or via proton and neutron capture processes in supernova explosions.
In almost every case, these nuclear fusion processes happen in a hot, dense plasma environment.
Until now, Measurements of nuclear fusion reactions using accelerators do not include these plasma backgrounds.
Significant discrepancies exist between the calculated abundance of elements based on the measured cross sections and those observed in astronomical measurements \cite{Sparta2020}.
These discrepancies influence our understanding of stars and their evolution, as well as the use of these processes in fusion technologies.

\begin{figure}[h]
\centering 
\includegraphics[width=.9\textwidth]{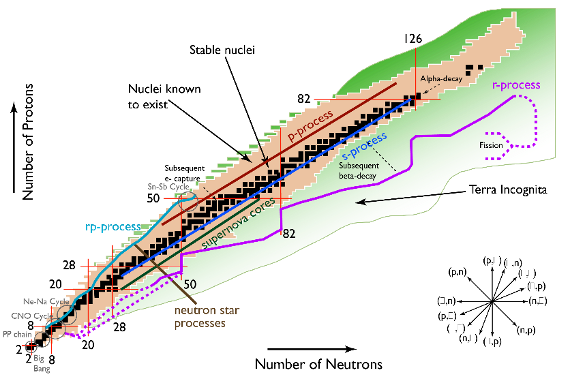}
\caption{\label{fig:SQ3_fig1} The chart of the nuclides in the (Z,N) plane  \cite{NNDC1995}. Stable nuclides indicated by squares along with pathways of different astrophysical processes. The rapid neutron capture r-process drives nuclear matter far to the neutron-rich side and is interrupted by fission. The rapid proton-capture rp-process on the neutron-deficient side produces nuclides close to the proton drip-line; p-process deals with $\gamma$n-processes \cite{Blaum_2006} }
\end{figure}

In nuclear astrophysics, the \textbf{rapid neutron-capture process, also known as the r-process}, is a set of nuclear reactions responsible creating approximately half of all the ``heavy elements'' with atomic nuclei heavier than iron. The \textbf{proton ``p-process''} and another slow neutron-capture ``s process'' account for the other heavy elements. The r-process also contributes to abundances of some lighter nuclides up to about the tin region and occurs in supernova explosions, while the heaviest elements are produced in binary neutron-star mergers.
Figure \ref{fig:SQ3_fig1} highlights the isotopes produced by these nucleosynthesis processes.

\begin{figure}[h]
\centering
\includegraphics[width=.4\textwidth]{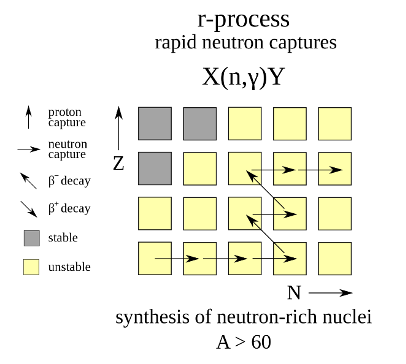}
\caption{\label{fig:SQ3_fig2} Schematic of nucleosynthesis r process. Successive rapid absorption of neutrons leads to unstable, neutron-rich nuclei with higher atomic weight, N. Beta decay increases the atomic number, Z, and lowers the neutron:proton ratio resulting in more stable nuclei. \cite{Rp-process2022}}
\end{figure}

The r-process \cite{Cowan_2021} proceeds at high temperatures (several giga-Kelvin, GK) and high neutron fluxes. Neutron captures occur on time scales comparable to the lifetime of excited nuclear quantum states. This sequence can continue up to the limit of stability of the increasingly neutron-rich nuclei, as governed by the short-range nuclear force. Figure \ref{fig:SQ3_fig2} illustrates schematically how the r-process and beta decay create new heavy isotopes and elements.
The r-process must occur where a high density of free neutrons exists, such as the material ejected from a core-collapse supernova \cite{Thielemann_2011} or decompression of neutron-star matter thrown off by a binary neutron star merger \cite{Kasen_2017}. The relative contribution to the astrophysical abundance of r-process elements proves a matter of ongoing research \cite{Kasen_2017}.

Quantitative modeling of the r-process path requires knowing the neutron-capture cross sections of these excited nuclear states, or theoretical modelling benchmarked with experimental data on nuclei where this can be done. Research facilities dedicated to the production of neutron-rich radioactive ions, such as the Facility for Rare Isotope Beams (FRIB) \cite{Balantekin_2014} or the Facility for Antiproton and Ion Research (FAIR), \cite{Fortov_2012} (WP-\ref{MP3_054}), will investigate the structure of r-process nuclei to provide direct or indirect information on neutron capture on the nuclear ground states, but neutron-capture cross sections on excited nuclear states will not be possible, except for a few very specific long-lived isomers.


\subsubsection{Roadmap: laser-driven nucleosynthesis using current facilities}

Experimental demonstrations can produce sequential neutron-capture reactions on short-lived nuclear excited states by exposing target materials of interest to short-pulse laser-driven neutron beams with neutron fluences exceeding $10^{20}$~n/(cm$^2$s) \cite{Treffert_2021, Chen_2019}. For ordinary neutron-capture cross sections (typically of the order of barns), a sufficient number of neutron-capture processes will produce isotopes with mass number A+1 in the ground state and in excited states and secondary-neutron captures on the unstable A+1 isotopes. These reactions will produce unstable A+2 isotopes that will eventually decay by beta- decay with typical half-lives of hours to their Z+1 isobars, typically odd-odd nuclides that are unstable by themselves and further decay to the Z+2 isobar, which will be stable.


After being exposed to the short-pulse neutron burst, the target will be retrieved and put into a low-background counting station to measure its radioactive decay to determine the production of consecutive two-neutron captures. The lifetimes of the longest-lived and lowest lying levels will be known. The sequential double-neutron capture cross section could be measured as a function of the temporal length of the neutron burst, which would vary the contribution of subsequent neutron captures on excited states. 

Current PW-class laser facilities can produce 10+ MeV-class proton beams suitable for high-flux neutron generation, including beams with more than $10^{11}$ protons above 1~MeV \cite{Kleinschmidt2018}. Achromatic divergence allows refocusing proton beams onto secondary proton-to-neutron converter targets with small spot sizes to generate intense neutron bursts \cite{Steinke2020}.

Experiments can scale up stepwise using laser systems listed in Table \ref{table_lasers_SQ3} ultimately leading to experiments using the 10-PW lasers coming online at ELI-Nuclear Physics (ELI-NP) in Romania near Bucharest and ELI-Beamlines (ELI-BL) in the Czech Republic near Prague.
\begin{center}
 \begin{tabular}{| m{0.18\linewidth} | m{0.23\linewidth} | m{0.5\linewidth} |}
\hline
\multicolumn{3}{|c|}{Table \ref{table_lasers_SQ3}:
Experimental steps moving from current to next-generation laser facilities}\\
\hline

\textbf{Laser Facilities} & \textbf{Laser Parameters} &  \textbf{Experimental Steps}\\

\hline \hline
BELLA (LBNL)  \newline VEGA-3 (CLPU) & 1~PW \newline (30~J, 30~fs, 1~Hz) & \multirow{2}{24em}{$\bullet$ Optimize proton, deuteron acceleration \& \newline neutron production\\ $\bullet$ Test neutron flux/angular dist.\ and source size \\ $\bullet$ Develop and test “rabbit” and nuclear \newline detection systems\\} \\
\cline{1-2} 
L3 (ELI-BL) \newline HPLS (ELI-NP) & 1 PW \newline (30~J, 30~fs, 10~Hz)  & \\
\hline
L4 (ELI-BL) \newline into E5 Hall & 10~PW \newline (1500 J, 150 fs, shot/1 min) & $\bullet$ Optimize ion acceleration and neutron production, scale for laser energy\\
\hline
HPLS (ELI-NP) & 10 PW (230~J, 23~fs, shot/1 min & $\bullet$ Optimize proton acceleration to cut-off energies $E_\mathrm{p}>100\,\mathrm{MeV}$ \\ 
\hline
Next-generation facilities \newline (Chapter \ref{sec-future}) & $\geq$ 25 PW \newline multi-kJ, multi-beam & $\bullet$ To be determined!\\
\hline

\end{tabular}
\label{table_lasers_SQ3}
\end{center}

These experiments will require improving plasma and neutron measurements on well-understood CD$_2$ and cryogenic deuterium targets at current petawatt-level facilities. The high instantaneous fluxes encountered in these laser-plasma interactions will require state-of-the-art neutron and plasma diagnostics (\S~\ref{sec-PAALS}). The L3 laser at ELI-Beamlines and the two 1-PW HPLS beamlines at ELI-NP offer performance similar to earlier PW-class experiments but at higher shot rates. Experiments using the two 10-PW HPLS beamlines at ELI-NP are expected to increase reaction rates by up to 10$\times$. Longer pulses at the ELI-BL L4 beamline will deliver more laser energy and generate more protons and neutrons.


Intense proton sources developed to drive r-process experiments can also drive p-process experiments, and the extensive suite of photon and neutron diagnostics available at ELI-NP may enable searching for secondary nuclear reactions. Theory suggests an intricate coupling of the ground-state in $^{26\!}\mathrm{Al}$ and its first excited isomer $^{26m\!}\mathrm{Al}$ via higher-lying excitation levels resulting in a dramatic reduction of the effective lifetime of $^{26\!}\mathrm{Al}$, which will influence the abundance of this isotope in our Galaxy. It is estimated that a short burst of laser-driven protons with energies $> 5$~MeV can induce the $^{26}\mathrm{Mg}$(p,n)$^{26\!}\mathrm{Al}$ reaction that proves out of reach with current DC- and RF-accelerator systems. A multi-PW laser-driven experiment could lead to high and comparable spatial and temporal yields of the three lowest-lying states in $^{26\!}\mathrm{Al}$ including the short-lived state at 417~keV. After such a p-process reaction, the yield ratios between the ground and the two lowest-lying excited states will represent thermodynamic imbalance. This condition can mimic thermodynamic equilibrium at high temperatures experienced in stellar conditions and serve as a first step towards investigating the interplay between those states in plasma environments that will inform studies of nuclei decay in astrophysical conditions. 

Tight laser focusing combined with techniques to enhance laser temporal contrast \cite{Dromey_2004} could enable new ion acceleration mechanisms, like magnetic vortex \cite{Esirkepov_1999,Bulanov_2010} and radiation pressure acceleration (light sail), that could scale to multi-PW experiments at higher particle fluxes and energies to potentially access proton energies $>100\,$MeV \cite{Hakimi_2022}.

Ultraintense lasers can create hot, dense plasmas in which to measure fusion processes. The laser will create a billion-degree hot plasma and accelerate atoms to the required energies to cause them to fuse. By measuring the different escaping particles, it is then possible to understand both the plasma conditions and how they influence the fusion reactions. This data can then be used to improve nuclear and stellar models and improve our understanding of the universe, and possibly even help to develop controlled fusion technology here on earth, providing a clean energy source.

\subsubsection{Roadmap: theoretical understanding}

Fusion reactions between light nuclei in far-from-equilibrium plasmas, such as deuteron-deuteron fusion [d(d,n)$^3$He] reactions using laser-driven neutron sources with flux $> 10^{25}$~neutrons/cm$^{2}$/s, carry information about the plasma in the output neutron spectrum. Experiments can provide observables from plasma conditions to compare and improve theoretical understanding and computational models used to compute corrections to nuclear reaction rates. Numerical simulations will optimize laser and target parameters to maximize energy transfer to deuterons, the distribution of electrons and deuterons as inputs to the fusion cross sections and their plasma-dependent corrections, as well as reaction volume and confinement time. Continuous theory support during experiments will analyze and understand the data gathered and connect it with existing stellar and nuclear models.



\subsubsection{Roadmap: flagship experiments requiring new capabilities}

Experiments with existing PW-class systems have yielded $10^{11}$~MeV protons in single shots. Multi-PW lasers will extend this capability and access the currently elusive Radiation Pressure Acceleration (RPA) or even the Single-Cycle Laser Acceleration (SCLA) regimes that could herald the advent of ``collective ion acceleration'' reaching GeV proton energies. Developing high-repetition-rate lasers and targetry at multi-Hz rates would enable highly productive experiments with good statistics. 
The two-beam, 10-PW HPLS system at ELI-NP drives development of nuclear detection and analysis tools for fast and slow neutrons, ion characterization (Thomson parabola, LaBr$_{3}$ scintillation detectors for delayed $\gamma$ photons, and activation methods), and $\gamma$-flash measurements with optical-based scintillator and image plate systems. Part of these detector systems are already in construction at PW-class sites such as ELI-NP.

\subsubsection{Roadmap: flagship experiments requiring next-generation facility capabilities}

\begin{figure}[h]
\centering 
\includegraphics[width=.3\textwidth]{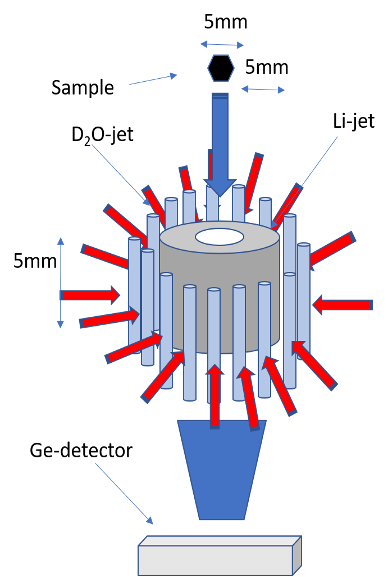}
\caption{\label{fig:SQ3_fig3} ``Multi-SHARC'' laser concept to produce a high-repetition-rate, high-flux neutron source for nuclear physics research.}
\end{figure}

Nuclear physics experiments benefit from high flux and high repetition rates for low-probability interactions to collect experimental statistics using ever larger single-beam laser systems.
Figure \ref{fig:SQ3_fig3} illustrates an alternate, modular approach for a next-generation facility that implements multiple state-of-the-art petawatt lasers.

The SHARC (Scalable High-power Advanced Radiographic Capability) laser concept \cite{Siders_2018} delivers 150-J, 150~fs laser pulses at 10~Hz.
A concentric arrangement of SHARC lasers focuses on a ring of deuterated water jets to generate protons that interact with an inner-ring jet of liquid-lithium jets to produce a high-flux source of neutrons for irradiating samples that pass through the central void. The lithium ring also works as a neutron reflector to confined the neutrons. Scaling from existing laser systems and experimental demonstrations leads to estimates of average neutron flux of $10^{24}$~n/s/cm$^2$ at 10~Hz with $10^{13}$~n/s in the inner void to irradiate samples before delivering them where detectors count decay reactions. 





\subsubsection{Broader Impacts}

Laser-driven nuclear physics experiments described above entail scientific and technological advances with potential broader impacts. Understanding and controlling excited nuclear states and isomers offers the potential to identify and create unique signatures for identifying nuclear material by active interrogation, as well as understanding and realizing transmutation of actinides for nuclear waste disposal.

The broader impacts of laser-driven nucleosynthesis research can inform applied and fundamental research topics with societal impact that includes: 
\begin{itemize}
    \item the quest for a net-zero carbon emission energy source to combat global warming, such as novel approaches to inertial fusion energy that can profit from high-field effects \cite{Imasaki2008}, high-energy proton and ions \cite{Zhou2016}, and isochoric heating \cite{Sakata2018};
    \item new methods for nuclear waste characterization \cite{Tanaka2020}; and
    \item enhanced by in-situ studies of the FLASH effect for efficient cancer treatments \cite{Aymar2020}.
\end{itemize}







\subsubsection{Recommendations}

Multi-petawatt laser systems offer new and powerful tools for nuclear astrophysics research. They provide ultrashort pulses that can recreate or at least mimic conditions that exist in the inner core of stars and were present at the time of the Big Bang. Plasma conditions will be far from equilibrium, but corrections can be applied to deduce the interaction to be measured experimentally in plasma for the first time. Experimental programs can start now with existing petawatt lasers and scale with the laser intensity to shed much-needed light (pun intended!) on the abundance of heavy elements by evaluating the r-process path, their creation and decay within the first moments at high $\mathrm{k_{B}}T$ values after the Big Bang and stellar lifecycles (e.g., $^{26\!}$Al). The study of multi-neutron absorption processes spanning over the relevant energy windows seems in reach due to the high temporal and spatial confinement of the produced neutrons. Recommendations include:
\begin{itemize}
    \item Experimentally develop high-flux, high-energy particle (proton, neutron, deuteron) generation and characterization methods using existing laser facilities.
    \item Conceive and implement next-generation facilities (Chapter \ref{sec-future}) with kJ laser pulse energies that increase particle energy and flux rates for experiments and applications. 

\end{itemize}

\subsection{Science Question 3B: How can high-flux gamma sources generated from multi-PW lasers be used to explore Hadronic Physics (Low-Energy QCD)?}
\label{subsec-SQ3B}


\subsubsection{Introduction: Nuclear physics with precision photon sources}

The US and EU Nuclear Physics Long-Range Plans \cite{NSAC_2022,Jenas_2022} both identify the following as one of the central questions of nuclear physics: \textit{``How does subatomic matter organize itself and what phenomena emerge?''}.
Answering this question requires high-intensity 
gamma-ray beams with narrow energy spread ($\Delta$E$_\gamma$/E$_\gamma <$  0.03 FWHM) and high beam polarization, both linear and circular. A recent white paper \cite{Howell_2020} identifies the required beam parameters 
for pursuing the low-energy QCD research described in this section.  Gamma-ray beams in two distinct energy ranges are needed: (1) E$_\gamma <$ 12 MeV for studying hadronic parity violation in light nuclei via photonuclear reactions, and (2) 60 $< E_\gamma <$ 300 MeV for nucleon structure measurements via Compton scattering and charge-symmetry investigation using photo-pion production.  In both energy ranges beam intensities greater than $10^{10}$ photons/s/cm$^2$ on target are needed.  This required beam intensity is about a factor of $\times$ 100 the capability of current gamma-ray beam technology. The research opportunities and estimated beam parameters are summarized below. 
\\
\\
\noindent
\textbf{Nucleon Low-Energy Electromagnetic Structure}\\
In real Compton scattering on nucleons and light nuclei, electric and magnetic fields induce radiation multipoles by displacing a hadron's electric charge constituents and currents. The energy- and angle-dependence of the emitted radiation encodes in the form of the nucleon polarizabilities the strengths and symmetries of the interactions 
with the incident photon and between them. 
That is, the nucleon polarizabilities parametrize the stiffness of the responses of the nucleon's internal charge and magnetic constituents to applied electromagnetic fields.
The difference in the magnetic polarizabilities between proton and neutron values is a salient contribution which narrows the mass gap between protons and neutrons. Additionally, the spin-polarizabilities describe hadronic birefringence caused by the electromagnetic field of the spin degrees, like in the classical Faraday effect. They therefore parametrize the response of the nucleon itself, complementing experiments at Jefferson Lab. Thus far, these quantities are not well known, often with error bars of 40\% to 100\%. Reference \cite{Howell_2020} identified that the parameters above allow extractions from data to the 20\% level or better from ``aspirational'' uncertainties for the cross section of 3\% and beam-target asymmetries of 0.03. This suffices to answer the fundamental questions raised.
\\
\noindent
\textbf{The QCD Origin of Charge-symmetry Breaking}\\
The mechanism for charge-symmetry breaking (CSB) in the nucleon-nucleon interaction is well established in terms of hadronic degrees of freedom, e.g., $\rho-\omega$ mixing in the meson-exchange interaction between identical nucleons \cite{Cohen_1995}. Steven Weinberg postulated that the fundamental cause of CSB is the differences in the masses of the up and down quarks and in their electromagnetic interactions \cite{Weinberg_1977}. In this paper, Weinberg showed that the difference is the masses of the down and up quarks is proposal to the difference in the s-wave $\pi^0n$ and $\pi^0p$ scattering lengths.  
High-accuracy pion-photoproduction on the proton and light nuclei at energies above about 140 MeV will answer the question of how CSB emerges from QCD. The most relevant amplitudes are the photon-pion charge-transfer reactions (at 1\% accuracy), as well as neutral pion photoproduction (to 7\%). Data on the proton and the lightest nuclei at energies up to 300~MeV will elucidate the interplay of chiral physics and resonance physics in the transition from non-perturbative to perturbative quarks, and hence the emergence of the complex structures and interactions of nuclear physics from the deceptively simple QCD interactions between quarks and gluons.
\\
\\
\noindent
\textbf{Hadronic parity violation in few-nucleon systems} \\
Hadronic parity violation (HPV) in light nuclei provides an important probe of two phenomena that are not well understood: \textit{neutral-current nonleptonic weak interactions} and \textit{nonperturbative strong dynamics}. While PV is well understood in quark-quark weak interactions, it is ultimately the interplay of different forces at different length scales that is responsible for hadronic PV phenomena. Neutral-current interactions are suppressed in flavor-changing hadronic decays, making hadronic PV between nucleons the best place to study neutral-current effects. Because parity-violating, nucleon-nucleon (NN) interactions manifest the interplay of nonperturbative strong effects and short-range weak interactions between quarks, they are sensitive to short-distance, quark-quark correlations inside the nucleon.

Weak NN interactions at low energy are suppressed by six to seven orders of magnitude com- pared to strong NN interactions making them difficult to observe. Unambiguous extraction of information about weak interaction among nucleons requires parity violating measurements in very light nuclei (e.g., the deuteron, the triton, and $^3$He) where theory is under good control. These light nuclei can be calculated using two- and three-nucleon interactions formulated in terms of effective field theories (EFTs) that systematically incorporate the symmetries of QCD \cite{Schiavilla_2021}. The EFT approach can consistently be applied to strong and weak nucleon interactions as well as to external currents. 

Initial experiments will measure parity violation in photodisintegration reactions of few-nucleon systems, such as $^2$H and $^3$He. 
These experiments put stringent demands on the photon source-performance parameters because the parity violating asymmetry is extremely small ($\sim10^{-7}$) and these measurements are optimally performed near the reaction threshold energy where sensitivity to parity-nonconserving interactions is greatest but the cross section is changing rapidly with energy. The most recent DOE-NSF Nuclear Science Advisory Committee (NSAC) Long-Range plan lists P-odd deuteron photodisintegration \cite{NSAC_2022} near threshold as a priority NN weak interaction measurement. Table \ref{table_SQ3B_beam} lists the beam requirements to carry out such measurements – no photon beam facility currently can provide these parameters. 
\begin{center}
\begin{table}
 \begin{tabular}{| p{0.35\linewidth} | p{0.35\linewidth} |}
\hline
\multicolumn{2}{|c|}{Table \ref{table_SQ3B_beam}: P-odd deuteron photodisintegration beam requirements}\\
\hline
Photon Beam Parameter & Value \\ \hline \hline
Energy	&	$< 12$~MeV\\
\hline
Flux \newline $\Delta E/E$ \newline Polarization \newline Diameter \newline Time Features	& $10^{10}$ photons/s \newline $< 1\%$ FWHM	\newline	Circular ($> 97\%$)	\newline	$< 12$~mm on target \newline	$> 10$~Hz pol. flip	\\	
\hline
\end{tabular}
\end{table}
\label{table_SQ3B_beam}
\end{center}
\subsubsection{Roadmap: progress using current facilities}
Precision nuclear physics in the area of low-energy QCD requires beams of the range of $10^10$~ph/cm$^2$/s at 10- to 300-MeV photon energies are needed with percent-level energy spreads.  Measurements aimed at advancing topics in low-energy QCD described in this section have been performed during the last decade using conventional accelerator-driven laser Compton and tagged bremsstrahlung sources. The Compton-scattering experimental programs using the tagged bremsstrahlung photon beam at the Mainz Microtron ($E_\gamma >$ 150 MeV) \cite{Cividini_2022, Mornacchi_2022, Paudyal_2020, Martel_2015} and the quasi-monoenergetic phton beam from Compton source at the High Intensity Gamma-ray Source (HI$\gamma$S), $E_\gamma <$ 110 MeV \cite{Li_2022, Li_2020}, have significantly contributed to improving the accuracy of the electric and magnetic dipole polarizabilities of the proton.  Also, asymmetries for elastic Compton scattering from a polarized proton target using a circularly polarized photon beam were measured at the Mainz Microtron.  These asymmetries were used to determine spin-dependent polarizabilties of the proton.  In addition, photo-pion production measurements are carried out at the Mainz Microtron.  

There are currently no research programs in low-energy QCD using photons produced by high-power fast pulsed laser systems. However, there is potential on the horizon for developing such photon beam sources.  Compton backscattering gamma sources based on plasma accelerators have advanced with recent experiments at Lawrence Berkeley National Laboratory (LBNL) \cite{Gonsalves_2019} and University of Texas, Austin \cite{WangDowner2013} that have demonstrated electron beams of few GeV energies with 10$^8$ to 10$^9$ electrons/pulse, suitable for the required photon energies. Separate experiments at lower energy at University of Nebraska, Lincoln \cite{Golovin_2016}, University of Texas, Austin \cite{Tsai_2015}, Max-Planck-Institute for Quantum Optics (MPQ) \cite{Khrennikov_2015}, and LBNL \cite{Geddes_2015} have demonstrated photon production efficiencies approaching a photon per electron, and electron energy spreads at the percent level \cite{Geddes_2015, Faure_2006,Lin_2012}. Research and development underway aims to reach the required low divergence via either refocusing or advanced injection techniques \cite{Rykovanov_2014, Grote_2021}. Integration of these results at current facilities could result in a source at the range of a few 10$^7$ photons/shot with the required energy spread and energy. The required areal flux could be achieved in mm-scale spots at the 1- and 10-Hz repetition rates available at current facilities or at higher laser powers in the multi-PW range at Hz rates. Multi-PW systems could also offer high single-shot intensities if multi photon processes are important.


\subsubsection{Roadmap: Theoretical understanding}

New theory efforts focused on this science will provide important input to experimental planning and ensure quick and efficient analysis of data. Such efforts are well-aligned with the recommended ``Theory Initiative'' of the 2015 NSAC Long-Range Plan and continue the strong tradition of synergy between theory and experiment in this area. A range of initiatives would ensure that the strong international theory community working on this physics stays fully engaged:
\begin{enumerate}
    \item Workshops with small lead times and duration of up to a month.
    \item Funding for long-term theory visitors, including sabbatical and summer stays. 
    \item Support for off-site Postdoctoral Researchers and Graduate Students to ensure continued scientific progress and workforce development.
\end{enumerate}
In addition, computing resources must address the high-level computational needs of related theory projects. This contributes to the Long-Range Plan’s recommendation of ``new investments in computational Nuclear Theory that exploit US leadership in high-performance computing''.


%
%
%


\section{Particle-acceleration and high-energy-photon sources}
\label{sec-PAALS}

Observing the emitted particles and photons or probing interactions with secondary beams of particles or photons, provides a way to understand and diagnose the physics involved in a laser-plasma interaction. Historically, as laser-pulse intensities were increased, certain laser-plasma interactions gave rise to accelerating structures that created beams of particles, and secondary sources of photons. Through careful tailoring of the interactions -- choosing suitable laser pulse properties and target conditions -- a wide variety of particle and photon sources have been engineered with many unique properties. These sources are compact compared to conventional sources, and have a huge range of potential applications across science, technology, engineering and medicine. They are also extremely useful, essential even, to enable and probe the new science questions presented in this report. Indeed, sometimes the physics investigated in the science questions may lead to new radiation sources. 
This makes the development of laser-driven sources of high-energy particles and photons a core area of multi-PW laser research. 
The ``2020 roadmap on plasma accelerators'' gives a forward-looking overview of these advanced accelerator concepts \cite{Albert_2021}.

In response to the solicitation for white papers (WPs) the Particle Acceleration and Advanced Light Sources (PAALS) working group received thirty-six WPs that in part or in whole addressed the question of high-energy particle and radiation sources using the next generation of multi-PW lasers.
The WPs covered all of the major sub-areas of research and addressed topics such as electron and proton acceleration, generation of neutron and positron beams, as well as approaches to generate keV to MeV energy x-rays.
Additionally, several WPs focused on the diagnostic aspects.
The need for high-repetition rate laser facilities and diagnostics to enable large data sets for statistics and to cover parameter space, as well as for active feedback and machine learning was highlighted in WP-\ref{MP3_022} and WP-\ref{MP3_011} as a need for high quality science.

In this chapter, the main mechanisms for generating these particle (electron, positron, proton/ion, neutron) and high-energy photon sources will be reviewed and the current state-of-art summarized, before introducing the future work that may be enabled by the multi-PW lasers. On the low-frequency side, instead, the interaction of kJ class lasers with solid and gaseous targets have been proposed in the WP-\ref{MP3_088} to generate unprecedented J-class THz pulses, which in turn can open up a new regime of relativistic optics. Diagnostic development goes hand-in-hand with the source development, and some of the ideas are described in chapter \ref{sec-strategies}.


\subsection{Particle sources: Relativistic electron beams and heating}
\label{electrons}

Multi-PW laser systems will drive compact, high-energy electron accelerators that are predicted to enable energies up to $20$--$30$~GeV for physics applications and also enable new sources of ultrafast, high-brightness x-rays for applications \cite{Albert_2016}.
The electrons may gain energy from laser-driven plasma waves in a scheme known as Laser WakeField Acceleration (LWFA) \cite{Tajima_1979,Esarey_2009} that can generate multi-GeV, quasi-monoenergetic, high-charge, small-emittance, few-femtosecond duration electron beams, or directly from the laser fields (Direct Laser Acceleration or DLA) \cite{Pukhov_1999,Gahn_1999,Shaw_PPCF14,zhang_prl15} that are well-suited for generating broadband x-rays with high-efficiency. 
DLA will likely be a key mechanism in transferring laser energy to the electrons for creating the extremely hot dense plasma in very strong magnetic fields that are required for addressing SQ1A.
For other applications, such as the colliding beam experiments that will be the first steps to addressing SQ1B, the high-energy, small energy-spread LWFA electron sources are preferable. 
Other laser-to-electron heating mechanisms are important in different configurations, such as for solid target interactions typically used for producing laser-driven proton, ion and neutron beams.

\subsubsection{Laser wakefield acceleration (LWFA)}
In LWFA, an electron bunch ``surfs'' on the electron plasma wave (the ``wakefield'') generated by the ponderomotive force of an intense laser \cite{Tajima_1979, Esarey_2009}.
The plasma wave has a strong longitudinal electric field that stays in phase with the relativistic driver so that relativistic particles may remain in phase with the accelerating field over long distances and gain ultra-relativistic energies.
The accelerating electric field strength that the plasma wave can support can be many orders of magnitude higher than that of conventional metallic RF accelerators, which makes laser wakefield acceleration an exciting prospect as an advanced accelerator concept.
LWFAs using PW-class, short pulse lasers are able to generate $\sim 8$~GeV electron beams in underdense plasmas \cite{Gonsalves_2019}.
Not only can these compact, centimeter-scale accelerators generate high-energies, they can also have small energy spread (less that a few percent \cite{Faure_2006,Schmid_2010,Buck_2013,WangDowner2013}), small transverse emittance ($0.1$~mm~mrad, \cite{Brunetti_2010, Kneip_2012, Plateau_2012}), short duration (few femtoseconds) bunch length \cite{Lundh_2011}, and high-charge (5~pC for the current record energy beam \cite{Gonsalves_2019}); considerably higher for smaller energies \cite{Couperus_Draco2017,gilljohann2019}). 
Despite these successes, it is still challenging to optimize all of the beam properties simultaneously. The rational for using multi-PW laser pulse is that, owing to the scaling of the critical beam power as $P_{\rm crit} \propto 1/n_p \propto \lambda_p^2$ (where $\lambda_p$ is the plasma wavelength), such pulses can be self-guided in a fairly tenuous plasma ($n_p < 10^{16} cm^{-3}$), resulting in a longer dephasing distance $L_{\rm deph} \propto \lambda_p^3$ and higher beam energy $\gamma_b mc^2 \propto 1/n_p$~\cite{Esarey_2009}.

Once high-quality electrons beams have been formed, secondary sources of radiation, such as positron beams or high-energy photons can be generated. Of particular note, the electrons undergo transverse oscillations in the multi-PW-laser-driven plasma wave during and after their acceleration. These so-called betatron oscillations result in the emission of energetic (keV-MeV) photons (see \S~\ref{sec_betatron}) that could be used for a variety of scientific applications.


\vspace{11pt}
\noindent
\textbf{LWFA using multi-Petawatt lasers}\\
The MP3 science questions and realizing next-generation, all-optical light sources (WP-\ref{MP3_045}) both require improvements to  LWFA source brilliance, shot-to-shot reproducibility, conversion efficiency and repetition rate.
A number of interesting new concepts were proposed in this area of research.
Multi-PW laser pulses are expected to generate beams with energy exceeding $10$~GeV (WP-\ref{MP3_048} and WP-\ref{MP3_023}), i.e.\ the electron beams have large $\gamma$, such that the $\chi$ of an electron beam-laser counter-propagating interaction will be greatly increased (SQ 1B).
Novel concepts were also proposed to overcome the limitations of traditional LWFA to enable even more compact accelerators.
Approaches to overcome dephasing, where the electron beam outruns the accelerating structure and thereby limits the energy gain, include the combination of two obliquely incident pulses with tilted phase fronts using cylindrical mirrors to create line foci along the accelerated electron trajectory \cite{Debus_2019} or controlling the spatiotemporal structure of the laser pulse \cite{Palastro_2020} (WP-\ref{MP3_003}, and WP-\ref{MP3_095}).
Post-acceleration of LWFA beams is also proposed, through adding acceleration stages either optical, by transitioning to a beam driven stage (WP-\ref{MP3_077}), or by combining electron bunch and laser pulse drivers~\cite{Wang_2020}.
WP-\ref{MP3_010} proposes using vacuum post-acceleration of relativistic electrons by a combination of THz electromagnetic pulses and constant magnetic fields to boost the energy of the electron beams.

Study of LWFA using multi-PW facilities promises more stable and better control of the electron beam characteristics, and higher energy beams beyond current scaling limits that would advance the understanding of laser-plasma acceleration for future high-energy physics colliders.
The interaction of these multi-GeV electron beams with high intensity laser pulses will enable the study of nonlinear quantum electrodynamics (QED) addressed in SQ 1B and 2C.
Use of these electron beams to generate positrons and x-rays are covered in \S~\ref{positrons} and \S~\ref{photons}.

\subsubsection{Direct Laser Acceleration (DLA) and electron heating mechanisms}

The mechanisms through which a laser pulse transfers energy to electrons can be complex and depends on the target and laser pulse conditions. 
Vacuum heating \cite{Brunel_1987} or j $\times$ B heating \cite{Kruer_1985} prove to be key mechanisms for solid target interactions at laser intensities $>10^{18}$~W/cm$^2$, but intensity to electron temperature scalings are tricky because of the dynamic energy partition.
For a laser pulse interacting with a solid-density target, heating of plasma electrons is the first and foundational step for several interesting secondary processes, such as positron generation, proton, ion and neutron beam generation, compact x-ray sources, or the creation of extreme field or pressure conditions.
Understanding the key energy transfer step of laser-pulse energy to the electrons is therefore of vital importance.

For a laser pulse duration longer than the inverse of the plasma frequency in a low density plasma, or at higher densities with ultra-intense laser conditions, the plasma electrons quickly respond to the ponderomotive force of the laser to create a channel. Within this channel strong transverse electric fields are present, and as an electron beam is driven forward, large azimuthal magnetic fields form.
These channel fields can enable considerable energy gain \cite{Hussein_2019}.


\vspace{11pt}
\noindent
\textbf{DLA and electron heating using multi-Petawatt lasers}\\
DLA in underdense plasmas could be used to demonstrate high-energy, high-charge electrons beams for a variety of potential applications (WP-\ref{MP3_051}).
At ultra-high-intensities, DLA may play an ever increasing role in the energy transfer process as even solid-density targets will become transparent due to the relativistic heating \cite{Ridgers_2012} and radiation reaction will also influence DLA \cite{Jirka_2020b}.
At high enough intensities, the radiation reaction force alters the electron trajectories, both as a loss mechanism into a bright photon beam (see below, WP-\ref{MP3_008}) and by interfering with the dephasing from the laser fields \cite{Gong_2019}.
For picosecond, kilojoule-class, multi-petawatt pulses, the energy distribution evolution will be affected by multiple interactions with the laser and self-generated fields, stochastic effects including collisions, re-circulation and radiation loss effects (WP-\ref{MP3_037}).
WP-\ref{MP3_043} and WP-\ref{MP3_046} both consider how the hot electron generation can be enhanced by magnetizing the plasma.
WP-\ref{MP3_044} also noted that understanding how fast electrons behave in highly magnetized environments is of significance to laboratory astrophysics and may be beneficial for advanced inertial fusion schemes.
WP-\ref{MP3_050} considers using nanowire foam targets to form uniformly heated, relativistic electron temperature conditions \cite{Kemp_2019}.
Schemes with multiple multi-petawatt pulses with overlapping focal spots are considered as a way to enhance the intensity and the laser energy coupling to the electrons and consequently proton acceleration (WP-\ref{MP3_085}).
WP-\ref{MP3_085} identifies that laser-driven magnetic reconnection \cite{Raymond_2018} may play a role in this enhancement.

\subsection{Particle sources: Positron generation}
\label{positrons}

Addressing SQ1A requires a macroscopic relativistic plasma created from electrons and positrons is required.
The current route to creating positrons uses high-Z targets and produces an electron-positron jet.
Relativistic electrons are either produced through laser-solid interactions \cite{Chen_2009}, or via LWFA \cite{Sarri_2015}.
As the relativistic electrons propagate through the high-Z material (typically gold), the high-energy electrons produce high-energy Bremsstrahlung photons. These photons interact with the high-Z nucleus and the photons decay into electron-positron pairs via the Bethe-Heitler process.
These methods have been shown to generate almost charge neutral jets \cite{Sarri_2015, Chen_2015}, and confining the pairs in a magnetic bottle has been demonstrated \cite{Peebles_2021, vonderLinden_2021}, although not yet at sufficient density and scale to be considered a pair plasma.

SQ1B has the ultimate aim of using strong fields to create a plasma fireball of matter (electrons), anti-matter (positrons) and photons.
The diagnostic techniques developed for characterizing the positrons using the high-Z targets will benefit the experiments designed to study the routes to positron generation via strong fields.

\vspace{11pt}
\noindent
\textbf{Positron generation using multi-petawatt lasers}\\
The ultimate goal of creating a pair plasma fundamentally requires sufficient positrons to be created, which translates back to the laser energy as well as the conversion efficiencies involved in each step of the process.
The goal is to achieve a high pair density, volume and lifetime (WP-\ref{MP3_027}).
For example, the duration of the pair jet needs to be greater than the typical instability growth time of interest. For a Weibel instability, this means for a 1~ps duration requires a pair density $> 10^{15}$~/cm$^3$.
Experimental data using the direct heating of a high-Z target has shown a quadratic scaling of positron yield with laser energy \cite{Chen_2015}, making a case for high-energy multi-Petawatt lasers.
Optimization and control of the pair plasma promises to be a unique laboratory platform to study extreme astrophysical phenomena (WP-\ref{MP3_065}).


Beyond the scope of the science discussed in this report, future laser-driven colliders for high-energy-physics experiments are envisioned \cite{Leemans_2009}, requiring synchronized electron and positron beams.
Multi-PW laser systems generating ultra-relativistic positron beams with femtosecond duration, and high spatial and spectral quality via LWFA would be an important step towards building plasma-based colliders.
The initial stage of positron generation experiments will involve characterization of the positron beam properties such as the emittance, divergence and energy spread (WP-\ref{MP3_040}).
An important next step would show that these positron beams can subsequently be accelerated using plasma waves.
Standard plasma waves defocus positrons, so WP-\ref{MP3_023} proposed creating azimuthally symmetric ``donut-shaped'' wakefields by manipulating the laser focus.


\subsection{Particle sources: Proton and ion beams}
\label{protons}

The relative inertia of protons and ions compared to electrons means the laser fields do not directly interact with them over a single laser cycle.
Once the laser heats the plasma electrons, the electrons move and expand to form electric sheath fields, the resulting currents form magnetic fields that can launch a collisionless electrostatic shock.
These high-gradient ($\sim$TV/m), longer timescale secondary fields can accelerate protons and ions.

A number of ion-acceleration mechanisms that depend on target and laser conditions have been identified \cite{daido.rpp.2012,Macchi_2013,bulanov.pu.2014}.
The most robust and well-studied is target normal sheath acceleration (TNSA) \cite{Hatchett_2000}.
A solid foil is irradiated with the laser pulse, a hot electron cloud expands in all directions out into the vacuum.
The sheath field on the rear surface of the target rapidly ionizes and then accelerates protons and ions from a thin layer typical composed of hydrocarbons.
Protons are preferentially accelerated due to their higher charge-to-mass ratio, but special surface cleaning may allow improved acceleration of higher-Z ions.
The energy spectra is broad and Maxwellian-like, with maximum proton energies up to $\sim100$~MeV achieved \cite{kim.pop.2016,wagner.prl.2016,Higginson_2019}.
The proton beams have a divergence of $<30^{\circ}$ and a very small transverse emittance \cite{Cowan_2004}, making the beams suitable for imaging applications \cite{Borghesi_2002}.
Typical energy-conversion efficiencies from laser-energy into proton-beam energy are a few percent \cite{Fuchs_2006}.

To boost maximum energies, the rear surface fields can be enhanced.
The Break-Out Afterburner (BOA) acceleration mechanism uses relativistically induced transparency in exploding thin foil targets to enhance the sheath field through a relativistic Buneman instability \cite{Yin_2007}.
Carbon ions with greater than 1 GeV have been accelerated in this way \cite{Jung_2013}.
Magnetic Vortex Acceleration (MVA) \cite{Esirkepov_1999,Bulanov_2010} is another mechanism where the rear surface field is enhanced by the laser pulse relativistically channeling through the target, confining and heating electrons so that they emerge as a huge current from a small region at the rear.
The exiting electron current induces a magnetic field on the rear surface that acts to enhance the accelerating fields.

For many applications, a proton or ion beam requires a narrow energy spread.
It is possible to perform a post-acceleration energy selection step on the TNSA proton beam, but this is not efficient.
Alternative schemes have been developed to tune the accelerated ion spectra.
Extensive theoretical and numerical simulation studies indicate that Radiation Pressure Acceleration (RPA) has the potential to achieve energies per nucleon beyond the GeV range \cite{Esirkepov_2004,Klimo_2008, Macchi_2009,Qiao_2009,Chen_2009_proton, Bulanov_2010}.
The required targets are very thin ($\sim10$’s nanometers thick), so the light pressure of the high-intensity laser pulse moves the whole volume of electrons forward, and generates a large electric field that then accelerates the whole ion volume in a uniform potential, to create a quasi-monoenergetic spectra with high efficiency \cite{Robinson_2008}.
Experimentally realizing results predicted by numerical modeling proves extremely difficult for several reasons:

(1) RPA requires exceptionally high-temporal-contrast laser pulses to avoid pre-plasma formation expansion/destruction of the ultra-thin targets, even on a picosecond timescale. This proves particularly tricky to attain for the ultra-high intensities required. Enhancing the contrast usually requires plasma mirrors \cite{Dromey_2004} that add significant complexity to the experimental set up.

(2) The ultra-thin foil targets are susceptible to instabilities, such as Rayleigh-Taylor, that are detrimental to the acceleration \cite{Palmer_2012}. Various remedies, such as using multi-species or variable-thickness targets, are being explored~\cite{Yu_2010,Wang_PRL2021}.  Furthermore, minimal heating of electrons is required, so circular polarization of the laser pulse is superior to linear. 

(3) To achieve high intensities, tight focusing is currently required and the rapid deformation of the target enables efficient electron heating reducing the effectiveness of the acceleration \cite{Dollar_2012,bulanov.pop.2016}.

The laser interaction at the front surface provides at least two other possible ion acceleration mechanisms: skin-layer ponderomotive acceleration \cite{Maksimchuk_2000} or shock acceleration \cite{Silva_2004, Habara_2004}.
The ponderomotive force of the laser exerts a pressure on the target electrons driving them away from regions of highest intensity into the target. The ions are unaffected by the laser fields directly, but respond to the electric field due to the electron displacement.
Under optimum density conditions, a high-Mach-number electrostatic shock can be driven into the target that is capable of accelerating ions. Proof-of-principle experiments for this  ion beam acceleration mechanism, Collisionless Shock Acceleration (CSA) \cite{Silva_2004, Fiuza_2012, Fiuza_2013}, have accelerated quasi-monoenergetic ion beams \cite{Haberberger_2012,Palmer_2011}.

\vspace{11pt}
\noindent
\textbf{Proton and ion beams using multi-petawatt lasers}\\
Multi-PW laser pulses bring significant improvements to the stability, efficiency and achievable energies of proton and ion beams.
Enhancements to TNSA through temporally shaping the laser pulse and secondary effects, like relativistic transparency, are predicted to boost the maximum energy by a factor of 2.5-3 (WP-\ref{MP3_047}).
Multiplexing petawatt-class laser beams in different configurations \cite{Markey_2010, Morace_2019, Raymond_2018} would enhance the laser coupling, matter heating, and particle acceleration (WP-\ref{MP3_085}).
Ideas for using laser-generated proton beams to produce isotopes that might be used as a nuclear battery were proposed (WP-\ref{MP3_060}) which requires high-flux proton beams.
Using the high-intensity and energy density laser pulses provides a potential route to creating highly ionized, high-Z plasmas and accelerated ions (WP-\ref{MP3_006}).

Multi-PW laser pulses promise to better match the criteria required for RPA: high-intensity, high-contrast, with a sufficiently large focal spot.
Recent theoretical work demonstrated that proton beams can be simultaneously accelerated and focused to a tiny volume when non-uniform thickness targets are used, and that  ultra-high laser power (tens of PW) improves target stability with respect to Rayleigh-Taylor instability~\cite{Wang_PRL2021}.
Likewise, dual RPA in a colliding ion beam geometry is proposed for a compact hadron collider (WP-\ref{MP3_074}), or to generate TeraBar pressures (WP-\ref{MP3_093}).
How the QED process that will likely occur at the highest intensities will affect the ion acceleration mechanisms is also of considerable interest (WP-\ref{MP3_064}). There is an experimental need for ultra-thin (possibly as thin as tens of nanometers) foils, and one leading contender for enabling relatively high repetition rates is liquid crystal targets \cite{Poole_2014}. WP-\ref{MP3_067} considered using helical laser beams to impart unique orbital angular momentum to ion beams, and it was suggested that the hollow intensity profile could be beneficial for RPA.

\subsection{Particle sources: Neutron generation}
\label{neutrons}


Once high-energy electrons, protons, ions or photons (Secs.\ \ref{electrons}, \ref{protons}, \ref{photons}) have been generated, nuclear reactions may take place that create neutrons. This opens the door to compact, high-flux, high-energy neutron sources. Neutron beams can be generated through nuclear spallation, when beams of particles (typically GeV protons) pass through a ``catcher'' target made from a material with a large cross-section for the neutron generating reaction of interest.
The beam-target reaction can upshift the neutron energy from the reaction center-of-mass energy. Examples of reactions that have been used are $^2$d(d,n)$3$He \cite{Norreys_1998, Pretzler_1998}, $^7$Li(p,n)$^7$Be \cite{Lancaster_2004}, $^7$Li(d,n)$^8$Be \cite{Higginson_2011}, $^9$Be(d,n)$^{10}$B \cite{Roth_2013}.
Electron-beam-induced nuclear reactions via Bremsstrahlung in copper (reactions $^{65}$Cu($\gamma$, n)$^{64}$Cu and $^{63}$Cu($\gamma$,n)$^{62}$Cu) can create a neutron beam with short pulse duration ($< 50$~ps) and high peak flux ($>10^{18}$ n/cm$^2$/s)\cite{Pomerantz_2014}.

\vspace{11pt}
\noindent
\textbf{Neutron generation using multi-petawatt lasers}\\
Scaling to multi-PW lasers opens the possibility to combine the high-flux of a conventional spallation source, the ultrashort nature of the laser pulse with the ability to collocate a compact facility with existing infrastructure (WP-\ref{MP3_002}).
The potential exists to produce directional, short-duration neutron sources with energies of up to 100's MeV.

\subsection{High-energy photon sources}
\label{photons}
 
A variety of approaches exist to create compact and ultrashort XUV to x-ray light sources based on laser-plasma accelerators.
Laser-based, high-energy photon sources have made tremendous progress and have huge potential.
LWFA-based sources now produce keV-MeV photon sources with unprecedented spatial, temporal and spectral characteristics \cite{Albert_2016}.
They are becoming standard tools for high-energy-density physics experiments and used for imaging \cite{Kneip_2010}, absorption spectroscopy \cite{Kettle_2019}, or diffraction (WP-\ref{MP3_049}).
These photon sources may have a much broader potential across STEM disciplines, with medical imaging \cite{Cole_2018} and material science \cite{Hussein_2019} applications already demonstrated.
This section briefly reviews the basic mechanisms for high-energy photon generation -- high-harmonic generation, Bremstrahlung, emission of betatron radiation, radiation reaction, inverse Compton scattering (ICS), x-ray free electron lasers (XFELs) -- along with their photon source properties and their potential using Multi-PW laser pulses.

\subsubsection{Relativistic high-order harmonic generation (RHHG)}
\label{RHHG}
A high-intensity laser pulse focused onto an overdense target with a sharp density gradient induces the surface electrons to rotate in relativistic orbits \cite{Gordienko_2004, Dromey_2006}.
The relativistic motion causes them to emit temporally coherent photons at high-harmonics of the driving laser pulse fundamental frequency.
The superposition of these coherent high-harmonics forms attosecond pulse trains.
The photon spectrum can extend into the extreme ultraviolet and even x-ray energies.
Science Question 2C discusses how RHHG or this can be considered a relativistic plasma mirror could be used to intensify the light energy (see \S \ref{SQ2C_new_capabilities}).

A key challenge for creating efficient relativistic high-harmonic generation at high-intensity is preserving a steep density gradient on the target. Chirped Pulse Amplification-based laser systems have non-compressible energy component due to Amplified Spontaneous Emission (ASE) that forms a $\sim$nanosecond pre-pulse pedestal that the high-intensity short pulse sits on.
A measure of the size of the pre-pulse is the intensity temporal contrast, the ratio between the pedestal intensity to the peak intensity.
Even with advanced pulse cleaning techniques in the laser systems, best contrast ratios would be $\sim 10^{-11}$.
To remove this pre-pulse laser energy, plasma mirrors can ``clean'' the pulse \cite{Dromey_2004}.
Double plasma mirrors typically can improve the temporal contrast by several orders of magnitude or more at the expense of some of the pulse energy.

\vspace{11pt}
\noindent
\textbf{RHHG using multi-petawatt lasers}\\
Work by Edwards and Mikhailova examines how HHG might scale with intensity (excluding QED effects) \cite{Edwards_2020}. Science Question 2C (\S \ref{SQ2C_new_capabilities}) also requires the focusing properties of the naturally curved interaction surface to be tested and characterized, which will be challenging.

\subsubsection{Bremsstrahlung radiation sources}
High-energy electrons passing through a dense material, particularly a higher-Z material, can pass near to nuclei such that they are scattered by the fields and generate Bremsstrahlung photons. The Bremsstrahlung photon spectrum is continuous, up to a maximum photon energy being about that of the maximum electron energy. The Bremsstrahlung method is a straightforward technique for generating a lot of high-energy photons and has been demonstrated to have applications for imaging of high-energy-density materials. Using LWFA to generate the Bremsstrahlung source, spatial resolutions of better than $30\; \mu$m have been achieved \cite{Lemos_2018}.

\vspace{11pt}
\noindent
\textbf{Bremsstrahlung sources using multi-petawatt lasers}\\
The brightness of Bremsstahlung sources increase with increasing laser-pulse peak power and LWFA electron beam energy that will translate directly to the highest photon energies achievable.

\subsubsection{Betatron radiation sources}
\label{sec_betatron}
LWFA accelerates a highly relativistic electron bunch in a moving plasma wave, with longitudinal electric fields accelerating the beam forward. Transverse electric fields act to center the electron bunch and the electron bunch undergoes betatron oscillations that generate an x-ray beam.
The properties of the x-rays closely resemble synchrotron radiation, a well collimated broadband beam in the 1-100~keV range. The duration of the LWFA electron bunch, and therefore the emitted x-rays, are a few-to-tens femtoseconds in duration, making them $\sim 1000$-fold shorter pulse duration compared to conventional synchrotron facilities.
Additionally, the betatron radiation has $\mu$m source size giving a high degree of spatial coherence that enables phase-contrast imaging \cite{Kneip_2010}.

\vspace{11pt}
\noindent
\textbf{Betatron sources using multi-petawatt lasers}\\
Moving to multi-PW laser pulses will increase the energy of the electron beam created during LWFA. The synchrotron-like betatron spectrum peaks at the critical energy, $E_{crit} \propto \gamma^2$.
Therefore the increase in electron $\gamma$ expected from increasing laser power will shift the critical energy of the photon spectra to higher energies.
WP-\ref{MP3_004} proposed a LWFA based scheme to create a tunable, ultra-bright $\gamma$ beam ($>10^{12}$ $\gamma$ photons per shot with a brilliance of $10^{26}$~photons/s/mm$^2$/mrad$^2$~per 0.1\% bandwidth) with energy conversion efficiency above 10\% for photons above 1~MeV. A double stage process is used; first an efficient laser-driven stage, then a second higher density beam driven stage to increase the trapped charge, enhancing the betatron photons \cite{Zhu_2020}.

\subsubsection{Photon sources from radiation reaction effects}
A bright, directional photon beam is also created during during direct laser acceleration (DLA) of electrons \cite{Kneip_2008} in a mechanism similar to the betatron source. 
The laser-induced oscillation of the electrons coupling with channel fields in the plasma act to confine the electrons in the laser focal region.
Increasing the intensity of the laser pulses eventually gives rise to the onset of a regime where a large fraction of the energy gained by electrons, usually directly from the laser, is radiated into photons as the electrons oscillate.

\vspace{11pt}
\noindent
\textbf{Radiation reaction effects using multi-petawatt lasers}\\
Numerical modeling studies have shown that plasma channels can sustain extremely strong azimuthal magnetic fields ($\sim$mega-Telsa for multi-PW conditions) that efficiently convert the electron energy into a dense $\gamma$-ray pulse (WP-\ref{MP3_008}).
Furthermore, the intensity increase for multi-PW conditions mean radiation reaction will begin to influence the DLA mechanism (see above).
Similarly, optimizing the density profile of a solid target should lead to very efficient gamma beam sources (WP-\ref{MP3_042}).
These intense $\gamma$ beams are a potential source for testing two-photon Breit-Wheeler pair production \cite{Drebot_2017,Wang_2020}, and photon-photon scattering \cite{Homma_2016}.


\subsubsection{Inverse Compton Scattering (ICS) sources}

Inverse Compton Scattering of laser photons from a relativistic electron beam leads to an upshift in photon energy by a factor $4 \gamma^2$, where $\gamma$ is the electron relativistic factor.
ICS promises collimated, ultrafast, tunable, and narrowband MeV and GeV photon beams.
Even for modest electron beam energies, ICS sources are attractive because the scattered photon energies can be $> 100$~keV.
A laser pulse scattering off a conventional RF-accelerator-generated electron beams is a well established x-ray source for many applications.
All-optical approaches, using high-quality LWFA electron beams to perform ICS experiments have been demonstrated.
Challenges for ICS include the need for a very high-quality (dense) electron beam, and overlapping the micron-scale electron bunch with the comparably sized laser pulse in space and time.
The quasi-monochromatic photon energy can be tuned by controlling the electron and laser parameters \cite{Khrennikov_2015}.
Increasing the laser intensity requires additional efforts to retain the narrow energy spread. It has been shown that a chirped laser pulse can compensate for the ponderomotive line-broadening in the scattered radiation \cite{Ghebregziabher_2013, Terzic_2014, Seipt_2015, Rykovanov_2016, Maroli_2018} (WP-\ref{MP3_015}).
Recent experiments have shown ICS schemes showing evidence for radiation reaction with photons with a critical energy $> 30$~MeV \cite{Cole_PRX}.

\vspace{11pt}
\noindent
\textbf{Inverse Compton Scattering using multi-petawatt lasers}\\
Prospects for ICS using multi-PW laser pulses look excellent. 
Not only can LWFA potentially produce higher quality electron beams with higher $\gamma$ values, but multi-PW lasers can deliver them with higher peak powers.
Properties of the ICS output beam may be better controlled through the properties of the laser pulse, such as the spectral chirp (WP-\ref{MP3_015}, WP-\ref{MP3_016}), manipulating the focusing properties to create caustics \cite{Kharin_2018}, or using polarization gating techniques to generate MeV frequency combs \cite{Valialshchikov_2021} (WP-\ref{MP3_016}).
Using electron beam with a correlated energy spread may reduce the spectral bandwidth \cite{Ruijter_2021} (WP-\ref{MP3_016}).
WP-\ref{MP3_034} \cite{Jirka_2020} proposed an alternative scheme using ultrahigh-intensity laser pulses with 10s of femtosecond duration to irradiate near-critical-density targets.
Prospects for ICS beyond all-optical schemes include coupling a multi-PW laser to a meter-long, x-band linac to generate 20-keV x-rays for probing warm dense matter states (WP-\ref{MP3_053}).
WP-\ref{MP3_017} looks to ICS experiments as a way to inspect the conditions of the coherently emitting region of pulsars and fast radio bursts.

\subsubsection{X-ray free electron lasers (XFELs)}
X-ray free electron lasers (XFELs) have proven a revolutionary tool to probe ultrafast, and ultra small phenomena across STEM fields.
An electron beam in an XFEL passes through an undulator (periodic magnets) that induces the beam to oscillate and emit radiation.
The radiation then continues to interact with the electron beam so it forms mircobunches at the wavelength of the emitted radiation.
These microbunches then begin to emit coherent, mono-chromatic radiation that reinforces the microbunching process. The energy extracted from the electrons generates a coherent x-ray beam.
Conventional XFEL facilities require kilometer-scale electron accelerators, which limits their availability due to size and expense.
This makes compact LWFA electron sources a desirable alternative to drive an XFEL and may even open the door to attosecond x-ray pulses. Making small-scale XFEL systems widely available can potentially transform science and applications (WP-\ref{MP3_092}), but this requires an electron beam of exceptional quality with a narrow energy spread, small emittance, and high current. This challenges the limits of what LWFA can achieve, but a breakthrough experiment has recently demonstrated the use of a LWFA to drive a 27~nm wavelength FEL \cite{Wang_2021}.

\vspace{11pt}
\noindent
\textbf{XFEL sources using multi-Petawatt lasers}\\
Potential improvements at multi-PW laser facilities in electron beam current and stability will directly translate to the success of future LWFA-driven XFELs.
Improvements to modeling and simulation are another tool that can further the optimize scheme for specific configurations (WP-\ref{MP3_005}). 
A further challenge for LWFA driven XFELs is increasing the repetition rate to compete with conventional facilities, with the laser technology being a bottleneck (WP-\ref{MP3_005}).
Schemes for all-optical free-electron-lasers are being developed, that use LWFA electron beams, and they replace the magnetic undulator with the electromagnetic fields of a laser pulse \cite{Gea-Banacloche_1987,Gallardo_1988,Steiniger_2016}. This reduces the undulator period to the laser wavelength of the and shrinks the overall size of the FEL.
One all-optical scheme uses traveling-wave Thomson Scattering (WP-\ref{MP3_094}).
A relatively long laser pulse with a few 100 to several 1000 cycles is required for FEL operation. The traveling-wave Thomson Scattering scheme creates the electron-laser overlap by passing the electron beam at angle to the a laser pulse with a compensating phase-front tilt.

\subsection{Broader Impacts of laser based particle and high-energy-photon sources}

New laser-driven particle acceleration technology could lead to better understanding of potential screening and high-field effects that may affect fusion cross sections relevant to fusion processes for future energy sources.
Intense, high-charge proton beams would enable fast ignition and warm-dense matter studies, and multi-100-MeV ion beams for cancer radiotherapy treatments.
Fundamental ion acceleration studies could develop sources using radiation pressure acceleration, magnetic vortex, single-cycle laser acceleration to generate ion beams with $>1$\,GeV particle energies that could access even more extreme regimes of nuclear physics.

Enabling advanced lights sources represents one of the most promising applications of laser-wakefield acceleration of electrons, with applications in medicine, high-energy-density and material sciences, and national security \cite{Corde_2013, Albert_2016}.

\section{MP3 joint strategies for needed diagnostic capabilities}
\label{sec-strategies}

The new capabilities needed to answer the MP3 science questions with flagship experiments identified in Chapters \ref{sec-ST1}-\ref{sec-ST3} using high-energy particle and photon sources identified in Chapter \ref{sec-PAALS} will require concerted efforts. Advancing ultrahigh-intensity scientific research depends on establishing a stable ecosystem that sustains existing laser facilities, upgrades them, and builds new ones to pursue new frontier science. Often, these efforts will exceed the capacity of any individual research institutes, or even national science and international organizations, so a key charge for the MP3 workshop included identifying common interests and joint strategies for developing \textbf{diagnostics needed for the flagship experiments}. 

This chapter addresses this charge, as well as \textbf{new theoretical constructs and computational capabilities} needed to guide and interpret the flagship experiments, and \textbf{novel targets} that can relieve some laser constraints by tailoring laser-matter interactions to achieve desired results. Chapter \ref{sec-future} addresses next-generation lasers needed for advancing multi-petawatt laser-based science that could benefit from coordinated efforts. This MP3 workshop report does not explicitly treat next-generation experimental systems, ultrahigh vacuum, and other support systems, since these advances typically require engineering to meet requirements at each specific facility.

\textbf{Research networks} offer excellent opportunities to pursue joint strategies. \textbf{Laserlab Europe} \cite{Laserlab}, an integrated initiative of European laser research infrastructures, fosters new research and developments in a flexible and coordinated fashion beyond the national scale. It brings together leading organizations in laser-based inter-disciplinary research from 18 countries and its main objectives include: maintaining a sustainable inter-disciplinary network of European laboratories; strengthening research through Joint Research Activities; and offering access to state-of-the-art laser research facilities to researchers from all fields of science and from any laboratory in order to perform world-class research. \textbf{LaserNetUS} \cite{LaserNetUS} the \textbf{Extreme Light Infrastructure European Research Infrastructure Consortium (ELI ERIC)} \cite{ELI_ERIC} provide similar benefits in the United State/Canada and across Europe, respectively.

The US National Science Foundation (NSF) recently announced funding of to develop an international ``network of networks'' to apply “extreme light” to advance the frontiers of science and engineering. The Ohio State University will lead development of this \textbf{Extreme Light in Intensity, Time, and Space (X-\textit{lites})} collaboration initiative. X-\textit{lites} aims to promote collaboration among researchers around the world to make full use of new laser facilities to improve communications among, and promote broad participation at the frontiers of laser-driven science.

\textbf{Adopting standards and sharing best practices} promises a way to reduce both capital and operating costs, while improving performance and maximizing compatibility, interoperability, safety, repeatability, and quality. Adopting standards depends on setting performance specifications and timelines that can meet broad needs, as outlined in this report. Road maps can identify facilities required to meet established scientific research needs, as well as needed technologies that can benefit by taking a standards-based approach to research and development. Developing and implementing technical standards requires consensus and compromise among parties involved, including both users and suppliers. All parties can realize mutual gains by making mutually consistent decisions.

One of the greatest benefits from research networks would be enhanced \textbf{network effects}. In economics, a network effect (also called network externality or demand-side economies of scale) is the phenomenon by which the value or utility a user derives from a good or service depends on the number of users of compatible products. Network effects are typically positive, resulting in a given user deriving more value from a product as more users join the same network. Upon reaching critical mass, a ``bandwagon effect'' can result. \cite{Net_Eff} 

\textbf{Standards and coordination} increase compatibility and interoperability, allowing information to be shared within a larger network and attracting more users of new technologies, further enhancing the network effects. Standardization can facilitate commoditization of formerly custom products and processes, which can lead to broader markets where only niche markets might otherwise exist. \textit{De facto standards} can arise organically, but coordinating research and development to satisfy the broader needs of the community and aligning it with a coordinated strategy generally prove more effective. Nonrecurring engineering costs can be spread across multiple units and users. Proven designs can be easily adapted and extended to new applications while minimizing cost and development time.

Near-term upgrades to existing facilities can serve two important purposes: (1) they meet the ever-increasing demands of frontier science at established research centers, and (2) they provide platforms for developing technology required in new facilities. Pursuing upgrades at existing infrastructure proves essential to realize future opportunities, and cooperation among university, national laboratory and industry stakeholders, as well as retain, renew and grow the talent base.


\subsection{Diagnostics}

The diagnostics for multi-petawatt physics experiments present excellent opportunities for exercising joint strategies and realizing the benefits of network effects. Some diagnostics can port to multiple facilities by defining standard interfaces or just by providing adaptable space and services. In other cases that demand substantial diagnostic installations, such as those requiring heavy shielding, multiple facilities can leverage common designs and technology to minimize cost and schedule with the added benefits of improving the ability to compare experimental results and to support the equipment with common logistics.

\begin{itemize}
\item \underline{Laser pulse peak intensity on shot at full energy} - measuring the intensity at focus represents a critical need for practically every ultraintense laser used for experiments. 
Typically measurements of the energy on target, the focal spot profile (taken at low power), and pulse duration by autocorrelation measurement are combined to estimate the peak laser intensity.
WP-\ref{MP3_035} discusses the prospects and challenges of measuring extreme intensities using Nobel gas ionization states.
WP-\ref{MP3_033} proposed a method using Thomson scattering of protons from within the focal volume to achieve an in-situ measurement for the intensities of $\sim 10^{24}$~Wcm$^2$ \cite{He_2019}.
WP-\ref{MP3_014}  proposed methods to generate electrons or electron-positron pairs with the desired density distribution at the surface of a thin flat foil target and detecting the number and angular distribution of electrons or positrons generated. 
The multiphoton Breit-Wheeler process, the dominant process responsible for pair creation, has a very sharp intensity threshold above I = 10$^{22}$ W/cm$^2$. 
Below this threshold, no pairs are created and above the number of pairs increases with laser intensity. 
This sensitivity of detectable positron number on laser intensity makes this diagnostic very precise. Joint and/or coordinated development of this approach can start with demonstrations at PW lasers using electrons and proceed to MPW lasers that exceed the threshold.

\item \underline{Colliding beam overlap} - The colliding electron beam and laser pulse experiments required to study SQ1B will be challenging. Both beams will need to have extremely stable pointing stability to overlap the micrometer scale electron beam with the tightly focused laser beam. WP-\ref{MP3_086} considered the development of a performance metric to assist with the optimization process.

\item \underline{High-energy X-ray diagnostics} – Multi-petawatt lasers being developed or commissioned around the world will generate copious multi-MeV photons associated. WP-\ref{MP3_019} describes diagnostics to measure  photons with energies greater than 100 keV. It also summarizes the requirements for successfully implementing these diagnostics on upcoming facilities. 
WP-\ref{MP3_039} considers the feasibility of using di-muon production as a diagnostic of the high-energy photons.

\item \underline{Phase-based X- and gamma ray diagnostics} – 
Novel phase-based diagnostics for laser-produced X-ray and gamma ray sources can measure pre-plasma electron density profiles. Phase-based (refraction and diffraction) measurements prove much more accurate and feasible at high photon energies than attenuation-based methods, because the real part of the complex index of refraction that determines the phase effects is orders of magnitude larger than the imaginary part determining the attenuation effects ($1/E^2$ scaling, versus $1/E^4$ energy scaling).
Phase-based X-ray measurements are well suited for diagnosing the extreme gradients expected in the dense pre-plasmas produced by high-temporal-contrast, multi-PW laser pulses (gradient scale lengths of a few microns), as refraction-based diagnostics measure directly the plasma electron density gradients. 
One phase-based X-ray method proposed for diagnosing multi-PW laser experiments is grating-based Talbot-Lau X-ray Deflectometry (TXD). (WP-\ref{MP3_007})


\item \underline{Laser-generated x-ray sources for diffraction experiments} – probing the lattice spacing of matter at extreme conditions often uses diffraction techniques that prove limited. Lasers heat a backlighter foil to generate helium-like, quasi-monoenergetic x-rays \cite{park_2021}. These sources convert only $\sim$1\% of the laser energy into x-rays that usually include emission lines with the spectral resolution $\Delta E_x/E_x\lesssim 0.6\,\%$, where $E_x$ is the backlighter x-ray energy, along with continuum bremsstrahlung components. Current capabilities limit these x-ray sources to only 10 keV photons, which prevents probing very high-pressure planetary core conditions. Diffraction by up to $\sim$MeV electrons or $\sim$20-100 keV x-rays would provide information about the long-range order of phase transitions at TPa pressures needs to study complex material properties and quantum phenomena that emerge at atomic pressures and temperatures relevant to planetary cores (\S~\ref{subsec-SQ2A}).

\item \underline{Positron detection} – the strong fields used to create a plasma fireball of matter (electrons), anti-matter (positrons) and photons described in \S~\ref{subsec-SQ1B} will require diagnostics to study the routes to positron generation via strong fields. 

\item \underline{Intense proton and neutron sources for probes} – Intense proton sources developed to drive r-process experiments with intense neutron fluxes described in \S~\ref{subsec-SQ3A} can also probe magnetic fields in plasmas studied in Chapters \ref{sec-ST1} and \ref{sec-ST2}, while the neutron sources can enable non-destructive imaging of dense materials and active interrogation for identifying nuclear materials. Joint and/or coordinated development of these sources at existing and upgraded PW and MPW facilities would advance many scientific areas.

\item \underline{Electromagnetic pulse (EMP)} – EMP generated by MPW laser-matter interaction will create a hostile environment and render many diagnostics vulnerable to disruption or even damage. An ongoing Expert Group Laser-generated EMP Workshop supported by Laserlab Europe and held five times since 2018 provides a venue for collaboratively addressing these issues.
\item \underline{High-count-rate diagnostics} – many existing experimental approaches depend on single-photon or single-particle counting techniques to analyze and differentiate outcomes. Some MPW experiments enable new regimes that require operating in a high-intensity, low-shot-rate mode cannot apply these techniques because the experimental signatures comprise copious outputs. These will require diagnostic strategies that can record and distinguish signals with potentially large backgrounds both coincident and time-delayed with respect to the signal.
\item \underline{Real-time data collection and analysis} – fast and accurate data processing can provide improved experimental productivity and flexibility when coupled with high shot-rate or high repetition rate lasers, especially if machine learning can be applied to the large data sets that result. 
\item \underline{Machine learning for 3D reconstruction of high-density laser plasmas} – 
WP-\ref{MP3_057} proposed to develop and demonstrate machine learning (ML) algorithms as a tool to reconstruct the 3D plasma plume using a high speed x-ray imaging from different lines of sight. The formation and time evolution of X-ray emission from the plasma formed when the laser pulse hits the target would be diagnosed so that the 2D radiographs are converted to 3D models. This will enable to identify the corresponding ion and proton acceleration phases based on plasma dynamics.
\end{itemize}

\subsection{Opportunities for joint development strategies}
Specific examples of diagnostics that would benefit from joint development strategies:
\begin{itemize}
\item Laser focal spot intensity
\item X-ray and gamma-ray light sources for probes
\item Gamma spectrometers for radiation up to E = 50~MeV via the conversion of gamma radiation by Li target into electron-positron pairs.
\item Triggered LaBr3 arrays to measure delayed gamma radiation from produced isotopes with lifetimes above the ns-regime.
\item Intense proton and neutron sources for probes
\item Thomson parabolas for proton energies above 200~MeV.
\item Time-of-flight MeV-neutron detection systems to measure neutron energy spectra.
\item Electromagnetic pulse (EMP) detectors and mitigation schemes
\end{itemize}

\section{MP3 vision for the optimal next-generation facility}
\label{sec-future}

\begin{center}
\fbox{\begin{minipage}{15cm}
\begin{center}
\underline{\textbf{The Adam Project (2022)}}\\
\textbf{``It's not meant to be easy. That is the beauty of physics ... that}\\
\textbf{is the beauty of life. We are meant to work on problems that our}\\
\textbf{children will solve. You will die before your life's work is done.''}
\end{center}
\end{minipage}}
\end{center}
\vspace{0.5cm}

The science questions identified and developed in Chapters \ref{sec-ST1}-\ref{sec-ST3} present grand challenges in three research themes – high-field physics, laboratory astrophysics and planetary physics, and nuclear physics – enabled by a new generation of ultra-intense and powerful lasers. Chapter \ref{sec-PAALS} presents research and development required to produce high-energy particle and photon sources using multi-petawatt lasers needed for the grand-challenge experiments. Research in the three research themes can start using existing petawatt and a few multi-petawatt laser facilities now coming online, some with upgrades to provide additional capabilities, but all would benefit from a next-generation facility with performance possible beyond the current state.
 
Science enabled by multi-petawatt lasers depends on several important laser parameters: pulse peak power (or peak intensity), pulse energy, pulse duration, focal spot size, and repetition rate. These parameters relate in relatively simple ways physically, but advancing all of these parameters simultaneously presents challenges that will entail compromises and optimizing across them for different experimental applications.

\begin{itemize}
\item \underline{Peak power (intensity)}: Laser power and focusing conditions determine the attainable laser intensity and field strengths. The record intensity from CoReLS using a 4-PW laser with tight focusing (56~J, 20~fs, 0.9~$\mu$m focus, 0.68  Strehl ratio) stands at just over $10^{23}$~W/cm$^2$ \cite{Yoon_2021}. Achieving significantly higher focused intensities requires higher pulse energy, shorter pulse durations, and/or tighter focusing. 
\item \underline{Pulse energy}: Kilojoule-class, multi-petawatt lasers promise 10$\times$ to 100$\times$ improvements in pulse energy in a single beamline. Combining the output of multiple beamlines may provide a path to even higher energies. Higher pulse energy can yield higher peak intensities with ultrashort pulses for some physics experiments to achieve threshold conditions, like producing electromagnet cascades (\S~\ref{sec-SQ1B-roadmap-new-cap}, Fig.\ \ref{fig:RoadmapSFQED}); or longer pulses with larger focal spots where peak intensity or power is not so important as the total number of secondary particles generated and accelerated, like solid-density neutron beams for laser-driven nucleosynthesis (\S~\ref{subsec-SQ3A}).
\item \underline{Pulse duration}: Shorter pulses require laser systems that can deliver broader bandwidth and/or employ nonlinear compression techniques. Demonstrated nonlinear compression techniques have shown to up to 6$\times$ improvements, as described below, but realizing this kind of performance at very high pulse energies with large beams poses challenges. 
\item \underline{Focal spot size}: Tighter focusing requires exquisite control and promises less than a factor of two given the CoReLS result; not much improvement is expected using conventional focusing schemes.
\item \underline{Repetition rate}: Increased experimental repetition rates provide a way to study low-probability events and reducing measurement uncertainties by producing large data sets and using statistical methods. Realizing higher shot rates and ultimately multi-Hz and even multi-kHz repetition rates requires technological advances in thermal management and high-average-power optics.
\end{itemize}

\subsection{Opportunities to advance laser performance}
The following highlights opportunities to advance laser parameters and offers a vision for next-generation, high-intensity lasers to address the frontier science goals.

\medspace

\textbf{Kilojoule-class, multi-petawatt lasers}: Most of the petawatt and the few multi-petawatt lasers operating now employ titanium-doped sapphire (Ti:Sa) laser amplification, which has likely reached its maximum output energy due to technical limitations. Optical parametric chirped pulse amplification (OPCPA) pumped by high-energy laser systems promise to surpass this limit and potentially increase the bandwidth to produce even shorter pulses. 

\begin{figure}[h]
\begin{center}
\includegraphics{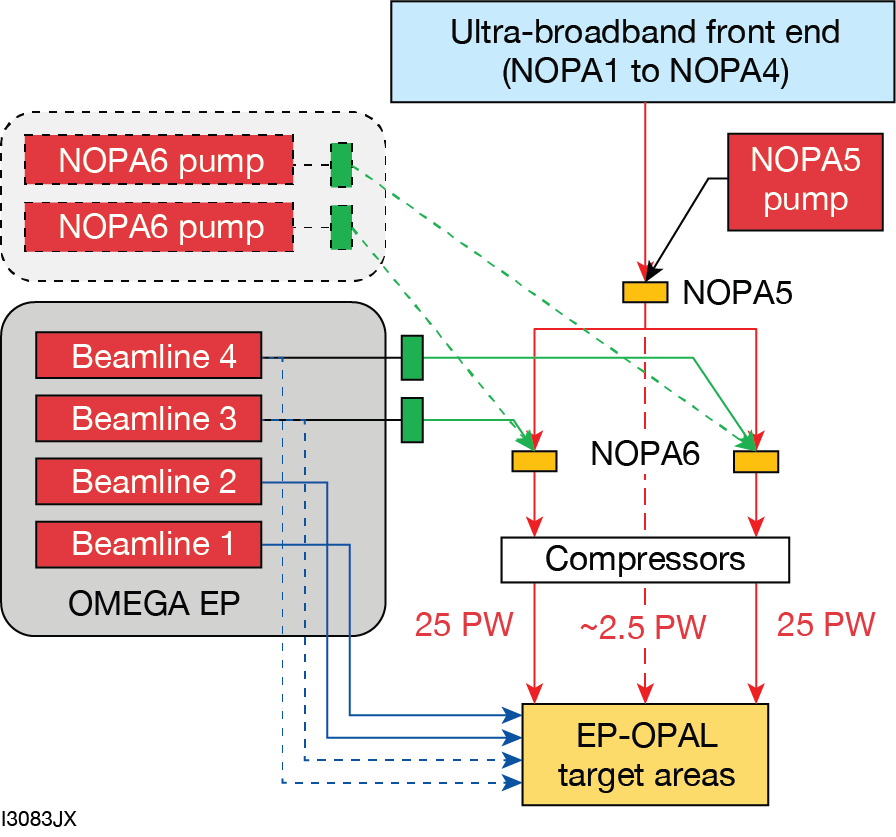}
\end{center}
\caption{Schematic for EP-OPAL facility proposed by UR/LLE.}
\label{Fig_7.1a}
\end{figure}

\begin{figure}[h]
\begin{center}
\includegraphics[scale=0.25]{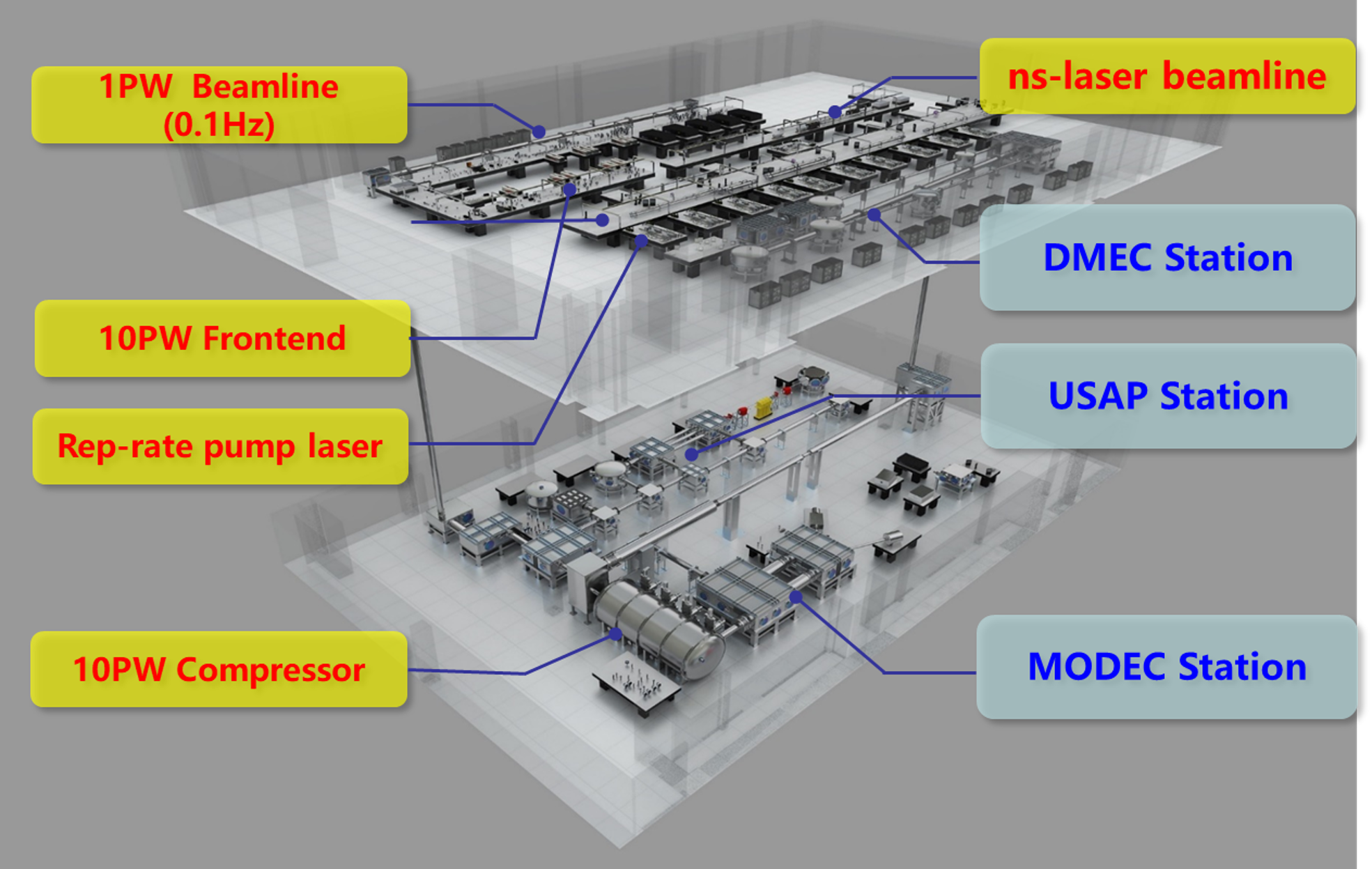}
\end{center}
\caption{CAD rendering of SEL-100PW project underway at SIOM \cite{Li_Poster}}
\label{Fig_7.1b}
\end{figure}

Several proposed facilities aim to deliver beamlines each delivering up to 25~PW (e.g.\ 500~J in 20~fs) and even possibly to 100 PW with advances in large optics required for larger beams. Figure \ref{Fig_7.1a} illustrates a concept proposed for an optical parametric amplifier line pumped by the OMEGA EP laser system (EP-OPAL) at the University of Rochester Laboratory for Laser Energetics (UR/LLE) \cite{Zuegel_Poster}. 
This all-OPCPA system approach promises to achieve high energy and ultrahigh temporal contrast by leveraging multi-kilojoule beamlines to pump large-aperture, highly deuterated dihydrogen potassium phosphate (DKDP) crystals in a ultrabroadband phase matching configuration \cite{Freidman_2002,Lozhkarev_2005}. 
Figure \ref{Fig_7.1b} shows a similar approach being pursued at the Shanghai Institute of Optics and Fine Mechanics (SIOM) for the Station of Extreme Light (SEL-100PW) that targets a 100-PW, 20-fs, 2 kJ beamline \cite{Li_Poster}. 
SEL-100PW will serve three end stations: Dynamics of Materials under Extreme Conditions (DMEC); Ultrafast Sub-atomic Physics (USAP); and Molecular Dynamics and Extremely fast Chemistry (MODEC).

\textbf{Plasma amplification and plasma optics:} As noted above, technical limitations of the optic size that can be feasibly manufactured is a challenge for creating higher power pulses. An alternative is to consider plasma amplification methods of the laser pulse and plasma optic, to replace conventional gratings for example. WP-\ref{MP3_036} highlighted the possibility of using multi-PW laser facilities to explore pushing the potential of plasma optics beyond the academic and apply to real systems.

\begin{figure}[h]
\begin{center}
\includegraphics{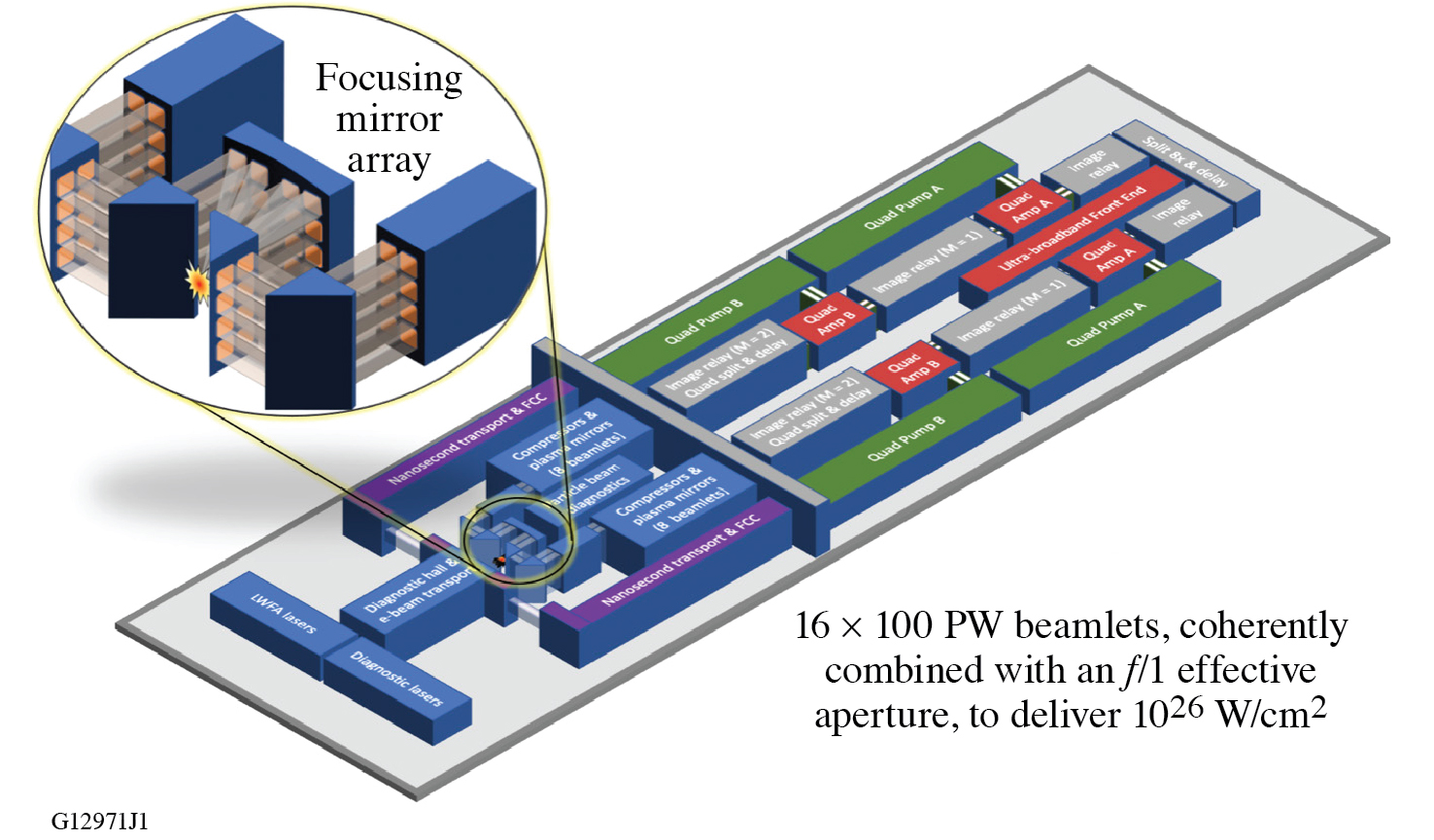}
\end{center}
\caption{Concept layout for an exawatt laser facility with an array of coherently combined 100-PW beams with a 4$\times$4 array of beamlines arranged to provide low-effective f-number focusing.}
\label{Fig_7.2}
\end{figure}

\textbf{Coherent combination of multi-PW modules}: Figure \ref{Fig_7.2} shows a concept developed at the 2019 Brightest Light Initiative (BLI) workshop. %
It envisions an approach to realize exawatt peak powers with an array of coherently combined 100-PW beams with low effective f-number focusing. The higher energy could be capable of reaching $10^{26}$~W/cm$^2$ with near-diffraction-limited focusing to an approximately 1~$\mu$m$^2$ spot. 

\begin{figure}[h]
\begin{center}
\includegraphics{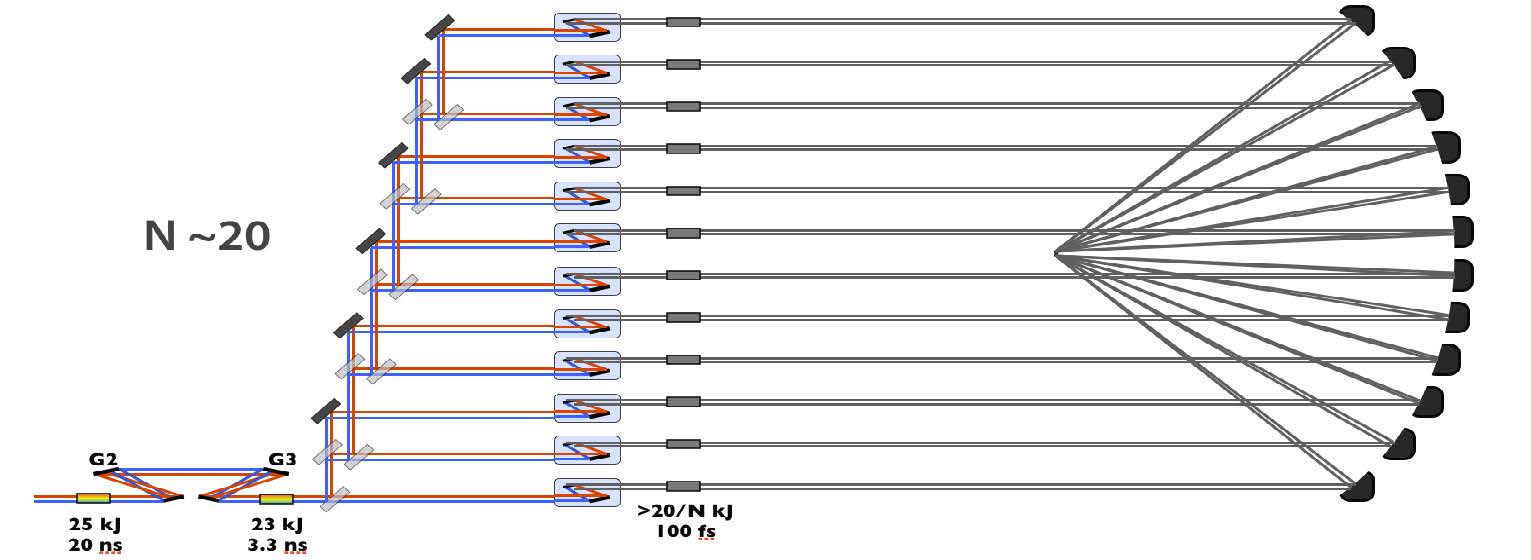}
\end{center}
\caption{Schematic of Nexawatt concept \cite{Barty_2016}.}
\label{Fig_7.3}
\end{figure}

Figure \ref{Fig_7.3} illustrates the ``Nexawatt'' concept \cite{Barty_2016} that coherently combines the output of high-energy, chirped-pulse amplification where grating and focusing optic size and damage thresholds typically limit the maximum compressed pulse energy. This scheme divides final pulse compression into multiple channels where each channel completes compression of the pulse and focuses it on target. Maximizing performance depends on matching precisely the dispersion and timing of all the channels. 

\begin{figure}[h]
\begin{center}
\includegraphics{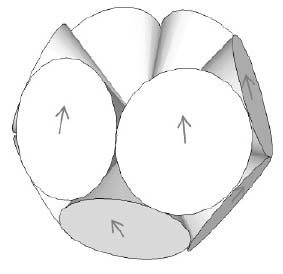}
\end{center}
\caption{A ``double-belt'' focusing configuration using coherent beam combination and f/1 focusing optics can approach the ultimate intensity or field strengths possible with a given amount of laser power.}
\label{Fig_7.4}
\end{figure}

Maximizing the electric field at a focusing point requires coherently combining multiple laser beams to reproduce a phase conjugated dipole radiation field. Figure \ref{Fig_7.4} illustrates a ``dipole focusing'' concept \cite{XCELS,Korzhimanov_2011,Afimenko_2018}, where a number of tightly focused laser pulses mimic such a dipole-like wave structure that can reduce the total power needed to reach a desired field intensity with a minimum dipole focusing volume, $V_{dp}\approx 0.032\,\lambda^3$ or $(0.32\,\lambda)^3$. 
A ``double-belt'' configuration with 12 beams using $f/1$ focusing optics can theoretically simulate a 10~PW electric-dipole wave with only 12~PW of laser power. This advanced form of coherent combination approaches the ultimate intensity or field strength possible with a given amount of laser power.

\textbf{Higher repetition rates}: Running any of the schemes noted above at higher repetition rates poses significant but surmountable technical challenges that can be met with appropriate research and development efforts. Lasers generate heat due to inherent quantum defects and this heat load can degrade performance or even damage the lasers without proper thermal management. Minimizing heat loads by using lower-quantum defects laser materials \cite{Phillips_2016}, and actively cooling the laser gain media \cite{Bayramian_2011} have enabled 10-Hz repetition rates at 100-J pulse energies that deliver kilowatt average powers. Scaling to higher pulse energies and repetition rates looks feasible but it will require investments and careful engineering. Figure \ref{fig:SQ3_fig3} illustrates another approach to realize multi-petawatt lasers that can operate at high repetition rates. Many petawatt lasers working simultaneously can essentially run multiple experiments in parallel to produce high-energy particles and drive nuclear reactions that do not depend on laser pulse coherence with experimentally useful yields. This ``multi-SHARC'' concept could lead to otherwise unreachable peak neutron fluxes of $10^{21}$ n/s/cm$^2$ and average fluxes of $10^{13}$ n/s.

\begin{figure}[t]
\begin{center}
\includegraphics[width=0.5\textwidth]{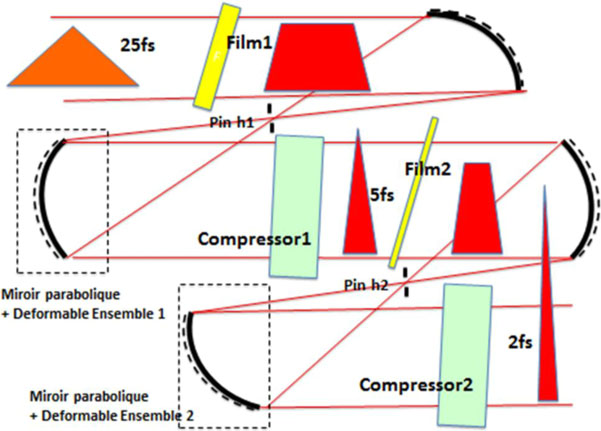}
\end{center}
\caption{Nonlinear pulse compression: Thin-film compressor}
\label{Fig_7.5a}
\end{figure}

\textbf{Shorter pulses using nonlinear mechanisms:} Figure \ref{Fig_7.5a} shows an approach to shorten pulse durations using self-phase modulation of ultrashort pulses to broaden their spectrum and subsequently compress them \cite{Mourou_2014}. 
It employs two stages of “thin-film” compressors that each might achieve several factors of compression to approach single-cycle pulses. 

\begin{figure}[h]
\begin{center}
\includegraphics{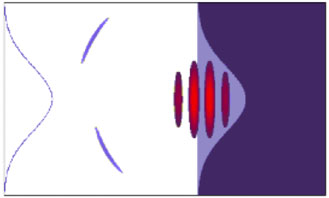}
\end{center}
\caption{Nonlinear pulse compression: Relativistic mirror/focusing}
\label{Fig_7.5b}
\end{figure}

Figure \ref{Fig_7.5b} illustrates another approach that implements a relativistic mirror with focusing, where the light pressure of a few-cycle pulse interaction with a solid target distorts the target \cite{Mourou_2014}.
The plasma takes the form of a shaped mirror that reflects and compresses the pulse by a factor proportion to $a_0$, the vector potential, into the attosecond or even the zeptosecond regime.

%
%
%
%

\subsection{Vision for a next-generation facility}
A multi-petawatt laser facility would ideally combine every ``trick in the book'' noted above to achieve the highest peak intensity, pulse energy and repetition rate possible to address all of the MP3 science questions. Achieving this ultimate goal with a single facility looks unlikely because the laser requirements vary greatly, but a flexible and comprehensive approach could make significant progress towards it. The first next-generation facility would implement at least two multi-10+ PW laser beamlines required for science questions that involve colliding electron-laser or laser-laser beams. These beamlines would provide adjustable pulse widths and focusing geometries in different target areas optimized for specific types of studies. A prudent approach would also make provisions for coherently combining beamlines, as well as applying nonlinear pulse compression and advanced focusing schemes to extend their performance. The facility would implement state-of-the-art technologies, like actively cooled kJ lasers with shot per minute shot rates, and provide clear upgrade pathways to provide multi-Hz and even higher repetition rates. Each of these steps depend on commensurate improvements in target and experimental diagnostic capabilities.




\addcontentsline{toc}{part}{References Cited}
\bibliographystyle{unsrt}
\bibliography{MP3-bib}

\includepdf[pages={-}]{figures/Blank_Page.pdf}

\part*{Appendices}

\appendix
\addappheadtotoc

\section{Workshop Participants and Additional Report Contributions}
\label{sec-WorkPart}

This appendix contains a list (alphabetical) of the registered participants of the workshop, both in-person participants (IP) and virtual participants (VP), as well as additional contributors to the final report.

\subsection{Workshop Participants}

In-Person participants:\\
Alexey Arefiev (UC San Diego, USA),
Sudeep Banerjee (Arizona State University, USA),
Tom	Blackburn (University of Gothenburg, SWEDEN),
Sergei V. Bulanov (ELI-Beamlines, CZECH REPUBLIC),
Stepan S. Bulanov (Lawrence Berkeley National Laboratory, USA),
Federico Canova (ELI ERIC, CZECH REPUBLIC),
John Collier (Central Laser Facility, UK),
Antonino Di Piazza (Max Planck Institute for Nuclear Physics, Heidelberg, GERMANY),
Domenico Doria (ELI-NP/ IFIN-HH, ROMANIA),
Roger Falcone (UC Berkeley, USA),
Cameron Geddes (Lawrence Berkeley National Laboratory, USA),
Evgeny Gelfer (ELI-Beamlines, CZECH REPUBLIC),
Anna Grassi (Sorbonne Universitè (LULI), FRANCE),
Mickael Grech (LULI, CNRS, Ecole Polytechnique, FRANCE),
Gianluca Gregori (University of Oxford, UK),
Harald Griesshammer (George Washington University, USA),
Gabriele Maria Grittani (ELI-Beamlines, CZECH REPUBLIC),
Wendell Hill (University of Maryland-College Park, USA),
Anton Ilderton (University of Edinburgh, UK),
Subhendu Kahaly (ELI-ALPS, HUNGARY),
Masaki Kando (National Institutes for Quantum Science and Technology, JAPAN),
Felix Karbstein (Helmholtz Institute Jena, GERMANY),
Karl M. Krushelnick (University of Michigan, USA),
G. Ravindra Kumar (Tata Institute of Fundamental Research, INDIA),
Sebastien Le Pape (LULI, FRANCE),
Rodrigo Lopez-Martens (ELI-ALPS, HUNGARY),
Stuart P. D. Mangles (Imperial College London, UK),
Mattias Marklund (University of Gothenburg, SWEDEN),
Judith McGovern (University of Manchester, UK),
Arseny Mironov (LULI, Sorbonne University, FRANCE),
Karoly Osvay (University of Szeged, HUNGARY),
Hye-Sook Park (Lawrence Livermore National Laboratory, USA),
Rajeev Pattathil (Rutherford Appleton Laboratory, UK),
Norbert Pietralla (TU Darmstadt, GERMANY),
Ishay Pomerantz (Tel Aviv University, ISRAEL),
Caterina Riconda (LULI - Sorbonne Université - CNRS, FRANCE),
Hans Rinderknecht (Laboratory for Laser Energetics, USA),
Carl Schroeder (Lawrence Berkeley National Laboratory, USA),
Luis O. Silva (Instituto Superior Tecnico, PORTUGAL),
Klaus Steiniger (Helmholtz-Zentrum Dresden-Rossendorf, GERMANY),
Matteo Tamburini (Max Planck Institute for Nuclear Physics, Heidelberg, GERMANY),
Elizabeth Tolman (Institute for Advanced Study, Princeton, USA),
Ion C. Turcu (Rutherford Appleton Laboratory, UK),
Dmitri Uzdensky (University of Colorado Boulder, USA),
Henri Vincenti (Commissariat à l'Energie Atomique, FRANCE),
Stefan Weber (Institute of Physics of the Czech Academy of Sciences \& ELI Beamlines, CZECH REPUBLIC),
Louise Willingale (University of Michigan, USA),
Jonathan Zuegel (Laboratory for Laser Energetics, USA).\\
\\
\\
Virtual Participants:\\
Naser Ahmadiniaz (Helmholtz-Zentrum Dresden-Rossendorf, GERMANY),
Reinhard Alkofer (University in Graz, AUSTRIA),
Anabella Araudo (ELI-Beamlines, CZECH REPUBLIC),
Stefan Ataman (ELI-NP, ROMANIA),
Christopher Barty (UC Irvine, USA),
Shikha Bhadoria (University of Gothenburg, SWEDEN),
Robert Bingham (Rutherford Appleton Laboratory, UK),
Federico Canova (ELI-ERIC),
Mihail Cernaianu (ELI-NP, ROMANIA),
Zenghu Chang (University of Central Florida, USA),
Hui Chen (Lawrence Livermore National Laboratory, USA),
Kyle Chesnut (UC Irvine, USA),
Thomas Cowan (Helmholtz-Zentrum Dresden-Rossendorf, GERMANY),
Xavier Davoine (CEA, FRANCE),
Matthias Diez (University in Graz, AUSTRIA),
Emmanuel d'Humières (University of Bordeaux, FRANCE),
Tilo Doeppner (Lawrence Livermore National Laboratory, USA),
Cesim Dumlu (ELI-NP, ROMANIA),
James Edwards (University of Plymouth, UK),
Frederico Fiuza (SLAC, USA),
Kirk Flippo (Los Alamos National Laboratory, USA),
Martin Formanek (Max Planck Institute for Nuclear Physics, Heidelberg, GERMANY),
Chad Forrest (Laboratory for Laser Energetics, USA),
Andrew Forsman (General Atomics, USA),
Henry Freund (University of New Mexico, USA),
Matthias Fuchs (University of Nebraska, Lincoln, USA),
David Garand (Sydor Technologies, USA),
Elias Gerstmayr (Stanford, USA),
Petru Ghenuche (ELI-NP, ROMANIA),
Holger Gies (Friedrich Schiller University Jena, GERMANY),
Siegfried Glenzer (SLAC, USA),
Thomas Grismayer (Instituto Superior Tecnico, PORTUGAL),
Stefan Haessler (Institut Polytechnique de Paris, ENSTA Paris, FRANCE),
Karen Hatsagortsyan (Max Planck Institute for Nuclear Physics, Heidelberg, GERMANY),
Calin Hojbota (Institute for Basic Science, SOUTH KOREA),
Axel H\"{o}rhager (European Investment Bank),
Calvin Howell (Duke University, USA),
Chengkun Huang (Los Alamos National Laboratory, USA),
Roman Hvezda (ELI-Beamlines, CZECH REPUBLIC),
Liangliang Ji (Shanghai Institute of Optics and Fine Mechanics, CHINA),
Chan Joshi (UCLA, USA),
Igor Jovanovic (University of Michigan, USA),
Christos Kamperidis (ELI-ALPS, HUNGARY),
Kiyong Kim (University of Maryland, USA),
Ben King (University of Plymouth, UK),
James Koga (National Institutes for Quantum Science and Technology, JAPAN),
Christian Kohlfürst (Helmholtz-Zentrum Dresden-Rossendorf, GERMANY),
Santhosh Krishnamurthy (Indian Institute of Technology, Hyderabad, INDIA),
Pawan Kumar (Raj Kumar Goel Institute Of Technology, INDIA),
Dario Lattuada (INFN, ITALY),
Zsolt Lécz (ELI-ALPS, HUNGARY),
Seong Ku Lee (Gwangju Institute of Science and Technology, SOUTH KOREA),
Feiyu Li (New Mexico Consortium),
Yan-Fei Li (Xi'an Jiaotong University, CHINA),
Jose Lopez (NSF, USA),
Misha A Lopez-Lopez (Instituto de F\'{i}sica y Matem\'{a}ticas, Universidad Michoacana de San Nicol\'{a}s de Hidalgo, MEXICO),
John Luginsland (Air Force, USA),
Vyacheslav Lukin (NSF, USA),
Maxim Lyutikov (Purdue University, USA),
Bertrand Martinez (Instituto Superior Tecnico, PORTUGAL),
Paul McKenna (University of Strathclyde, UK),
Sebastian Meuren (Stanford, USA),
Howard Milchberg (University of Maryland, USA),
Mohammad Mirzaie (Institute for Basic Science, SOUTH KOREA),
Elena Mosman (Tomsk Polytechnic University, RUSSIA),
Chang Hee Nam (Institute for Basic Science, SOUTH KOREA),
Eric Nelson (UC Irvine, USA),
Karoly Osvay (University of Szeged, HUNGARY),
John Palastro (Laboratory for Laser Energetics, USA),
Vishwa Bandhu Pathak (Institute for Basic Science, SOUTH KOREA),
Francesco Pegoraro (University of Pisa, ITALY),
Daniel Phillips (Ohio University, USA),
Tobias Podszus (Max Planck Institute for Nuclear Physics, Heidelberg, GERMANY),
Brian Reville  (Max Planck Institute for Nuclear Physics, Heidelberg, GERMANY),
Xavier Ribeyre (University of Bordeaux, FRANCE),
Luis Roso (Centro De Laseres Pulsados, SPAIN),
Markus Roth (TU-Darmstadt, GERMANY),
Chang-Mo Ryu (Pohang University of Science and Technology, SOUTH KOREA),
Antonin Sainte-Marie (CEA, FRANCE),
Subir Sarkar (University of Oxford, UK),
Gianluca Sarri (Queens University Belfast, UK),
Hiroshi Sawada (University of Nevada, Reno, USA),
Derek Schaeffer (Princeton University, USA),
Christian Schubert (RWTH Aachen University, GERMANY),
Daniel Seipt (Helmholtz Institute Jena, GERMANY),
Jessica Shaw (Laboratory for Laser Energetics, USA),
Ronnie Shepherd (Lawrence Livermore National Laboratory, USA),
Yuan Shi (Lawrence Livermore National Laboratory, USA),
Gennady Shvets (Cornell University, USA),
Christophe Simon-Boisson (Thales, FRANCE),
Klaus Spohr (ELI-NP, ROMANIA),
Dan Stutman (ELI-NP, ROMANIA),
François Sylla (SourceLAB, FRANCE),
Kazuo Tanaka (ELI-NP, ROMANIA),
Balsa Terzic (Old Dominion University, USA),
Peter Thirolf (Ludwig-Maximilians-Universit\"{a}t M\"{u}nchen, GERMANY),
Toma Toncian (Helmholtz-Zentrum Dresden-Rossendorf, GERMANY),
Csaba Toth (Lawrence Berkeley National Laboratory, USA),
Marija Vranic (Instituto Superior Tecnico, PORTUGAL),
Mingsheng Wei (Laboratory for Laser Energetics, USA),
Scott Wilks (Lawrence Livermore National Laboratory, USA),
Yuanbin Wu (Max Planck Institute for Nuclear Physics, Heidelberg, GERMANY),
Weipeng Yao (Institut Polytechnique de Paris, FRANCE),
Kaikai Zhang (Amplitude Laser, USA).

\subsection{Additional Report Contributors}
Henri Vincenti (CEA, France),
Charles Wang (University of Aberdeen, UK)

\includepdf[pages={-}]{figures/Blank_Page.pdf}

\section{Acronym Glossary}
\label{sec-glossary}

\begin{itemize}
\item AGN: Active galactic nucleus 
\item ASE: Amplified Spontaneous Emission
\item BH: Black hole
\item BLI: Brightest Light Initiative
\item BOA: Break-Out Afterburner
\item CPA: Chirped pulse amplification
\item CR: Cosmic ray(s)
\item CSA: Collisionless Shock Acceleration
\item DC: Direct-current
\item DKDP: deuterated dihydrogen potassium phosphate
\item DLA: Direct Laser Acceleration
\item DMEC: Dynamics of Materials under Extreme Conditions
\item DOE: Department of Energy
\item EFT: effective field theories
\item ELI: Extreme Light Infrastructure
\item ELI-BL: ELI-Beamlines
\item ELI ERIC: Extreme Light Infrastructure European Research Infrastructure Consortium
\item ELI-NP: ELI-Nuclear Physics
\item EM: Electromagnetic
\item EMP: Electromagnetic pulse
\item EU: European Union
\item EW: exawatt
\item FAIR: Facility for Antiproton and Ion Research
\item FRIB: Facility for Rare Isotope Beams
\item GRB: Gamma ray burst
\item HED: High-energy-density
\item HFP/QED: high-field physics and quantum electrodynamics
\item ICS: inverse Compton scattering
\item LAPP: laboratory astrophysics and planetary physics
\item LBNL: Lawrence Berkeley National Laboratory
\item LCFA: Local constant field approximation
\item LDNP: laser-driven nuclear physics
\item LWFA: Laser wakefield acceleration
\item MCLP: multiple colliding laser pulses
\item ML: machine learning
\item MODEC: Molecular Dynamics and Extremely fast Chemistry
\item MP3: Multi-Petawatt Physics Prioritization
\item MPQ: Max-Planck-Institute for Quantum Optics
\item MPW: Multi-Petawatt
\item MVA: Magnetic Vortex Acceleration
\item NIF: National Ignition Facility
\item NN: nucleon-nucleon
\item NS: Neutron star
\item NSAC: Nuclear Science Advisory Committee
\item NSF: National Science Foundation
\item OPCPA: optical parametric chirped-pulse amplification
\item PAALS: particle acceleration and advanced light sources
\item PIC: Particle-in-cell
\item PV: parity violation
\item PW: Petawatt
\item QCD: Quantum chromodynamics
\item QED: Quantum electrodynamics
\item RF: Radio-frequency
\item RHHG: Relativistic high-order harmonic generation
\item RPA: Radiation Pressure Acceleration
\item SEL-100PW: Station of Extreme Light
\item SCLA: Single-Cycle Laser Acceleration
\item SF: Strong-field
\item SF-QED: Strong-field quantum electrodynamics
\item SHARC: Scalable High-power Advanced Radiographic Capability
\item SIOM Shanghai Institute of Optics and Fine Mechanics
\item SN: supernovae
\item SQ: Science Question
\item STEM: Science technology engineering and medicine
\item Ti:Sa: titanium-doped sapphire
\item TNSA: target normal sheath acceleration
\item TXD: Talbot-Lau X-ray Deflectometry
\item UHECR: ultra-high-energy cosmic ray(s) 
\item UR/LLE: University of Rochester Laboratory for Laser Energetics
\item US: United States
\item USAP: Ultrafast Sub-atomic Physics
\item WG: Working group
\item WP: White Paper
\item XFEL: x-ray free electron laser
\item X-lites: Extreme Light in Intensity, Time, and Space
\item XUV: Extreme Ultraviolet
\item ZEUS: Zettawatt-Equivalent Ultrashort pulse laser System 




\end{itemize}


\includepdf[pages={-}]{figures/Blank_Page.pdf}

\section{White Papers}
\label{sec-WP}

This appendix contains a list of the white paper author(s) and titles that were received for the MP3 workshop. Listed in the order they were received. The list is linked to the full text of each white paper as they were submitted.

\begin{enumerate}
   
\item Markus Roth, Marc Zimmer, Stefan Scheuren, Christian R\"{o}del, \textit{Laser Neutron Production entering the Spallation Mode}
\label{MP3_002}

\item J.P. Palastro, A. Arefiev, J. Bromage, M. Campbell, A. Di Piazza, M. Formanek, P. Franke, D.H. Froula, B. Malaca, W. Mori, J. Pierce, D. Ramsey, J.L. Shaw, T.T. Simpson, J. Vieira, M. Vranic, K. Weichman, J. Zuegel, \textit{Novel Plasma Physics Regimes enabled by Spatiotemporally Shaped Laser Pulses}
\label{MP3_003}

\item Xing-Long Zhu, and Zheng-Ming Sheng, \textit{Ultra-brilliant GeV gamma-ray emission with multi-petawatt lasers}
\label{MP3_004}

\item Henry P. Freund, \textit{Advanced Light Sources with Polarization Control Based upon Laser-Based Accelerators}
\label{MP3_005}

\item Yasuhiko Sentoku, Natsumi Iwata, Mamiko Nishiuchi, Alexey Arefiev, \textit{Dynamics of multi-petawatt laser driven high Z radiative plasmas}
\label{MP3_006}

\item D. Stutman, M.P. Valdivia, D. Kumar, M. O. Cernaianu, D. Doria, P. Ghenuche, N. Safca, \textit{Phase-based X and gamma ray diagnostics for multi-Petawatt laser research}
\label{MP3_007}

\item Hans Rinderknecht, A. Arefiev, T. Toncian, M. Wei, T. Wang, G. Bruhaug, A. Laso Garcia, K. Weichman, J. Palastro, D. Doria, K. Spohr, M. Cernaianu, P. Ghenuche, J. Williams, A. Haid, T. Ditmire, H. Quevedo, and Dan Stutman, \textit{Relativistically transparent magnetic filaments: a platform for high-field science and efficient gamma-ray sources}
\label{MP3_008}

\item Kathleen Weichman, M. Murakami, J.J. Honrubia, S.V. Bulanov, J.K. Koga, M.A.H. Zosa, A.V. Arefiev, and J. Palastro, \textit{Microcavity implosions for the generation of extreme electric fields and ultrahigh magnetic fields}
\label{MP3_009}

\item Oleg B.\ Shiryaev, Michael Yu.\ Romanovsky, and Vladimir V.\ Bukin, \textit{Vacuum post-acceleration of relativistic electrons by combinations of THz electromagnetic pulses and constant magnetic fields}
\label{MP3_010}

\item Stephane Branly, Franck Falcoz, Catalin Neacsu, Kaikai Zhang and Pierre-Mary Paul, \textit{Status and future of high energy high average power lasers}
\label{MP3_011}

\item James K. Koga, Masaki Kando, Sergei V. Bulanov, Timur Zh. Esirkepov, Alexander S. Pirozhkov, Petr Valenta, Tae Moon Jeong, Georg Korn, Stepan S. Bulanov, \textit{Relativistic Flying Mirrors for Fundamental Science and Applications}
\label{MP3_012}

\item T. Heinzl, A. Ilderton, and B. King, \textit{Laser-Matter Interactions at the Intensity Frontier}
\label{MP3_013}

\item Zsolt L\'{e}cz1, and Alexander Andreev, \textit{Positron detection for plasma and field diagnostics in high-power laser-solid interactions}
\label{MP3_014}

\item Bal\v{s}a Terzi\'{c}, and Geoffrey Krafft, \textit{Improving Performance of Inverse Compton Sources in High-Field Regime via Laser Chirping}
\label{MP3_015}

\item D. Seipt, M. Zepf, and S. G. Rykovanov, \textit{Advanced Concepts for Monoenergetic $\gamma$-Ray Generation using Nonlinear Compton Scattering}
\label{MP3_016}

\item Shuta J. Tanaka, Keita Seto, Yasuhiro Kuramitsu, Yuji Fukuda, and Youichi Sakawa, \textit{Experimental observation of induced Compton scattering}
\label{MP3_017}

\item Felix Karbstein, Holger Gies, Elena A. Mosman, Gerhard
G. Paulus, Kai S. Schulze, and Thomas St\"{o}hlker, \textit{Co-location of a multi-petawatt laser with an x-ray free electron laser}
\label{MP3_018}

\item Deepak Kumar, Chris Armstrong, Florian Condamine, Thomas Cowan, Alejandro Laso Garcia, Tae Moon Jeong, Ond\v{r}ej Klimo, Paul McKenna, Alexander Pirozhkov, Sushil Singh, Dan Stutman, Vladimir Tikhonchuk, Roberto Versaci, June Wicks, \textit{High energy X-ray diagnostics for interaction of a high repetition rate multi-petawatt laser with solid targets}
\label{MP3_019}

\item Holger Gies, Felix Karbstein, J\"{o}rg Schreiber, and Matt Zepf, \textit{Quantum vacuum signatures in laser pulse collisions}
\label{MP3_020}

\item Hiroshi Sawada, \textit{Relativistic Electron Isochoric Heating with Petawatt Femtosecond Lasers for Creation of Warm Dense Matter}
\label{MP3_021}

\item Peter V. Heuer, Hans G. Rinderknecht, Derek B. Schaeffer, Scott Feister, \textit{The Role of High Repetition Rate Experiments in Advancing HEDP Science}
\label{MP3_022}

\item Karl Krushelnick, \textit{Laser wakefield acceleration with Multi Petawatt lasers}
\label{MP3_023}

\item Matteo Tamburini, Antonino Di Piazza, Christoph H. Keitel, \textit{The quest for precision measurements of strong-field QED effects}
\label{MP3_024}

\item Yan-Fei Li, Wei-Min Wang, Yu-Tong Li, \textit{The interaction of a multi-petawatt laser pulse with an ultrarelativistic electron beam could be utilized to produce polarized positron-beam source or conduct confirmatory experiments on QED theory}
\label{MP3_025}

\item Bruce A. Remington, Hui Chen, \textit{Relativistic astrophysical jets from accreting massive black holes in galactic centers}
\label{MP3_026}

\item Hui Chen, Frederico Fiuza, and Bruce Remington, \textit{Laser produced relativistic pair plasmas in the laboratory}
\label{MP3_027}

\item Stephanie Hansen, and Brent Jones, \textit{MP3 Opportunities for Benchmark-quality Lab-Astro Experiments}
\label{MP3_028}

\item Reinhard Alkofer, Matthias Diez, and Christian Kohlf\"{u}rst \textit{What are the time-scales of particle formation in the Schwinger effect?}
\label{MP3_029}

\item Ronnie Shepherd, Mark Sherlock, Max Tabak, Steve Libby, Yuan Shi, Nathaniel Roth, Hui Chen, Howard Scott, \textit{Soft X-ray emission spectroscopy to search for Unruh radiation}
\label{MP3_030}

\item Wendell T Hill, III, Andrew Longman, Calvin Z He, Smrithan Ravichandran, José Antonio Pérez-Hernández, Jon I Apiñaniz, Roberto Lera, Luis Roso, and Robert Fedosejevs, \textit{Direct Measure of Quantum Vacuum Properties}
\label{MP3_031}

\item Luis Roso, Andrew Longman, Calvin Z He, Smrithan Ravichandran, Jose Antonio Pérez-Hernández, Jon I Apiñaniz, Roberto Lera, Robert Fedosejevs, and Wendell T Hill, III \textit{Radiation from the electron-positron virtual pairs accompanying laser driven electrons.}
\label{MP3_032}

\item Robert Fedosejevs, Andrew Longman, Calvin Z He, Smrithan Ravichandran, Jose Antonio Pérez-Hernández, Jon I Apiñaniz, Roberto Lera, Luis Roso, and Wendell T Hill, III, \textit{In situ intensity gauge adapted for extremely-high intensities}
\label{MP3_033}

\item Ondrej Klimo, Stefan Weber, \textit{Controlled $\gamma$-ray secondary source}
\label{MP3_034}

\item Marcelo Ciappina, Stefan Weber, \textit{Atomic high-field diagnostic using multiple sequential ionization}
\label{MP3_035}

\item Caterina Riconda, Stefan Weber, \textit{Plasma Amplification for High-Power Lasers}
\label{MP3_036}

\item N. Iwata, Y. Sentoku, A. J. Kemp, and S. C. Wilks, \textit{Stochastic laser-plasma interaction with kJ-class petawatt laser light}
\label{MP3_037}

\item Timur Zh. Esirkepov, Akito Sagisaka, Koichi Ogura, Hiromitsu Kiriyama, James K. Koga, Masaki Kando, Danila R. Khikhluha, Ilia P. Tsygvintsev, Chris D. Armstrong, Deepak Kumar, Stepan S. Bulanov, Georg Korn, Sergei V. Bulanov, Alexander S. Pirozhkov, \textit{Experimental Access to Laser-Driven Gamma Flare}
\label{MP3_038}

\item Calin Ioan Hojbota, Mohammad Rezaei-Pandari, Mohammad Mirzaie, Vishwa Bandhu Pathak, Doyeon Kim, Jeongho Jeon, Kiyong Kim, Chang Hee Nam, \textit{Muon production for the detection of high-energy photons from laser-electron collisions}
\label{MP3_039}

\item G. Sarri, M. Streeter, L. Calvin, N. Cavanagh, \textit{High-energy and high-quality positron beams from a laser wake-field accelerator}
\label{MP3_040}

\item G. Sarri, M. Streeter, L. Calvin, N. Cavanagh, K. Fleck \textit{High-field quantum electrodynamics in the field of an intense laser}
\label{MP3_041}

\item Sergei V. Bulanov, Prokopis Hadjisolomou, Gabriele Grittani, Tae Moon Jeong, Georg Korn, Danila R. Khikhlukha, Pavel V. Sasorov, Petr Valenta, Kirill V. Lezhnin, Ilia P. Tsygvintsev, Vladimir A. Gasilov, Timur Zh. Esirkepov, Alexander S. Pirozhkov, James K. Koga, Masaki Kando, Tetsuya Kawachi, \textit{High Power Gamma Flashes Generated in Multi-Petawatt Laser-Matter Interaction for Fundamental Science and Applications}
\label{MP3_042}

\item Thomas G. White, Matthew Oliver, and Tilo D\"{o}ppner, \textit{Measurements of Transport Properties in Warm Dense Matter with Multi-Petawatt Lasers}
\label{MP3_043}

\item K. Matsuo, M. Bailly-Grandvaux, S. Bolanos, D. Kawahito, C. McGuffey,
M. Dozières, J. Peebles, J. R. Davies, M.S. Wei, W. Theobald, P.-A. Gourdain, J. Honrubia, J.J. Santos, and F.N. Beg, \textit{Energy deposition in magnetized dense plasma by laser-driven relativistic electrons}
\label{MP3_044}

\item Matthias Fuchs \textit{Next-generation Lightsources and Applications}
\label{MP3_045}

\item Lili Manzo, Matthew Edwards, Yuan Shi, \textit{Enhancing hot electron generation via magnetized laser absorption}
\label{MP3_046}

\item J. Kim, D. Mariscal, B. Djordjevic, \textit{Efficient laser-driven ion acceleration using temporally shaped high-intensity pulses}
\label{MP3_047}

\item C. B. Schroeder, C. Benedetti, E. Esarey, C. G. R. Geddes, J. van Tilborg, \textit{Laser wakefield acceleration using multi-PW lasers}
\label{MP3_048}

\item F\'{e}licie Albert \textit{X-ray light sources from laser plasma acceleration and their applications}
\label{MP3_049}

\item Andreas Kemp, S.C.Wilks, G.Grim, E.Hartouni, S.Kerr, G.Cochran \textit{Development of a new laser-based platform capable of measuring nuclear reaction rates in hot plasmas}
\label{MP3_050}

\item Amina E. Hussein, \textit{Multi-petawatt electron acceleration and strong magnetic field generation}
\label{MP3_051}

\item G. Gregori, S. Sarkar, R. Bingham, C. H.-T. Wang, \textit{Going Beyond the Standard Model with High Power Lasers}
\label{MP3_052}

\item Pierre Gourdain, and Georg Hoffstaetter, \textit{Using inverse Compton scattering from a short pulse laser to measure the properties of warm dense matter produced by pulsed-power drivers}
\label{MP3_053}

\item P. Neumayer, V. Bagnoud, \textit{Research at the high-energy short-pulse laser system PHELIX user facility}
\label{MP3_054}

\item Keita Seto, Jian Fuh Ong, Yoshihide Nakamiya, Mihai Cuciuc, Stefan Ataman, Cesim
Dumlu, Madalin Rosu, Ovidiu Tesileanu, Takahisa Jitsuno, and Kazuo A. Tanaka, \textit{Radiation reaction experiment with 60 PW laser pulse and 10 GeV electron bunch}
\label{MP3_055}

\item Bruce A. Remington and Hui Chen, \textit{High power laser experiments probing plasma nuclear science}
\label{MP3_056}

\item Alexandru Măgureanu, Viorel Nastasa, Deepak Sangwan, Bogdan Diaconescu, Dan G. Ghiţă, Marius Gugiu, Theodor Asavei, Zhehui Wang, Cătălin M. Ticoş, \textit{Machine Learning for 3D Reconstruction of High Density Laser-Plasma}
\label{MP3_057}

\item Andrei Tudor Patrascu, \textit{Unruh Radiation Detection}
\label{MP3_058}

\item K. M. Spohr, D. Doria, K. A. Tanaka, G. Bruhaug, C. Forrest, and H. Rinderknecht, et al., \textit{Nuclear physics research in laser induced plasma with a multi-petawatt laser system: production and decay studies of cosmogenic 26Al}
\label{MP3_059}

\item K. M. Spohr, D. Doria, K. A. Tanaka, M. O. Cernaianu, D. O’Donnell, V. Nastasa, P. Ghenuche, P. A. Söderström, G. Bruhaug, C. Forrest, and H. Rinderknecht, \textit{Laser-driven population and release of the 2.4 MeV isomer in 93Mo, towards a `Nuclear Battery'}
\label{MP3_060}

\item Cesim K. Dumlu, Keita Seto, Stefan Ataman, Jian-Fuh Ong, Ovidiu Tesileanu and Kazuo A. Tanaka, \textit{A Proposal for the Observation of Pair Production in Intense Field Regime}
\label{MP3_061}

\item M. Vranic, T. Grismayer, A. AreBiev, H. Chen, S. Corde, R. A. Fonseca, M. Grech, C. Joshi, A. G. R. Thomas, B. Martinez, S. Meuren, W. B. Mori, C. Riconda, C. Ridgers, D. Seipt, L. O. Silva, T. Silva, J. Vieira, L. Willingale, \textit{Positron creation and acceleration with multi-petawatt lasers}
\label{MP3_062}

\item Stefan Ataman, Keita Seto, Cesim Dumlu, Jian-Fuh Ong, Andrei Berceanu, Ovidiu Tesileanu, Kazuo A. Tanaka, \textit{MP3 all-optical vacuum birefringence experiment proposal}
\label{MP3_063}

\item Domenico Doria, A. Arefiev, M. O. Cernaianu, P. Ghenuche, K. Spohr, K. A. Tanaka, \textit{Laser driven-acceleration of ion and energy scaling modification by competitive processes.}
\label{MP3_064}

\item Edison Liang, \textit{High-Energy Astrophysics Experiments Based on Ultra-intense Lasers Irradiating Solid Targets}
\label{MP3_065}

\item Sergei V. Bulanov, Yan Gun Gu, \textit{Magnetic field annihilation and charged particle acceleration in ultra-relativistic laser plasmas}
\label{MP3_066}

\item P. Ghenuche, M. O. Cernaianu, D. Stutman, F. Negoita, D.L. Balabanski, A. V. Arefiev, D. Schumacher, and, D. Doria, \textit{Enhanced laser-plasma experiments with helical laser beams}
\label{MP3_067}

\item Edmond Turcu, \textit{Electron-Positron Pair Production Interaction Chambers with Counter-propagating, Colliding multi-PW Laser Pulses}
\label{MP3_068}

\item M. Grech, L. Lancia, I. A. Andriyash, P. Audebert, A. Beck, S. Corde, X. Davoine, M. Frotin, A. Grassi, L. Gremillet, S. Le Pape, M. Lobet, A. Leblanc, F. Mathieu, F. Massimo, A. Mercuri-Baron, D. Papadopoulos, F. P\'{e}rez, J. Prudent, C. Riconda, L. Romagnani, A. Specka, C. Thaury, K. Ta Phuoc, T. Vinci, \textit{Investigating strong-field QED processes in laser-electron beam collisions at Apollon}
\label{MP3_069}

\item C. A. J. Palmer, B. Reville, A. R. Bell, A. F. A. Bott, H. Chen, G. Gregori, D. Lamb, C. K. Li, J. Matthews, J. Meinecke, H-S.Park, R. Petrasso, S. Sarkar, A. Schekochihin, L. Silva, A. Robinson, P. Tzeferacos, and M. Vranic, \textit{Exploring cosmic-ray physics with high-energy, high-intensity lasers}
\label{MP3_070}

\item Jian-Xing Li, Yue-Yue Chen, Yan-Fei Li, Karen Z. Hatsagortsyan, Christoph H. Keitel, \textit{Generation of polarized high-energy lepton and photon beams}
\label{MP3_071}

\item Sergei V. Bulanov, Gabriele Grittani, Tae Moon Jeong, Martin Jirka, Georg Korn, Pavel V. Sasorov, Timur Zh. Esirkepov, James K. Koga, Stepan S. Bulanov, Francesco Pegoraro, \textit{Nonlinear Electromagnetic Waves in Quantum Vacuum}
\label{MP3_072}

\item Matteo Tamburini, Brian Reville, Jim Hinton, Christoph H. Keitel, \textit{Laboratory gamma-electron-positron plasma – a new frontier in strong-field QED and relativistic laboratory astrophysics}
\label{MP3_073}

\item Sergei V. Bulanov, Timur Z. Esirkepov, Stepan S. Bulanov, \textit{Compact Hadron Collider with the Ion Accelerator Using the Laser Radiation Pressure}
\label{MP3_074}

\item J. Fuchs, W. Yao, V. Horny, K. Burdonov, C. Ruyer, L. Gremillet. X. Davoine, E. d’Humières, X. Ribeyre, S.N. Chen, D. Schaeffer, W. Fox, P. Heuer, \textit{Towards investigating astrophysically relevant fast collisionless shocks}
\label{MP3_075}

\item Sebastian Meuren, David A. Reis, Roger Blandford, Phil H. Bucksbaum, Nathaniel
J. Fisch, Frederico Fiuza, Elias Gerstmayr, Siegfried Glenzer, Mark J. Hogan, Claudio
Pellegrini, Michael E. Peskin, Kenan Qu, Glen White, and Vitaly Yakimenko, \textit{Research Opportunities Enabled by Co-locating Multi-Petawatt Lasers with Dense Ultra-Relativistic Electron Beams}
\label{MP3_076}

\item Vishwa Bandhu Pathak, Calin Ioan Hojbota, Mohammad Mirzaie, and Chang Hee Nam, \textit{Electron acceleration beyond the single-stage laser wakefield acceleration}
\label{MP3_077}

\item J. Fuchs, V. Horny, K. Burdonov, V. Lelasseux, W. Yao, L. Gremillet. X. Davoine, E. d’Humi\`{e}res, X. Rib\`{e}re, S.N. Chen, F. Negoita, P-A S\"{o}derstr\"{o}m, I. Pomerantz, \textit{Opening the experimental investigation of heavy elements nucleosynthesis through extreme brightness neutron beams generated by multi-PetaWatt-class lasers}
\label{MP3_078}

\item P. Zhang, S. S. Bulanov, D. Seipt, A. V. Arefiev, and A. G. R. Thomas, \textit{Relativistic Plasma Physics in Supercritical Fields}
\label{MP3_079}

\item D. Seipt, C. P. Ridgers, M. Vranic, T. Grismayer, and A. G. R. Thomas, \textit{Spin and Polarization in High-Intensity Laser-Matter Interactions}
\label{MP3_080}

\item Chris Fryer, Shane Coffing, Suzannah Wood, Chris Fontes, Todd Urbatsch, \textit{Understanding Transients: Needs from the Laboratory Astrophysics Community}
\label{MP3_081}


\item Brandon K. Russell, Marija Vranic, Paul T. Campbell, Kevin M. Schoeffler, Alexander G. R. Thomas, Dmitri Uzdensky, and Louise Willingale, \textit{Electron-only radiation dominated magnetic reconnection}
\label{MP3_083}


\item J. Fuchs, W. Yao, V. Horny, K. Burdonov, L. Gremillet, X. Davoine, E. d’Humi\`{e}res, X. Rib\`{e}yre, S.N. Chen, \textit{Enhanced laser coupling, matter heating, and particle acceleration by multiplexing PetaWatt-class lasers}
\label{MP3_085}   
   
\item B. Manuel Hegelich, Lance Labun, Ou Zhang, \textit{Performance metrics for laser-electron colliders to aid design and discovery}
\label{MP3_086}

\item James P. Edwards, Misha A. Lopez-Lopez, Christian Schubert, \textit{Photon-photon scattering in a plane-wave background}
\label{MP3_087}
   
\item G. Bruhaug, H. G. Rinderknecht, J.R. Rygg, G. Collins, Y. E, K. Garriga, X.C. Zhang, \textit{Extreme THz Pulse Generation and Interaction with Matter}
\label{MP3_088}

\item M. Lyutikov, \textit{Astrophysical analogues of laser-plasma interaction}
\label{MP3_089}

\item M. C. Marshall, D. N. Polsin, G. Bruhaug, G. W. Collins, and J. R. Rygg, \textit{Ultrafast phase transformation studies in planetary materials to terapascal pressures using x-ray and electron diffraction}
\label{MP3_090}

\item Gilbert Collins, Tom Duffy, Danae Polsin, Scott Wilks, Gianluca Gregori, Eva Zurek, \textit{Multi-Petawatt Laser Systems to Interrogate Matter at Atomic Pressures}
\label{MP3_091}

\item Henry Freund, \textit{Discussion of XFELs \& Laser-Driven FELs Petawatt Laser Initiative}
\label{MP3_092}

\item Ion Cristian Edmond Turcu, \textit{Colliding Relativistic Light-Sails accelerated by Multi-PW lasers}
\label{MP3_093}

\item Klaus Steiniger, Alexander Debus, and Ulrich Schramm, \textit{Petawatt Class Lasers for Realizing Optical Free-Electron Lasers with Traveling-Wave Thomson-Scattering}
\label{MP3_094}

\item Alexander Debus, Klaus Steiniger, Ulrich Schramm, \textit{Scaling Laser-Plasma Electron Accelerators to 100GeV-scale Energies using TWEAC and Multi-Petawatt Lasers}
\label{MP3_095}
   
\end{enumerate}



\includepdf[pages={-}]{figures/Blank_Page.pdf}
\includepdf[pages=-]{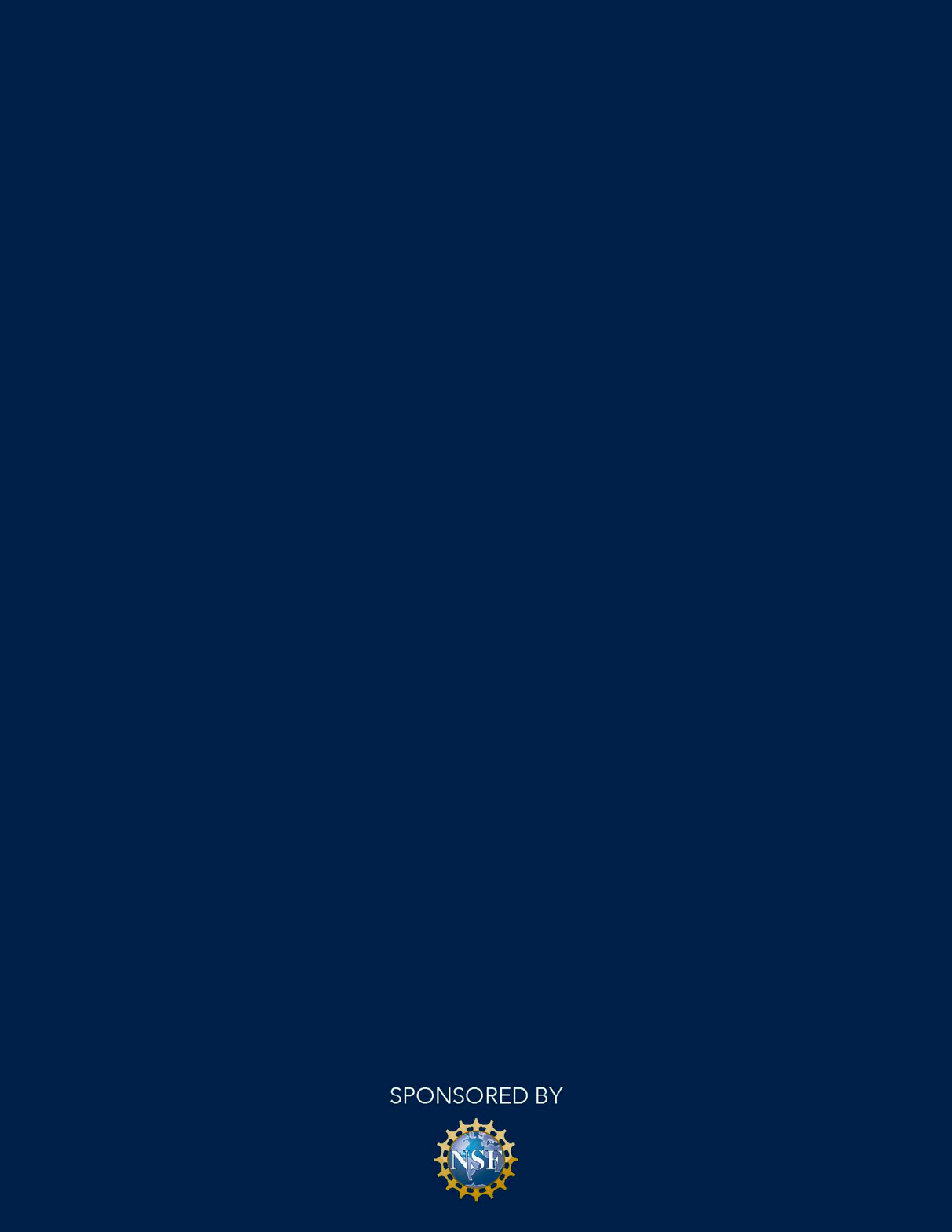}

\end{document}